\documentclass[reprint,preprintnumbers,longbibliography,onecolumn]{revtex4-2}

\usepackage{multirow}
\usepackage{siunitx}

\usepackage{tikz}
\usetikzlibrary{decorations.markings, arrows, positioning, calc}

\usepackage{graphicx,times}
\usepackage{latexsym}
\usepackage{mathtools}

\usepackage{amsmath,amssymb,amsbsy,amsfonts}
\usepackage{array}
\usepackage{bm}
\usepackage{bbold}

\usepackage{graphics}
\graphicspath{{./fig/}}

\usepackage{mathrsfs}
\usepackage{xcolor}
\usepackage{cancel}
\usepackage[normalem]{ulem}

\usepackage[unicode, pdfprintscaling=None, colorlinks]{hyperref}
\hypersetup{
    colorlinks,
    allcolors=blue!50!black,
}
\usepackage{inconsolata}
\usepackage{multirow}
\usepackage{makecell}
\usepackage{tabu}
\usepackage{tabularx}

\usepackage[capitalise]{cleveref}

\usepackage{relsize}

\usepackage{microtype}

\usepackage{xfrac}
\usepackage{orcidlink}

\usepackage{ulem} 
\usepackage{amssymb}
\usepackage[caption=false]{subfig}
\usepackage{stackengine}
\usepackage{floatrow}
\newfloatcommand{capbtabbox}{table}[][\FBheignt]

\newcommand\del\partial

\makeatletter
\def\blfootnote{\xdef\@thefnmark{}\@footnotetext}
\makeatother
\usepackage[hang,flushmargin]{footmisc} 
\usepackage{bbm}
\usepackage{textgreek}

\usepackage{xcolor}
\newcommand{\specnorm}[1]{\left|\left|#1\right|\right|}

\begin{document}

\begin{figure}
  \vskip -1.cm
  \leftline{\includegraphics[width=0.15\textwidth]{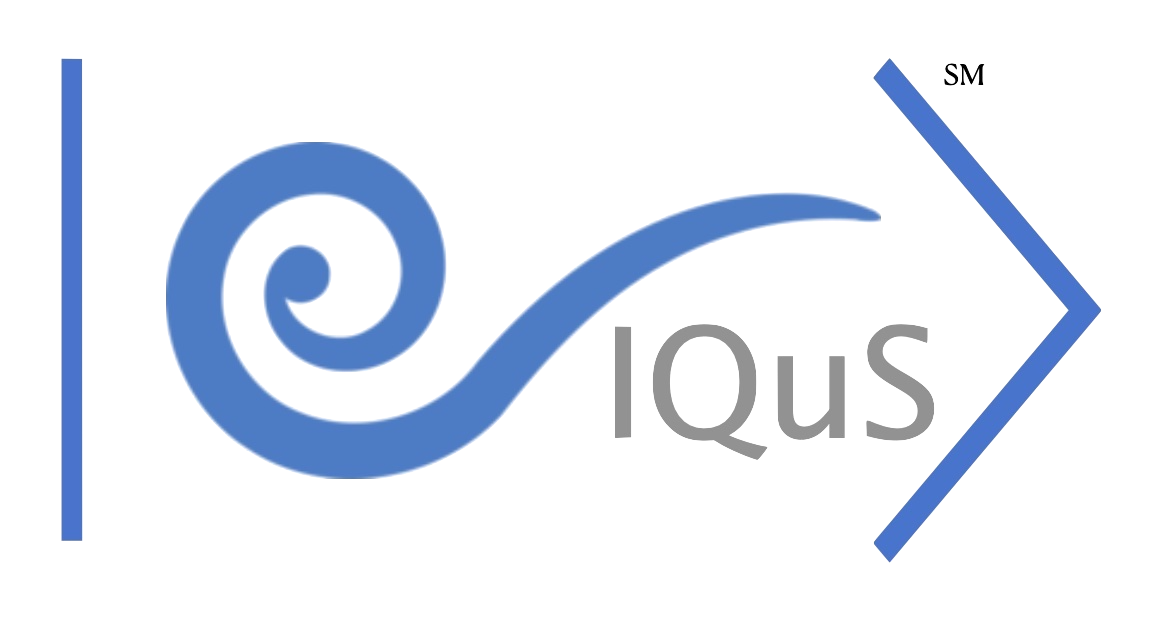}}
\end{figure}

\title{Optimization of Algorithmic Errors in Analog Quantum Simulations}

\author{Nikita A. Zemlevskiy\,\orcidlink{0000-0002-0794-2389}}
\email{zemlni@uw.edu}
\author{Henry F. Froland\,\orcidlink{0009-0008-4356-0602}}
\email{frolandh@uw.edu}
\author{Stephan Caspar\,\orcidlink{0000-0002-3658-9158}}
\email{caspar@uw.edu}
\affiliation{InQubator for Quantum Simulation (IQuS), Department of Physics, University of Washington, Seattle, WA 98195, USA.}

\preprint{IQuS@UW-21-059}
\date{\today}

\begin{abstract}
Analog quantum simulation is emerging as a powerful tool for uncovering classically unreachable physics such as many-body real-time dynamics. A complete quantification of uncertainties is necessary in order to make precise predictions using simulations on modern-day devices. Therefore, the inherent physical limitations of the device on the parameters of the simulation must be understood. This paper examines the interplay of errors arising from simulation of approximate time evolution with those due to practical, real-world device constraints. These errors are studied in Heisenberg-type systems on analog quantum devices described by the Ising Hamiltonian. A general framework for quantifying these errors is introduced and applied to several proposed time evolution methods, including Trotter-like methods and Floquet-engineered constant-field approaches.  The limitations placed on the accuracy of time evolution methods by current devices are discussed. Characterization of the scaling of coherent effects of different error sources provides a way to extend the presented Hamiltonian engineering methods to take advantage of forthcoming device capabilities.
\end{abstract}

\maketitle

\onecolumngrid

\section{Introduction}
Simulation of quantum many-body systems has been an outstanding goal of the scientific community for decades. Despite the enormous success of classical methods in determining static properties of quantum systems, they suffer from superpolynomial resource scaling for many real-time and finite-density phenomena \cite{preskill2018simulating, Jordan_2012, jordan2014quantum, jordan2019quantum, Bauer_Davoudi_Klco_Savage_2023,PhysRevD.86.105012}. Quantum computing (QC) has the potential to perform such calculations with realistic resource costs \cite{feynman_1982, benioff_1980}. A variety of noisy intermediate-scale quantum (NISQ) devices are becoming available for science, and while the applicability of these devices is limited by various factors, interesting results can be already obtained for modest calculations \cite{Bharti_2022,atas_zhang_lewis_jahanpour_haase_muschik_2021,Kim_Eddins_Anand_Wei_vandenBerg_Rosenblatt_Nayfeh_Wu_Zaletel_Temme,PhysRevD.107.054512,PhysRevD.107.054513, PhysRevD.103.094501}. In contrast to digital QC, which utilizes discrete logic gates to process information, analog QC operates by continuously modulating parameters of the experimental apparatus, e.g. a magnetic field. Analog simulation is possible in cases where the problem of interest can be mapped to a simulator's native Hamiltonian exactly, or with controllable error \cite{RevModPhys.86.153}. Hybrid digital-analog quantum simulation protocols have also been proposed \cite{Arrazola_Pedernales_Lamata_Solano_2016, Lamata_2018}. Although the analog approach is more limited than digital methods, current device capabilities show that beyond-classical results may be possible with analog simulation due to larger possible system sizes \cite{https://doi.org/10.48550/arxiv.2204.13644, scholl_schuler_williams_eberharter_barredo_schymik_lienhard_henry_lang_lahaye_et_al._2021, ebadi_wang_levine_keesling_semeghini_omran_bluvstein_samajdar_pichler_ho_et_al._2021, Daley_Bloch_Kokail_Flannigan_Pearson_Troyer_Zoller_2022}. 

Systems of many interacting spins are a common starting point for a number of the quantum simulations to date \cite{RevModPhys.93.025001, PhysRevLett.112.200501, Andrade_2022}, due to the ease of mapping physical problems to the effective Hamiltonians describing many NISQ devices. One of the most highly studied spin models is the Heisenberg model, which serves as a testing ground for condensed matter systems \cite{gong_zhu_sheng_2014, ma_dakic_naylor_zeilinger_walther_2011}, as well as a universal model for quantum computing \cite{cubitt_montanaro_piddock_2018}. Simulations of the Heisenberg model constitute a proposed step towards achieving quantum advantage \cite{childs_maslov_nam_ross_su_2018}; these simulations can be mapped to processes relevant for high-energy particle physics \cite{Nachman_2021, Bauer:2022hpo, Caspar:2022llo, 2a}, quantum gravity \cite{maldacena2023simple}, QCD processes \cite{florio2023realtime,PhysRevD.105.083020}, and other applications. Many of the analog platforms that are currently available can be naturally described by the Ising model, which has proven to be a flexible testbed for studying a broad variety of physics \cite{verresen2023quantum}, as well as a system where Heisenberg interactions can be realized \cite{https://doi.org/10.48550/arxiv.2209.09297, lukin_metrology_2020, tyler2023higherorder, zhou2023robust, Zhou:2023xnx, Geier:2021uxg, PRXQuantum.3.020303, choi_zhou_knowles_landig_choi_lukin_2020, richerme_gong_lee_senko_smith_foss-feig_michalakis_gorshkov_monroe_2014, jurcevic_lanyon_hauke_hempel_zoller_blatt_roos_2014,PhysRevA.95.013602, PhysRevA.97.023611, 1a}. Systems supporting Ising interactions include trapped ions \cite{RevModPhys.93.025001, PhysRevB.95.024431, PRXQuantum.2.020328}, neutral atoms \cite{Henriet_Beguin_Signoles_Lahaye_Browaeys_Reymond_Jurczak_2020, PRXQuantum.3.020303, Browaeys_Lahaye_2020, ebadi_wang_levine_keesling_semeghini_omran_bluvstein_samajdar_pichler_ho_et_al._2021}, nuclear spins \cite{MADI1997300, PhysRevB.75.094415}, and superconducting qubits \cite{PhysRevLett.112.200501, PhysRevX.5.021027, johnson_amin_gildert_lanting_hamze_dickson_harris_berkley_johansson_bunyk_et_al._2011}. 

Major limitations of all these devices are the numerous sources of error that accompany any particular simulation. Several error correction schemes have been proposed for analog quantum computation \cite{PhysRevLett.80.4088, PhysRevLett.119.180507}, but they have not yet been implemented on devices available today. Because of this, the ability to quantify these errors is crucial to make precise statements about a simulation. The three main sources of error that are important to consider for the purposes of simulation are the encoding error, the algorithmic error, and the hardware error \cite{PhysRevA.99.052335}. The encoding and algorithmic error come from the approximations that are made in translating physics to a form best suited for a particular device (e.g., Hilbert space truncations, commutator error for product formulas, etc.). The hardware error comes from experimental imperfections and noise on a device, which is usually uncontrolled and needs to be dealt with using error mitigation. Much work has been done examining Trotter errors \cite{Wiebe_2010,Childs_2019,childs_wiebe_2021,cai2022quantum,Endo_2018} and errors induced by analog devices \cite{Shaffer_Megidish_Broz_Chen_Häffner_2021, trivedi2022quantum, PRXQuantum.1.020308, PhysRevX.12.021049} individually, but little attention has been paid to the interplay of these different sources of error. As it turns out, there are particular choices of parameters that will minimize the cumulative error within and across these different categories.

\begin{figure}
\centering
\includegraphics[trim={1cm 3cm 2cm 0.5cm}, clip, width=\linewidth]{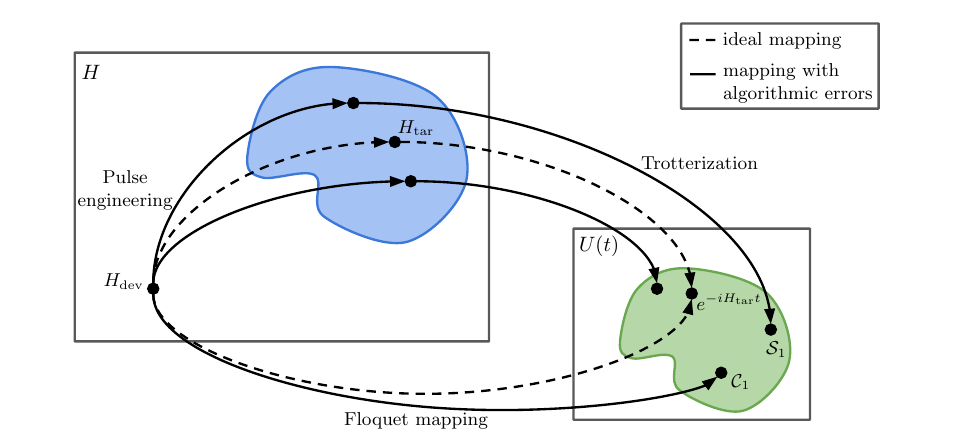}
\caption{The device Hamiltonian $H_\text{dev}$ can be used to emulate a target system of interest with Hamiltonian $H_\text{tar}$ using Trotter-like methods (top path) or continuous driving methods (bottom path). In the limit of perfect pulses (equivalently, infinite magnetic field), the dashed curves implement the target time evolution exactly. The solid curves represent the mappings that are approximate due to algorithmic errors. These errors increase the overall error when mapping between Hamiltonians due to idle errors and when implementing the time evolution operator $U(t)$ (Trotter errors). Algorithmic errors affect constant-field time evolution engineering methods as well. The constant-field method $\mathcal{C}_1$ and one of the Trotter methods $\mathcal{S}_1$ are labeled as resulting time evolution methods within the space of possible $U(t)$ operators.}
\label{fig:error_schematic}
\end{figure}

With the advent of analog quantum simulation, an understanding of uncertainties stemming from mapping physical problems and implementations of simulation algorithms is now necessary. This paper investigates the interplay of two types of algorithmic errors occurring in digital quantum computations on analog systems, Trotter errors and idle errors. Trotter errors arise due to the decomposition of the evolution operator into a finite product formula of non-commuting unitaries. Idle errors, on the other hand, stem from the inability of analog systems to turn off interactions while applying local pulses (gates). Continuously driven fields have also recently been shown to implement Heisenberg-type evolution in Ising systems \cite{1a}; such methods incur errors similar to idle errors as well. The decomposition of the algorithmic error into Trotter and idle errors is shown in Fig. \ref{fig:error_schematic}.

On a perfect device, improving precision is straightforward if these two error sources are analyzed independently. In the absence of realistic device constraints, decreasing the Trotter step size is guaranteed to decrease the Trotter error, and decreasing the pulse width on the device lowers the idle error contribution. However, these error sources play against each other on real-world machines. For example, taking many short Trotter steps will incur a large error if the idle errors are high in the device. Furthermore, taking very small Trotter steps may be impossible due to limited pulse width. Therefore, it is necessary to analyze these effects collectively to find the parameters that maximize the precision of the output from simulations on a real device. In this paper, a general framework for analyzing the error scaling of engineered time evolution methods is presented and is used to compare the performance of the aforementioned methods. We compare with numerical results for an analog simulator with (quasi-) local Ising interactions and global longitudinal and transverse-field addressing. This analysis gives the optimal choice of simulation parameters as a function of device parameters. In Section \ref{sec:engineering_methods}, the general Hamiltonian engineering methods are described, Section \ref{sec:error_analysis} introduces the types of algorithmic error studied, Section \ref{sec:results} presents the results of the error analysis, and Section \ref{sec:discussion} discusses the implications of this papers's findings.

\section{Hamiltonian Engineering Methods} \label{sec:engineering_methods}
Consider a general Hamiltonian that may be implemented in various experimental systems: 
\begin{align}
    H_\text{dev}(t) &= H_\text{idle} + H_\text{drive}(t), &
    H_\text{idle} &= \sum_{i<j} J_{ij} \left(c_xX_iX_j + c_y Y_iY_j + c_z Z_iZ_j\right), &
    H_\text{drive}(t) &= \vec{B}(t) \cdot \left( \sum_i \vec{S_i} \right). \label{eq:h_general}
\end{align}
Using carefully engineered pulse sequences of the global magnetic field $\vec{B}(t)$, the coefficients $c_{x,y,z}$ in $H_\text{idle}$ can be modified into any target Hamiltonian $H_\text{tar} = H_\text{idle}'$ in the interaction picture, obeying $c_x+c_y+c_z = c_x'+c_y'+c_z'$ and $\min c_{x,y,z} \leq c_{x,y,z}' \leq \max c_{x,y,z}$ \cite{masanes2002timeoptimal}. The targeted Hamiltonian $H_\text{tar}$ arises through dynamical decoupling of the strong magnetic fields $|\vec{B}(t)| \leq \Omega \gg \mathop{\max}J_{ij}$
\begin{align}
    H_\text{tar} &= \frac{1}{\tau} \int_0^\tau d t ~ U_\text{drive}(t)^\dag ~ H_\text{idle} ~ U_\text{drive}(t), &
    i\partial_t U_\text{drive}(t) &= H_\text{drive}(t) U_\text{drive}(t), &
    U_\text{drive}(0) &= \mathbb{1},
    \label{eq:intpicture}
\end{align}
where an average is taken over a period $\tau$ of the pulse sequence with $\vec{B}(t+\tau)=\vec{B}(t)$ and $\Omega$ is the maximum field strength that can be implemented on the device. The rest of the paper focuses on the concrete case of emulating the isotropic Heisenberg $c_{x,y,z}'=\frac{1}{3}$ evolution from Ising interactions $c_z=1,c_{x,y}=0$
\begin{align}
    H_\text{idle} = H_\text{Ising} &= \sum_{i < j} J_{ij} Z_i Z_j
    \quad \to \quad
    H_\text{tar} = H_{XXX} = \sum_{i < j} \frac{J_{ij}}{3} \left(X_i X_j + Y_i Y_j + Z_i Z_j\right). \label{eq:h_heisenberg}
\end{align}

Consider two types of pulse engineering methods, intermittent and continuous driving. Intermittent driving is inspired by Trotterized time-evolution on digital quantum computers. The strong magnetic field allows approximate global $X$, $Y$, or $Z$ $\frac{\pi}{2}$ -rotations  through short pulses $\epsilon = \frac{\pi}{2\Omega} \ll J_{ij}^{-1}$ at the maximal field strength $\Omega$
\begin{align} 
\label{eq:AnalogRotations}
\begin{split}
    R^\pm_X(\epsilon) & = \exp{\left(-i\epsilon \sum_{i<j} J_{i,j} \hat{Z}_i \hat{Z}_j \mp \frac{i \pi}{4} \sum_j \hat{X}_j\right)} , \\
    R^\pm_Y(\epsilon) & = \exp{\left(-i \epsilon \sum_{i<j} J_{i,j} \hat{Z}_i \hat{Z}_j \mp \frac{i \pi}{4} \sum_j \hat{Y}_j\right)} , \\
    R^\pm_Z(\epsilon) & = \exp{\left(-i \epsilon \sum_{i<j} J_{i,j} \hat{Z}_i \hat{Z}_j \mp \frac{i \pi}{4} \sum_j \hat{Z}_j\right)}.
\end{split}
\end{align}
Note that at infinite $\Omega$ these are exact rotation gates. These pulses effectively permute the coefficients $c_{x,y,z}$, yielding a set of global gates which can be composed in the usual Trotter-like fashion. There have been several recent developments showing Trotter-like methods of Hamiltonian engineering where similar target systems are recovered \cite{https://doi.org/10.48550/arxiv.2209.09297, lukin_metrology_2020, tyler2023higherorder, zhou2023robust, Zhou:2023xnx, Geier:2021uxg, PRXQuantum.3.020303, choi_zhou_knowles_landig_choi_lukin_2020, richerme_gong_lee_senko_smith_foss-feig_michalakis_gorshkov_monroe_2014, jurcevic_lanyon_hauke_hempel_zoller_blatt_roos_2014,PhysRevA.95.013602, PhysRevA.97.023611}. For the native Ising interactions of Eq. \eqref{eq:h_heisenberg}, there are global evolutions generated by native $XX$, $YY$ and $ZZ$ gates
\begin{align} \label{eq:AnalogGateSet}
\begin{split}
    R_{ZZ}(t) & = \exp{\left(-i t H_\text{Ising}\right)}, \\
    R_{XX}^{\pm}(t,\epsilon) &= R^{\mp}_{Y}(\epsilon)R_{ZZ}(t)R^{\pm}_{Y}(\epsilon), \\
    R_{YY}^{\pm}(t,\epsilon) &= R^{\mp}_{X}(\epsilon)R_{ZZ}(t)R^{\pm}_{X}(\epsilon).
\end{split}
\end{align}
Taking the product of these three gates defines a sequence which is referred to as ``pseudo-first order"
\begin{align} \label{eq:firstpseudo}
    \mathcal{S}_{1/2}(\vec{t},\epsilon) &= R_{ZZ}(t_z) R^{+}_{YY}(t_y,\epsilon) R^{+}_{XX}(t_x,\epsilon).
\end{align}
$\mathcal{S}$ stands for the pulse sequence and the subscript indicates the error scaling with $\epsilon$. As in digital product formulas, order is defined by the scaling of the error with the parameters of the sequence. For instance, a first-order sequence has errors that only scale like $\mathit{O}(t^2, \epsilon t, \epsilon^2)$. The ``pseudo"-modifier indicates that this sequence generates unwanted errors at order $\mathit{O}(\epsilon)$ which contribute to $H_\text{tar}$ in Eq. \eqref{eq:intpicture}. Such errors are avoidable if the alternative gate set
\begin{align} \label{eq:AltAnalogGateSet}
\begin{split}
    \tilde{R}_{XX}^{\pm}(t,\epsilon) &= R^{\pm}_{Y}(\epsilon)R_{ZZ}(t)R^{\pm}_{Y}(\epsilon), \\
    \tilde{R}_{YY}^{\pm}(t,\epsilon) &= R^{\pm}_{X}(\epsilon)R_{ZZ}(t)R^{\pm}_{X}(\epsilon),
\end{split}
\end{align}
is used to define a ``true-first order" method
\begin{align} \label{eq:firsttrue}
    \mathcal{S}_1(\vec{t},\epsilon) &= R_{ZZ}(t_z)\tilde{R}^{+}_{YY}(t_y,\epsilon)\tilde{R}^{+}_{XX}(t_x,\epsilon).
\end{align}
Applying $\frac{\pi}{2}$-pulses in the same direction twice leads to destructive interference of the error terms plaguing $\mathcal{S}_{1/2}$. This comes at the cost of a residual global $Z$ rotation by $\pi$. This overall rotation is easy to keep track of and correct for in subsequent stages of a simulation. To extend this Trotter-like approach, we present a ``pseudo-second order" sequence, $\tilde{\mathcal{S}}_1$, which carefully rearranges the gates in $\mathcal{S}_1$ to remove errors at order $\mathit{O}(t^2)$
\begin{align} \label{eq:secondpseudo}
    \tilde{\mathcal{S}}_1(\vec{t},\epsilon) &= R_{ZZ}(t_z/2)R^{+}_{X}(\epsilon)R_{ZZ}(t_y/2,\epsilon)\tilde{R}^{+}_{XX}(t_x, \epsilon) R_{ZZ}(t_y/2) R^{-}_{X}(\epsilon)R_{ZZ}(t_z/2).
\end{align}
Finally, there is the fully symmetrized version of $\mathcal{S}_1$ according to the formula $\mathcal{S}_2(\vec{t},\epsilon) = \mathcal{S}_1(\vec{t},\epsilon) \mathcal{S}_1(-\vec{t},-\epsilon)^\dagger$ \cite{suzuki_1976}. As one would expect, all errors $\mathit{O}(\epsilon^2,\epsilon t,t^2)$ vanish due to the symmetry, making this a ``true-second order" sequence
\begin{align} \label{eq:secondtrue}
    \mathcal{S}_2(\vec{t},\epsilon) &= R_{ZZ}(t_z/2) \tilde{R}^{+}_{YY}(t_y/2,\epsilon) \tilde{R}^{+}_{XX}(t_x/2,\epsilon) \tilde{R}^{-}_{XX}(t_x/2,\epsilon) \tilde{R}^{-}_{YY}(t_y/2,\epsilon) R_{ZZ}(t_z/2).
\end{align}

\begin{figure}
\centering
\includegraphics[width=0.5\linewidth]{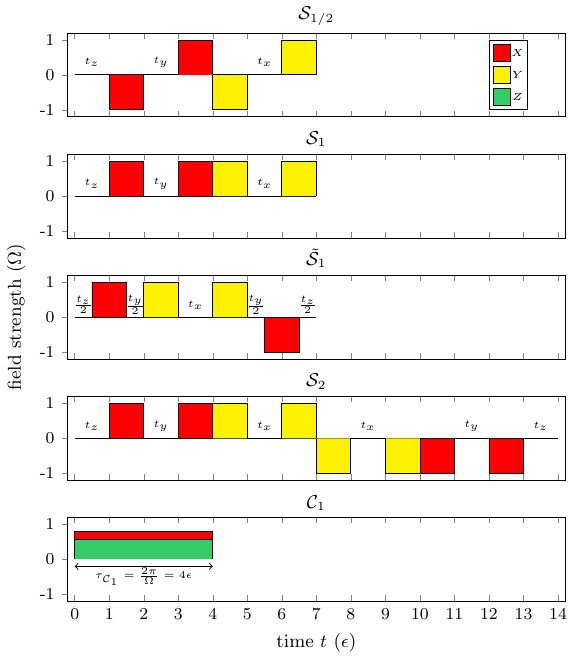}
\caption{Pulse sequences for generating Heisenberg evolution from $ZZ$ interactions, corresponding to Eqs. \eqref{eq:firstpseudo}, \eqref{eq:firsttrue} - \eqref{eq:bspiral}. For each pulse sequence, the magnetic-field strength (in units of $\Omega$) is given as a function of time (in units of $\epsilon = \frac{\pi}{2 \Omega}$). The length of each sequence $U$ is denoted by $\tau_U$. The idling times $t_i$ are shown in the graphics for each of the Trotter sequences.}
\label{fig:zz_pulse_sequences}
\end{figure}

These pulse sequences are illustrated in Fig. \ref{fig:zz_pulse_sequences}. The parameter $\vec{t} = (t_x,t_y,t_z)$ specifies the amount of time to evolve without a magnetic field in the respective frames $XX,YY,ZZ$. They obey $\tau c_i' = t_i + \mathit{O}(\epsilon)$ since the finite pulse width also contributes to these coefficients, and the $t_i$ need to be adjusted accordingly. It is important to reiterate that the sequences $\mathcal{S}_1$ and $\tilde{\mathcal{S}}_1$ reproduce evolution under $H_{XXX}$ up to a global $Z$ $\pi$-rotation, which is accounted for in this analysis. 

The constant drive field method of Refs. \cite{1a,1c} is an example of continuous driving. Similar methods have been explored in various contexts within simulations \cite{eckstein2023largescale}. This approach involves the constant driving field
\begin{equation} \label{eq:bspiral}
    \vec{B}(t) = \frac{\Omega}{\sqrt{3}}
    \begin{pmatrix}
    \sqrt{2}\\
    0\\
    1
    \end{pmatrix}.
\end{equation}
In contrast with finite pulses, this method has the effect of gradually rotating the interaction term around the Bloch sphere. It is straightforward to show that after a full period $\tau_{C_1} =  4 \epsilon = 2\pi/\Omega$, the generated $H_\text{tar}$ reproduces $H_{XXX}$ at leading-order in $\epsilon$. The error analysis for the constant field method is less involved than for the pulse sequences, since there is only one parameter $\epsilon$. The engineered Hamiltonian in Eq. \eqref{eq:intpicture} is nothing but the leading-order term in the Magnus expansion
\begin{align}
    U_\text{int}(t) &= U_\text{drive}(t)^\dag U(t) = e^{\sum_k \Omega_k(t)}, &
    \tau_{\mathcal{C}_1} H_\text{tar} &= i~\Omega_1(\tau_{\mathcal{C}_1}).
\end{align}
Due to the scaling $\Omega_k(\tau) = \mathit{O}(\tau^k)$, the expected error is of order $\mathit{O}(\tau^2)=\mathit{O}(\epsilon^2)$, which is equivalent to a first-order Trotter formula. This method is referred to as $\mathcal{C}_1$ with $\mathcal{C}$ standing for constant-field, and the subscript again indicating the error scaling with $\epsilon$. $\mathcal{C}_1$ is shown graphically in the bottom panel of Fig. \ref{fig:zz_pulse_sequences}.

The methods presented in this paper for approximating desired Hamiltonians using pulse engineering, as well as the methods described for analyzing the errors incurred, are general and can be applied to any case of Eq. \eqref{eq:h_general}. The next section describes how to analyze and quantify the errors associated with different time evolution methods.

\section{Characterization of Error Type and Error Rate}\label{sec:error_analysis}
Consider the problem of simulating evolution for a total time $T$, which is broken up into steps of length $\tau$ in physical time. As seen in Fig \ref{fig:zz_pulse_sequences}, the time it takes to implement a single step on the device varies for each sequence; these are labeled by $\tau_U$ for a pulse sequence $U$. This section analyzes the error accrued during a single step of the sequences described and shows how to optimize the length of this step to minimize the error. The contributions of different sources of error are isolated by considering a time evolution operator $U$ that is a smooth function of two parameters $t$ and $\epsilon$ that admits the following Taylor expansion:
\begin{align}\label{eq:pftaylor}
    U(t, \epsilon) &= \sum_{k_1=0}^{\infty}\sum_{k_2=0}^{\infty}t^{k_1}\epsilon^{k_2}U_{t^{k_1}\epsilon^{k_2}}, & U_{t^{k_1}\epsilon^{k_2}} = \frac{1}{k_1!k_2!}\partial_t^{k_1}\partial_{\epsilon}^{k_2}U(t,\epsilon)\big|_{t,\epsilon=0}.
\end{align}
The figure of merit that will be used to compare the error scaling of the different evolution methods is the error rate
\begin{align}\label{eq:ErrorRate}
    R_{U}(t,\epsilon) &= \frac{1}{\tau_U (t,\epsilon)}\specnorm{U(t,\epsilon)-e^{-i \tau_U (t,\epsilon) H_{tar}}}
    \leq \frac{1}{\tau_U (t,\epsilon)}\sum_{k_1=0}^{\infty}\sum_{k_2=0}^{\infty}t^{k_1}\epsilon^{k_2}\specnorm{\mathcal{E}_{U;t^{k_1}\epsilon^{k_2}}},
\end{align}
where 
\begin{equation}\label{eq:deriv_terms}
    \mathcal{E}_{U;t^{k_1}\epsilon^{k_2}} = U_{t^{k_1}\epsilon^{k_2}}-\frac{1}{k_1!k_2!}\partial_t^{k_1}\partial_{\epsilon}^{k_2}e^{-i\tau_U (t,\epsilon)H_{tar}}\big|_{t,\epsilon=0}
\end{equation}
are defined to be the error terms that contribute at $\mathit{O}(t^{k_1}, \epsilon^{k_2})$. As before, $H_\text{tar}$ describes the system to be simulated (in this paper it is $H_{XXX}$), and $\tau_U(t, \epsilon)$ is the optimal step size for a given time evolution $U$ and parameters $t$ and $\epsilon$. In principle, any norm can be used in Eq. \eqref{eq:ErrorRate}; throughout this paper the spectral norm induced by the Hilbert space two-norm is used, which is defined by: $||A|| = \mathop{\sup}\limits_{v}\frac{||Av||_2}{||v||_2}$. While the error for a single application of $U$ will always decrease with step size $\tau_U$, multiple applications may cause this error to grow unfavorably. The error accrued per step, the error rate $R_U$, is used to account for varying step size.

The different terms $\mathcal{E}_{U;t^{k_1}\epsilon^{k_2}}$ can be characterized as follows. Terms with $k_2=0$ are the standard Trotter error, as these come strictly from the algorithm used to simulate time evolution (i.e. commutator error from Trotter-Suzuki formulas). The magnitude of this type of error is controlled by $t$. Terms with $k_1=0$ are called ``idle errors" and come strictly from the inability to implement ideal single qubit rotations. For the device considered in this paper, the source of idle error is due to the persistent Ising interaction during $\frac{\pi}{2}$-pulses. The magnitude of this type of error is controlled by $\epsilon$, which is inversely related to the maximum magnetic-field strength. Terms that correspond to $k_1,k_2\neq 0$ are referred to as ``mixed errors", and these can be thought of as $\mathit{O}(\epsilon^{k_2})$ idle error that is propagated through the simulation by the $\mathit{O}(t^{k_1})$ action of the time evolution algorithm. 

In the Trotter-like digital methods there is freedom in the choice of $\tau_U$. The aforementioned sources of error can balance against one another to give overly pessimistic simulation results. This means that not all choices of $\tau_U$ are equally good, and some choices will give more error than others. To be precise, for a given time evolution method $U$, the optimal step size $\tau_U$ is split into 
\begin{align}
    \tau_U(t, \epsilon) &= 3(t_\text{rot} + t_\text{comp} + t), \label{eq:tau_u}
\end{align}
where $t_\text{rot}$ corresponds to the evolution due to the rotation gates, $t_\text{comp}$ is the compensation for unwanted evolution due to those gates, and $t$ is optional evolution time. The factor of 3 comes from mapping $H_\text{Ising} \rightarrow H_{XXX}$ in Eq. \eqref{eq:h_heisenberg}. The optimization of $\tau_U$ proceeds as follows. $t_\text{rot}$ is fixed by the sequence, and so $t_\text{comp}$ must be applied to match $H_\text{tar}$ evolution at leading-order in the expansion of Eq. \eqref{eq:pftaylor}. $t$ is then chosen to minimize the contribution of terms not canceled by $t_\text{comp}$ to $R_U$; the optimal choice of $t$ is labeled $t_*$. Here no assumptions are made about the relative smallness of $t$ and $\epsilon$, only that they are small enough to give systematically improvable expansions for each type of error such that Eq. \eqref{eq:pftaylor} may be truncated at finite-order.

The freedom in choosing $\tau_U$ for the Trotter sequences lies in the choice of $\vec{t}$ appearing in Eqs. \eqref{eq:firstpseudo}-\eqref{eq:secondtrue}. Specifically, $\vec{t}$ is chosen so that $t_i = t_{\text{comp},i} + t$\, where $t$ is the optional evolution time of Eq. \eqref{eq:tau_u} (this is expanded upon in Appendix \ref{app:first_order_error_analytics}). An important upshot of this relation is that all product formulas have a minimum value of $\tau_U$ where $t_i=0$ for some $i$ (e.g., corresponding to $\tau_{\mathcal{S}_1} = 6\epsilon$). In other words, for given experimental parameters and a given sequence, the step size $\tau_U$ cannot be made arbitrarily small. Therefore the optimization of $t$ must be constrained such that $\tau_U$ is always larger than its minimal value.

The optimal choice of $t$ (assuming this is greater than the minimal simulation time) is then given by
\begin{equation}
    t_{U*} = \mathop{\arg \min}\limits_{t_i>0} \left[R_U(t)\right]. \label{eq:step_size_formula}
\end{equation}
For the low-order product formulas considered in this paper, the solution becomes analytically tractable. Explicit expressions for these bounds are presented in subsequent sections, with derivations in Appendix \ref{app:first_order_error_analytics}. For the constant-field method $\mathcal{C}_1$, the step size is completely set by the Floquet period $\tau_{\mathcal{C}_1} = 4\epsilon$ and so this method only incurs (algorithmic) error as a function of $\epsilon$.

\begin{figure}
    \centering
    \includegraphics[scale=1]{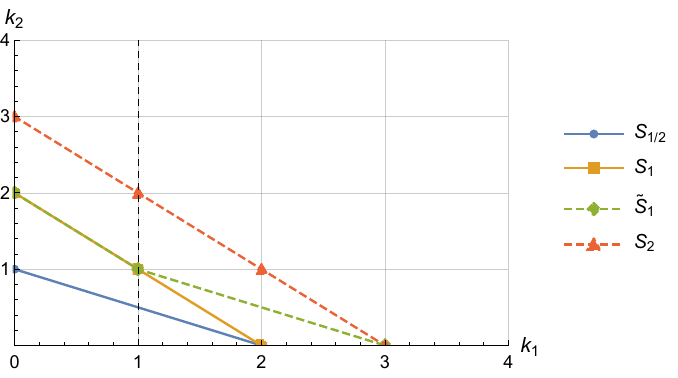}
    \caption{Different leading-order error terms contributing to the error rate analysis of the four Trotter methods considered in this paper. The optimal scaling is given by the negative slope of each line at $k_1=1$ and the corresponding error rate scaling is given by the $k_1=1$ intercept.}
    \label{fig:graphicMethod}
\end{figure}

There is a useful graphical view of the relevant error terms that makes the optimal scaling of $t$ and error rate with $\epsilon$ manifest. The error rate in Eq. \eqref{eq:ErrorRate} is expanded using Eq. \eqref{eq:pftaylor} and truncated based on the leading-order Trotter and idle errors (i.e. the lowest powers of strictly $k_2$ or $k_1$ respectively). This leaves a finite number of terms contributing to the error rate in unique powers of $t$ and $\epsilon$, which can be represented as a curve $k_2=\mathcal{B}(k_1)$ in the $k_1-k_2$ plane, as in Fig. \ref{fig:graphicMethod}. Specifically, the error rate series becomes

\begin{equation}
    R_U(t) \leq \frac{1}{\tau_U(t,\epsilon)}\left(t^{n+1}\specnorm{\mathcal{E}_{U;t^{n+1}}}+\epsilon^m\specnorm{\mathcal{E}_{U;\epsilon^{m}}}+\sum_{k=1}^{m-1}t^{\kappa}\epsilon^k\specnorm{\mathcal{E}_{U;t^{\kappa}\epsilon^{k}}}\right),\label{eq:errRateBound}
\end{equation}
where $\kappa$ is defined such that
\begin{equation}
    \specnorm{\mathcal{E}_{U;t^{\kappa}\epsilon^{k}}} \neq 0,\qquad \text{inf}\left(\specnorm{\mathcal{E}_{U;t^{\kappa-1}\epsilon^{k}}},\specnorm{\mathcal{E}_{U;t^{\kappa}\epsilon^{k-1}}}\right) = 0
\end{equation}
for some fixed $k$. By working in the asymptotic limit, which is defined as $\epsilon \ll 1$ with $t\propto\epsilon^{\alpha}$ for some number $\alpha$, the error rate scaling may be analyzed in terms of the individual powers of each of the contributing error terms. The leading-order $k_1$ error is $\alpha (k_1-1)+\mathcal{B}(k_1)$, where 1 is subtracted because $\tau_U(t,\epsilon)=\mathit{O}(t,\epsilon)$. The optimal error rate, for a given choice of $\alpha$, then scales with $\epsilon$ as 
\begin{equation}
    \mathop{\sup}\limits_{k_1} \alpha (k_1-1)+\mathcal{B}(k_1).
\end{equation}
This quantity then needs to be minimized over $\alpha$. This is a modified Legendre transform of $\mathcal{B}$, where $\alpha$ and $k_1$ form a conjugate pair. Because of this, the optimal scaling of the error rate is given by $\mathcal{B}(k_1=1)$ and the optimal scaling $\alpha$ is given by $-\left.\frac{d\mathcal{B}}{dk_1}\right|_{k_1=1}$. Comparing Fig. \ref{fig:graphicMethod} to the results below demonstrates these properties. An upshot of this analysis is that for curves whose slope at $k_1=1$ is undefined (e.g. $\tilde{\mathcal{S}}_1$), there are actually a range of optimal scalings between the slopes from the left and right side that give rise to the same optimal error rate.

\section{Results} \label{sec:results}
\subsection{Analytic Results} \label{sec:analytic_results}
This section gives the results of applying the analysis presented in the previous section to the Hamiltonian engineering methods of Section \ref{sec:engineering_methods}. Specifically, $U=\mathcal{S}_{1/2}, \mathcal{S}_1, \mathcal{C}_1$ are used to demonstrate the analysis described above, with details of the calculations contained in Appendices \ref{app:constant_field_drive_appendix} and \ref{app:first_order_error_analytics}. To isolate the contributions of different sources of error $\mathcal{E}$ to the error rate, the expansion of Eq. \eqref{eq:pftaylor} is applied to the two Trotter sequences. A bound on these contributions is then placed by separating the part of the expression depending on the geometry of the system from the operator content. The leading-order contributions to the error are found to be $\mathit{O}(N)$. These are given in Eqs. \eqref{eq:E1_tilde_eps_norm} - \eqref{eq:E1_t2_norm}.

As described in the previous section, the optimal step size $\tau_U$ is found by first choosing $t_\text{comp}$ to cancel leading-order error terms and optimizing $t$ to minimize the remaining terms at leading or next-leading order, the result of which is shown in Eqs. \eqref{eq:topt_S1t} and \eqref{eq:topt_S1}. $\tau_U$ is expected to vary for different sequences since the error made depends on the sequence used. The magnitudes of the different leading-order contributions to the error are used to find $t_*$. Subject to the constraints described in the previous section, the optimal $t_*$ values are (up to $\mathit{O}(1)$ factors determined by system geometry, see Eqs. \eqref{eq:topt_S1t_with_constraint} and \eqref{eq:topt_S1_with_constraint})
\begin{align}
    &t_{\mathcal{S}_{1/2}*} = \mathit{O}(\epsilon^{\frac{1}{2}}),
    &t_{\mathcal{S}_1*} = 0.
\label{eq:approximate_optimal_step_sizes}
\end{align}
The step size for $\mathcal{C}_1$ is fixed by $\tau_{\mathcal{C}_1}$, so there is no optimization to be done for the constant-field method. 

The error rate (Eq. \eqref{eq:ErrorRate}) is the main point of comparison between different pulse sequences. It is computed by using the individual error source contributions $\mathcal{E}_{U;\{\dots\}}$ and the optimal step sizes $t_*$ to find the error rate scaling as a function of $\epsilon$ for the sequences $\mathcal{S}_{1/2}$ and $\mathcal{S}_1$ 
\begin{align} \label{eq:approximate_Trotter_error_rates}
\begin{split}
    R_{\mathcal{S}_{1/2}} &\lesssim \epsilon^{\frac{1}{2}}\sqrt{\left(\sum_{i\neq j\neq k}J_{ik}J_{kj}\right)\left(\sum_{i\neq j}J_{ij}\right)}=\mathit{O}(\epsilon^{\frac{1}{2}} N),\\
    R_{\mathcal{S}_1} &\lesssim \epsilon \left(\sum_{i\neq j\neq k}J_{ik}J_{kj} + \sum_{i\neq j}J_{ij}^2\right)=\mathit{O}(\epsilon N).
\end{split}
\end{align}
Here $\mathit{O}(1)$ factors are again neglected for clarity. The full expressions can be found in Eqs. \eqref{eq:ER_S1t_with_constraint} and \eqref{eq:ER_S1_with_constraint}. To find the error rate for the constant-field method, the evolution implemented on the device is expanded in a Magnus series and the Heisenberg terms are subtracted off at leading-order to get a bound for $R_{\mathcal{C}_1}$. Dropping $\mathit{O}(1)$ factors, the error rate for this method is given by:
\begin{align} \label{eq:approximate_spiral_error_rate}
    R_{\mathcal{C}_1}
    &\lesssim \epsilon\left(\sum_{i\neq j\neq k}J_{ij}J_{jk}+\sum_{i\neq j}J_{ij}^2\right)=\mathit{O}(\epsilon N).
\end{align}
The error rates are found to be extensive quantities, which is not surprising: the number of terms in the Hamiltonian increases with the system size. For comparing systems of different sizes and analyzing local observables, the error rate density is useful: $\frac{R_U}{N}$.

The methods described in this paper can be used to map between device and target Hamiltonians obeying a simple relation described in Section \ref{sec:engineering_methods}. As an example of this, the setup of Ref. \cite{PRXQuantum.3.020303} can accommodate time evolution methods similar to those presented in previous sections. Their work assumes an experimental system with dipole $XX + YY$ interactions, and tunable $X$ and $Y$ magnetic fields. With this setup, the pulse sequences of Eqs. \eqref{eq:firstpseudo}, \eqref{eq:firsttrue} - \eqref{eq:secondtrue} implementing Trotterized time evolution carry over without change to emulate an isotropic Heisenberg system. In addition, if a tunable $Z$ magnetic field is assumed, the constant-field method of Eq. \eqref{eq:bspiral} works as well. Following the analysis for $R_U$, the method proposed in Ref. \cite{PRXQuantum.3.020303} has a $\mathit{O}(\sqrt{\epsilon})$ error rate scaling, similar to $\mathcal{S}_{1/2}$, with a slightly different prefactor.

There exist many sequences that have the same error scaling. Examples of this are the sequences $\mathcal{S}_1$ and $\tilde{\mathcal{S}}_1$, which are essentially reordered versions of each other. The method of Ref. \cite{PRXQuantum.3.020303} and $\mathcal{S}_{1/2}$ also both have $ER_U = \mathit{O}(\sqrt{\epsilon})$. This non-uniqueness comes from the fact that different combinations of $\pm\frac{\pi}{2}$-rotations may yield the same evolution and similar scaling but with different prefactors due to the combination of the errors incurred (see Appendix \ref{app:pi_2_pulse_error}). This paper shows low-lying sequences up to $\mathit{O}(\epsilon^2)$ in the error scaling. It is possible to extend these using the standard Trotter methods \cite{suzuki_1976} to stitch them together to create higher-order sequences. Furthermore, as the subscript on $\mathcal{C}_1$ suggests, it is possible to create higher-order variants of constant-field methods by stitching together lower-order methods in the appropriate manner, e.g., $\mathcal{C}_2(2\epsilon) = \mathcal{C}_1(\epsilon) \mathcal{C}_1(-\epsilon)$.

\subsection{Numerical Results} \label{sec:numerical_results}
The reliance of this analysis on inequalities rooted in the triangle inequality is a source of inflation of the derived upper bounds. In practice, cancellations between different terms give rise to a smaller error overall, compared to the worst-case bounds. Realistic values for the error rate can be found by evaluating Eq. \eqref{eq:ErrorRate} numerically to understand the extent to which our bounds ignore these cancellations.

For concreteness, the physical Hamiltonian underlying experimental systems of Rydberg atoms \cite{RevModPhys.82.2313, Wu_2021} is used for the rest of the paper, corresponding to $c_x = c_y = 0, c_z = 1$ in Eq. \eqref{eq:h_general}. Rydberg atoms have an all-to-all coupling mediated by a Van der Waals interaction (i.e. $J_{ij} = \frac{C_6}{|\vec{r}_i - \vec{r}_j|^6}$). This interaction only depends on the geometry of the Rydberg array which is assumed to be fixed for the duration of the simulation. In this paper, the values of $C_6 = \mathit{O}(10^6)$ \unit{MHz . \um^6 } and the lattice spacing $a = \mathit{O}(1)$ \unit{\um} are fixed to realistic values based on current hardware parameters \cite{wurtz2023aquila}. Changing these parameters is equivalent to rescaling the maximum field strength $\Omega$, so numerical calculations are performed with these constants fixed. In reality, devices require some time to change the magnetic fields; this time is determined by the slew rate. This analysis assumes the ideal case of an infinite slew rate for simplicity, although including realistic slew rates does not change the results qualitatively.

\begin{figure*}
\includegraphics[width=0.75\linewidth]{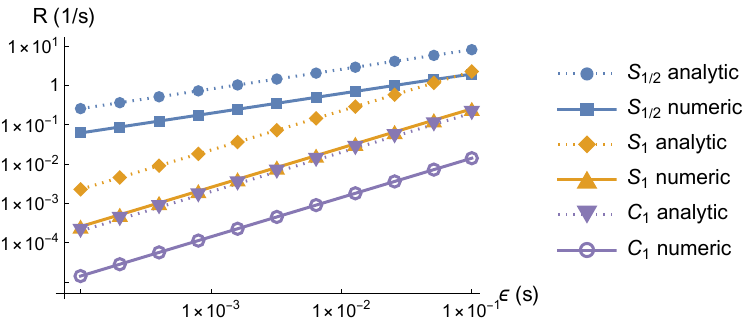}
\caption{Comparison of analytic bounds for the error rate for a single Trotter step from Eqs. \eqref{eq:approximate_Trotter_error_rates} and \eqref{eq:approximate_spiral_error_rate} to numerical results as a function of the device parameter $\epsilon$ for $2\times2$ systems. The analytic bounds consistently exceed the numerical results by factors of $\mathit{O}(1)-\mathit{O}(10)$, which is an expected consequence of using the triangle inequality to evaluate the bounds \cite{childs_wiebe_2021}.}
\label{fig:error_rates}
\end{figure*}

The individual contributions $\mathcal{E}$ are compared to the numerical values (obtained by directly evaluating the $\mathcal{E}$ terms in Eq. \eqref{eq:deriv_terms}) in Fig. \ref{fig:individual_term_scalings} as a function of system configuration. The results for the error rates of Eqs. \eqref{eq:approximate_Trotter_error_rates} and \eqref{eq:approximate_spiral_error_rate}, along with their numerical counterparts, can be seen in Fig. \ref{fig:error_rates}. In both Fig. \ref{fig:error_rates} and Fig. \ref{fig:individual_term_scalings}, we see the analytic bounds exceed the numerical values. This happens because the inequalities applied to derive the bounds effectively ignore the cancellations between various terms that happen in practice.

\begin{figure*}
\includegraphics[width=0.75\linewidth]{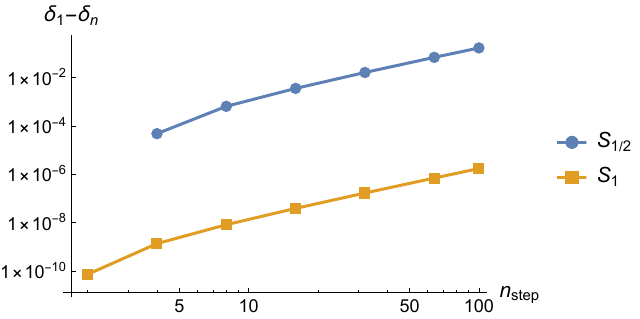}
\caption{The effect of multiple Trotter steps on errors for $2\times2$ systems with $\epsilon \sim 10^{-5}$ and pulse sequences $\mathcal{S}_{1/2}$ and $\mathcal{S}_1$. The quantity $\delta_1 - \delta_n = \specnorm{U_\text{trot}(T) - e^{-i H_{XXX} T}} - \specnorm{\prod_{n}U_\text{trot}(T/n) - e^{-i H_{XXX} T}}$ is positive when taking many small steps of $U_\text{trot}(T/n)$ replicates $e^{-iH_{XXX}T}$ better than taking one large step $U_\text{trot}(T)$. Here, $T = n \tau_U$ and $\tau_U$ is chosen optimally according to the Eq. \eqref{eq:approximate_optimal_step_sizes}. Taking many steps to reach a total time $T$ is beneficial for $\mathcal{S}_{1/2}$ and $\mathcal{S}_1$ compared to taking a single large step when the step size is chosen optimally. This demonstrates the importance of choosing the step size correctly to take advantage of the Trotter error scaling.}
\label{fig:crossovers}
\end{figure*}

So far, the error for a single Trotter step has been considered. Fig. \ref{fig:crossovers} investigates whether it is better to take many Trotter steps or a single large step by considering the quantity $\delta_1 - \delta_n = \specnorm{U_\text{trot}(T) - e^{-i H_{XXX} T}} - \specnorm{\prod_{n}U_\text{trot}(T/n) - e^{-i H_{XXX} T}}$. For a given Trotter time evolution method $U_\text{trot}$, measures how much closer $U_\text{trot}$ with $n$ steps is to the true time evolution than $U_\text{trot}$ with one large step. Considering this figure, the effect of the different error contributions on the overall error for each pulse sequence becomes evident: while decreasing $\epsilon$ will always decrease the error, the choice of the step size $\tau_U$ is of crucial importance to be able to take advantage of the scaling properties of Trotter formulas. In other words, a poorly chosen $\tau_U$ may inhibit successive application of the sequence and incur more error than a single large application as a result of constructive interference of different $\mathcal{E}$ contributions. Figs. \ref{fig:error_rates} and \ref{fig:crossovers} also demonstrate the order-of-magnitude difference between the rate at which errors accrue for $\mathcal{S}_{1/2}$ and $\mathcal{S}_1$.

\section{Discussion}\label{sec:discussion}

\begin{figure*}
\begin{floatrow}
\TopFloatBoxes
\centering
\ffigbox[][\FBheight][t]{
\caption{Numerical error rates corresponding to Eq. \eqref{eq:ErrorRate} for a single Trotter step (Floquet period for $\mathcal{C}_1$) for $2\times2$ systems. The slopes of lines on these log-log plots indicate the scaling of the error rate for each method (see Table \ref{tab:method_comparison}) and the y-intercepts give the scaling prefactor. The lines for $\mathcal{S}_{1/2}, \mathcal{S}_1$, and $\mathcal{C}_1$ are the same as the numerical ones in Fig. \ref{fig:error_rates}. Current device capabilities report pulse control to the ns level \cite{wurtz2023aquila}, meaning pulse sequences with $\epsilon \sim \mathit{O}(10^{-5})$ can be applied, indicated by the dashed vertical line.}
\label{fig:zz_norms}}{
\includegraphics[width=\linewidth]{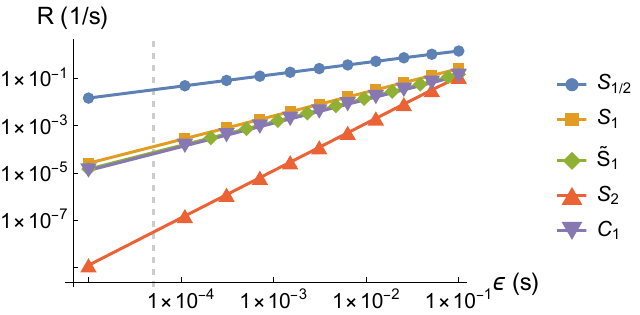}
}
\hfill
\centering
\renewcommand{\arraystretch}{1.2}
\capbtabbox[][\FBheight][t]{\caption{The error rates, step size, and step parameters for each method to simulate $H_{XXX}$ evolution. The three rightmost columns show the breakdown of $\tau_U$ into the evolution due to the rotations $t_\text{rot}$, the evolution to compensate for the rotations $t_\text{comp}$, and the optimal optional evolution time $t_*$. $t_*$ is given by Eqs. \eqref{eq:step_size_formula} and \eqref{eq:approximate_optimal_step_sizes}, and following similar analysis for the other Trotter sequences. For $\mathcal{C}_1$ the step size corresponds to the Floquet period $\tau_{\mathcal{C}_1} = 4\epsilon$. The essentially identical properties of $\mathcal{S}_1$ and $\tilde{\mathcal{S}}_1$ and the better error rate scaling of $\mathcal{S}_2$ at the cost of longer device time can be seen.}
\label{tab:method_comparison}}{
\begin{tabular}{||l|l|l|l|l|l||}
\hline
\multicolumn{6}{||c||}{Comparison of Time Evolution Methods} \\
\hline
 Method & \makecell{Error \\rate}& 
 \makecell{$\tau_U$} & \makecell{$\vec{t}_\text{rot}$} & \makecell{$\vec{t}_\text{comp}$} & \makecell{$t_*$} \\ 
\hline
\hline
$\mathcal{S}_{1/2}$ & $\mathit{O}\left(\sqrt{\epsilon}\right)$ & $\mathit{O}\left(\sqrt{\epsilon}\right)$ & $(\epsilon, \epsilon, 2\epsilon)$ & $(\epsilon, \epsilon, 0)$ & $\mathit{O}\left(\sqrt{\epsilon}\right)$ \\ 
\hline
$\mathcal{S}_1$ & $\mathit{O}(\epsilon$) & $6\epsilon$ & $(\epsilon, \epsilon, 2\epsilon)$ & $(\epsilon, \epsilon, 0)$ & 0 \\ 
\hline
$\tilde{\mathcal{S}}_1$ & $\mathit{O}(\epsilon)$ & $6\epsilon$ & $(\epsilon, 2\epsilon, \epsilon)$ & $(\epsilon, 0, \epsilon)$ & 0 \\ 
\hline
$\mathcal{S}_2$ & $\mathit{O}(\epsilon^2)$ & $12\epsilon$ & $(2\epsilon, 2\epsilon, 4\epsilon)$ & $(2\epsilon, 2\epsilon, 0)$ & 0 \\ 
\hline
$\mathcal{C}_1$ & $\mathit{O}(\epsilon)$ & $4\epsilon$ & - & - & - \\ 
\hline
\end{tabular}
\renewcommand{\arraystretch}{1}
}
\end{floatrow}
\end{figure*}

While simulation of real-life physical systems remains an outstanding goal, the onset of devices capable of supporting analog quantum simulation provides a path forward in the near-term. Because of the level of control and the large amount of qubits compared to digital quantum computers, these devices are attractive platforms for attempts at simulating classically inaccessible physics. Control and understanding of errors are important for any quantitative study. As demonstrated in this paper for the case of analog simulations, the error on real-life devices is not always minimized by trivially minimizing the sources of error individually. Instead, parameters of the simulation must be chosen specifically to maximize the precision of the simulation metrics of interest given the physical limitations of the device. This paper has outlined and shown examples of the analysis necessary to quantify the error incurred by methods of simulating Hamiltonians using analog quantum devices. Different sources of error and the interplay between them have been investigated, and the growth of these contributions with system parameters has been described. The analytic results provide worst-case guarantees for the methods presented; in practice, numerical simulations of the methods give more realistic reflections of the error. The error rate as a function of minimal $\frac{\pi}{2}$-pulse duration $\epsilon$ is used to compare the effectiveness of several time evolution methods. For methods emulating time evolution under $H_{XXX}$ starting from $H_\text{Ising}$, error rates are shown in Fig. \ref{fig:zz_norms}, and relevant features and parameters are given in Table \ref{tab:method_comparison}. Extensions of this analysis method may include using different norms, since typically only a small subset of the Hilbert space is of interest, such as low-energy effective field theory spaces \cite{_ahino_lu_2021}. Furthermore, physically-relevant observables may have additional protection from errors due to their locality \cite{doi:10.1126/sciadv.aau8342}, and properties of the physical system being simulated may combine with algorithmic and device errors to produce unexpectedly different results \cite{Chinni_2022}. Overall, the considerations presented in this paper are important for many analog quantum simulations on devices whose evolution is naturally described by Ising-type Hamiltonians such as $H_\text{dev}$. The discussion presented is an important step toward a complete quantification of uncertainties for quantum simulation on analog devices.

Current device precision capabilities may limit the ability to implement the time evolution methods that have been presented. To implement a sequence whose error grows as $\mathcal{O(\epsilon)}$, the device should have control of the pulse width at $\mathit{O}(\epsilon^2)$ or better precision. If such control is not possible, then the errors will wash out any algorithmic scaling assumed by this analysis. Because of this, it is important to consider the level of control that current devices have. Recently, nanosecond level control has been demonstrated on a Rydberg system \cite{wurtz2023aquila}. This corresponds to $\epsilon^2 \sim \mathit{O}(10^{-9})$, meaning sequences using $\epsilon \sim \mathit{O}(10^{-5})$ can be supported. This value of $\epsilon$ is shown by the dashed vertical line in Fig. \ref{fig:zz_norms}. Although constantfield methods such as $\mathcal{C}_1$ do not require rapid switching of the magnetic field, decreasing $\epsilon$ shortens the Floquet period, which makes the engineered time evolution more accurate. In other words, the level of control of the pulse width $\epsilon$ limits both types of time evolution engineering methods considered in this work. From Fig. \ref{fig:zz_norms}, it can be seen that at the level of control available today, $\mathcal{S}_2$ outperforms all other considered engineering methods. Unless the minimal pulse width $\epsilon \sim \mathit{O}(10^{-1})$, there is benefit to using higher-order formulas at the cost of rapid magnetic-field switching.

Many devices that implement Hamiltonians of the form of $H_\text{dev}$ are also able to modify the geometry of the system, such as allowing arbitrary placement of sites in the experimental plane of the apparatus using optical tweezers \cite{schlosser_ohl_demello_schaffner_preuschoff_kohfahl_birkl_2020}. The ability to run experiments with such configurations has been proposed to simulate interesting physical phenomena such as nontrivial $\theta$ dependence in non-linear $\sigma$-models \cite{Caspar:2022llo}, dynamical quantum phase transitions \cite{PhysRevLett.122.050403}, systems with topological order \cite{Samajdar_2021}, and optimizations for 2+1-dimensional gauge theory simulations \cite{müller2023simple}. These changes to the geometry would enter the error analysis in the form of modifying the sums in the bounds of Eqs. \eqref{eq:approximate_Trotter_error_rates} and \eqref{eq:approximate_spiral_error_rate}.

The interactions that can natively be supported on devices described by $H_\text{dev}$ are limited. Because of this, Hamiltonian engineering is an attractive way to emulate other models. While this paper constitutes an important step towards quantifying the errors involved in mapping a physical problem to an analog device, a collective study including other sources of error such as device imperfections and other limitations would be necessary for complete control of uncertainties of an analog quantum simulation. The dynamics of Heisenberg evolution are of broad interest in physics, and analog simulations are enabling their study today. Therefore, a full understanding of the impact of errors and approximations on analog quantum simulations of the Heisenberg model is vital in a path toward scientifically practical results. With the errors under control, data from simulations of systems beyond the reach of classical computers may be used to make predictions about nature.

\section*{Author Contributions}
HF, NZ and SC performed the theory analysis. NZ and HF performed the numerical calculations. The original idea was conceived by SC. Manuscript writing was carried out by HF, NZ, and SC. HF and NZ contributed equally.

\section*{Acknowledgments}
The authors thank Martin Savage, Anthony Ciavarella, Hersh Singh, Roland Farrell, William R. Marshall, and Murali Saravanan for helpful discussions and insightful comments. The authors would also like to acknowledge Pavel Lougovski, Peter Komar and Cedric Lin from the Amazon Braket team for contributions to this work. The views expressed are those of the authors and do not reflect the official policy or position of AWS. This work was supported in part by the U.S. Department of Energy, Office of Science, Office of Nuclear Physics, InQubator for Quantum Simulation (IQuS) \cite{iqus} under Award Number DOE (NP) Award DE-SC0020970 via the program on Quantum Horizons: QIS Research and Innovation for Nuclear Science, in part by the DOE
QuantISED program through the theory consortium “Intersections of QIS and Theoretical Particle Physics” at Fermilab
with Fermilab Subcontract No. 666484. This work was also
supported, in part, through the Department of Physics \cite{uw_phys} 
and the College of Arts and Sciences \cite{uw_artsci} at the University of
Washington. We have made extensive use of Wolfram {\tt Mathematica} \cite{Mathematica} and {\tt python} \cite{python} in this work.

\bibliography{refs}

\begin{thebibliography}{88}%
\makeatletter
\providecommand \@ifxundefined [1]{%
 \@ifx{#1\undefined}
}%
\providecommand \@ifnum [1]{%
 \ifnum #1\expandafter \@firstoftwo
 \else \expandafter \@secondoftwo
 \fi
}%
\providecommand \@ifx [1]{%
 \ifx #1\expandafter \@firstoftwo
 \else \expandafter \@secondoftwo
 \fi
}%
\providecommand \natexlab [1]{#1}%
\providecommand \enquote  [1]{``#1''}%
\providecommand \bibnamefont  [1]{#1}%
\providecommand \bibfnamefont [1]{#1}%
\providecommand \citenamefont [1]{#1}%
\providecommand \href@noop [0]{\@secondoftwo}%
\providecommand \href [0]{\begingroup \@sanitize@url \@href}%
\providecommand \@href[1]{\@@startlink{#1}\@@href}%
\providecommand \@@href[1]{\endgroup#1\@@endlink}%
\providecommand \@sanitize@url [0]{\catcode `\\12\catcode `\$12\catcode
  `\&12\catcode `\#12\catcode `\^12\catcode `\_12\catcode `\%12\relax}%
\providecommand \@@startlink[1]{}%
\providecommand \@@endlink[0]{}%
\providecommand \url  [0]{\begingroup\@sanitize@url \@url }%
\providecommand \@url [1]{\endgroup\@href {#1}{\urlprefix }}%
\providecommand \urlprefix  [0]{URL }%
\providecommand \Eprint [0]{\href }%
\providecommand \doibase [0]{https://doi.org/}%
\providecommand \selectlanguage [0]{\@gobble}%
\providecommand \bibinfo  [0]{\@secondoftwo}%
\providecommand \bibfield  [0]{\@secondoftwo}%
\providecommand \translation [1]{[#1]}%
\providecommand \BibitemOpen [0]{}%
\providecommand \bibitemStop [0]{}%
\providecommand \bibitemNoStop [0]{.\EOS\space}%
\providecommand \EOS [0]{\spacefactor3000\relax}%
\providecommand \BibitemShut  [1]{\csname bibitem#1\endcsname}%
\let\auto@bib@innerbib\@empty
\bibitem [{\citenamefont {Preskill}(2018)}]{preskill2018simulating}%
  \BibitemOpen
  \bibfield  {author} {\bibinfo {author} {\bibfnamefont {J.}~\bibnamefont
  {Preskill}},\ }\bibfield  {title} {\bibinfo {title} {{Simulating quantum
  field theory with a quantum computer}},\ }\href
  {https://doi.org/10.22323/1.334.0024} {\bibfield  {journal} {\bibinfo
  {journal} {PoS}\ }\textbf {\bibinfo {volume} {LATTICE2018}},\ \bibinfo
  {pages} {024} (\bibinfo {year} {2018})},\ \Eprint
  {https://arxiv.org/abs/1811.10085} {arXiv:1811.10085 [hep-lat]} \BibitemShut
  {NoStop}%
\bibitem [{\citenamefont {Jordan}\ \emph {et~al.}(2012)\citenamefont {Jordan},
  \citenamefont {Lee},\ and\ \citenamefont {Preskill}}]{Jordan_2012}%
  \BibitemOpen
  \bibfield  {author} {\bibinfo {author} {\bibfnamefont {S.~P.}\ \bibnamefont
  {Jordan}}, \bibinfo {author} {\bibfnamefont {K.~S.~M.}\ \bibnamefont {Lee}},\
  and\ \bibinfo {author} {\bibfnamefont {J.}~\bibnamefont {Preskill}},\
  }\bibfield  {title} {\bibinfo {title} {Quantum algorithms for quantum field
  theories},\ }\href {https://doi.org/10.1126/science.1217069} {\bibfield
  {journal} {\bibinfo  {journal} {Science}\ }\textbf {\bibinfo {volume}
  {336}},\ \bibinfo {pages} {1130} (\bibinfo {year} {2012})}\BibitemShut
  {NoStop}%
\bibitem [{\citenamefont {Jordan}\ \emph
  {et~al.}(2014{\natexlab{a}})\citenamefont {Jordan}, \citenamefont {Lee},\
  and\ \citenamefont {Preskill}}]{jordan2014quantum}%
  \BibitemOpen
  \bibfield  {author} {\bibinfo {author} {\bibfnamefont {S.~P.}\ \bibnamefont
  {Jordan}}, \bibinfo {author} {\bibfnamefont {K.~S.~M.}\ \bibnamefont {Lee}},\
  and\ \bibinfo {author} {\bibfnamefont {J.}~\bibnamefont {Preskill}},\
  }\href@noop {} {\bibinfo {title} {Quantum algorithms for fermionic quantum
  field theories}} (\bibinfo {year} {2014}{\natexlab{a}}),\ \Eprint
  {https://arxiv.org/abs/1404.7115} {arXiv:1404.7115 [hep-th]} \BibitemShut
  {NoStop}%
\bibitem [{\citenamefont {Jordan}\ \emph
  {et~al.}(2014{\natexlab{b}})\citenamefont {Jordan}, \citenamefont {Lee},\
  and\ \citenamefont {Preskill}}]{jordan2019quantum}%
  \BibitemOpen
  \bibfield  {author} {\bibinfo {author} {\bibfnamefont {S.~P.}\ \bibnamefont
  {Jordan}}, \bibinfo {author} {\bibfnamefont {K.~S.~M.}\ \bibnamefont {Lee}},\
  and\ \bibinfo {author} {\bibfnamefont {J.}~\bibnamefont {Preskill}},\
  }\bibfield  {title} {\bibinfo {title} {Quantum computation of scattering in
  scalar quantum field theories},\ }\href@noop {} {\bibfield  {journal}
  {\bibinfo  {journal} {Quantum Info. Comput.}\ }\textbf {\bibinfo {volume}
  {14}},\ \bibinfo {pages} {1014–1080} (\bibinfo {year}
  {2014}{\natexlab{b}})}\BibitemShut {NoStop}%
\bibitem [{\citenamefont {Bauer}\ \emph
  {et~al.}(2023{\natexlab{a}})\citenamefont {Bauer}, \citenamefont {Davoudi},
  \citenamefont {Klco},\ and\ \citenamefont
  {Savage}}]{Bauer_Davoudi_Klco_Savage_2023}%
  \BibitemOpen
  \bibfield  {author} {\bibinfo {author} {\bibfnamefont {C.~W.}\ \bibnamefont
  {Bauer}}, \bibinfo {author} {\bibfnamefont {Z.}~\bibnamefont {Davoudi}},
  \bibinfo {author} {\bibfnamefont {N.}~\bibnamefont {Klco}},\ and\ \bibinfo
  {author} {\bibfnamefont {M.~J.}\ \bibnamefont {Savage}},\ }\bibfield  {title}
  {\bibinfo {title} {Quantum simulation of fundamental particles and forces},\
  }\bibfield  {journal} {\bibinfo  {journal} {Nature Reviews Physics}\ }\href
  {https://doi.org/10.1038/s42254-023-00599-8} {10.1038/s42254-023-00599-8}
  (\bibinfo {year} {2023}{\natexlab{a}})\BibitemShut {NoStop}%
\bibitem [{\citenamefont {Unsal}(2012)}]{PhysRevD.86.105012}%
  \BibitemOpen
  \bibfield  {author} {\bibinfo {author} {\bibfnamefont {M.}~\bibnamefont
  {Unsal}},\ }\bibfield  {title} {\bibinfo {title} {{Theta dependence, sign
  problems and topological interference}},\ }\href
  {https://doi.org/10.1103/PhysRevD.86.105012} {\bibfield  {journal} {\bibinfo
  {journal} {Phys. Rev. D}\ }\textbf {\bibinfo {volume} {86}},\ \bibinfo
  {pages} {105012} (\bibinfo {year} {2012})},\ \Eprint
  {https://arxiv.org/abs/1201.6426} {arXiv:1201.6426 [hep-th]} \BibitemShut
  {NoStop}%
\bibitem [{\citenamefont {Feynman}(1982)}]{feynman_1982}%
  \BibitemOpen
  \bibfield  {author} {\bibinfo {author} {\bibfnamefont {R.~P.}\ \bibnamefont
  {Feynman}},\ }\bibfield  {title} {\bibinfo {title} {Simulating physics with
  computers},\ }\href {https://doi.org/10.1007/bf02650179} {\bibfield
  {journal} {\bibinfo  {journal} {International Journal of Theoretical
  Physics}\ }\textbf {\bibinfo {volume} {21}},\ \bibinfo {pages} {467}
  (\bibinfo {year} {1982})}\BibitemShut {NoStop}%
\bibitem [{\citenamefont {Benioff}(1980)}]{benioff_1980}%
  \BibitemOpen
  \bibfield  {author} {\bibinfo {author} {\bibfnamefont {P.}~\bibnamefont
  {Benioff}},\ }\bibfield  {title} {\bibinfo {title} {The computer as a
  physical system: A microscopic quantum mechanical hamiltonian model of
  computers as represented by turing machines},\ }\href
  {https://doi.org/10.1007/bf01011339} {\bibfield  {journal} {\bibinfo
  {journal} {Journal of Statistical Physics}\ }\textbf {\bibinfo {volume}
  {22}},\ \bibinfo {pages} {563–591} (\bibinfo {year} {1980})}\BibitemShut
  {NoStop}%
\bibitem [{\citenamefont {Bharti}\ \emph {et~al.}(2022)\citenamefont {Bharti}
  \emph {et~al.}}]{Bharti_2022}%
  \BibitemOpen
  \bibfield  {author} {\bibinfo {author} {\bibfnamefont {K.}~\bibnamefont
  {Bharti}} \emph {et~al.},\ }\bibfield  {title} {\bibinfo {title} {{Noisy
  intermediate-scale quantum algorithms}},\ }\href
  {https://doi.org/10.1103/RevModPhys.94.015004} {\bibfield  {journal}
  {\bibinfo  {journal} {Rev. Mod. Phys.}\ }\textbf {\bibinfo {volume} {94}},\
  \bibinfo {pages} {015004} (\bibinfo {year} {2022})},\ \Eprint
  {https://arxiv.org/abs/2101.08448} {arXiv:2101.08448 [quant-ph]} \BibitemShut
  {NoStop}%
\bibitem [{\citenamefont {Atas}\ \emph {et~al.}(2021)\citenamefont {Atas},
  \citenamefont {Zhang}, \citenamefont {Lewis}, \citenamefont {Jahanpour},
  \citenamefont {Haase},\ and\ \citenamefont
  {Muschik}}]{atas_zhang_lewis_jahanpour_haase_muschik_2021}%
  \BibitemOpen
  \bibfield  {author} {\bibinfo {author} {\bibfnamefont {Y.~Y.}\ \bibnamefont
  {Atas}}, \bibinfo {author} {\bibfnamefont {J.}~\bibnamefont {Zhang}},
  \bibinfo {author} {\bibfnamefont {R.}~\bibnamefont {Lewis}}, \bibinfo
  {author} {\bibfnamefont {A.}~\bibnamefont {Jahanpour}}, \bibinfo {author}
  {\bibfnamefont {J.~F.}\ \bibnamefont {Haase}},\ and\ \bibinfo {author}
  {\bibfnamefont {C.~A.}\ \bibnamefont {Muschik}},\ }\bibfield  {title}
  {\bibinfo {title} {{SU(2) hadrons on a quantum computer via a variational
  approach}},\ }\href {https://doi.org/10.1038/s41467-021-26825-4} {\bibfield
  {journal} {\bibinfo  {journal} {Nature Commun.}\ }\textbf {\bibinfo {volume}
  {12}},\ \bibinfo {pages} {6499} (\bibinfo {year} {2021})},\ \Eprint
  {https://arxiv.org/abs/2102.08920} {arXiv:2102.08920 [quant-ph]} \BibitemShut
  {NoStop}%
\bibitem [{\citenamefont {Kim}\ \emph {et~al.}(2023)\citenamefont {Kim},
  \citenamefont {Eddins}, \citenamefont {Anand}, \citenamefont {Wei},
  \citenamefont {van~den Berg}, \citenamefont {Rosenblatt}, \citenamefont
  {Nayfeh}, \citenamefont {Wu}, \citenamefont {Zaletel}, \citenamefont
  {Temme},\ and\ \citenamefont
  {et~al.}}]{Kim_Eddins_Anand_Wei_vandenBerg_Rosenblatt_Nayfeh_Wu_Zaletel_Temme}%
  \BibitemOpen
  \bibfield  {author} {\bibinfo {author} {\bibfnamefont {Y.}~\bibnamefont
  {Kim}}, \bibinfo {author} {\bibfnamefont {A.}~\bibnamefont {Eddins}},
  \bibinfo {author} {\bibfnamefont {S.}~\bibnamefont {Anand}}, \bibinfo
  {author} {\bibfnamefont {K.~X.}\ \bibnamefont {Wei}}, \bibinfo {author}
  {\bibfnamefont {E.}~\bibnamefont {van~den Berg}}, \bibinfo {author}
  {\bibfnamefont {S.}~\bibnamefont {Rosenblatt}}, \bibinfo {author}
  {\bibfnamefont {H.}~\bibnamefont {Nayfeh}}, \bibinfo {author} {\bibfnamefont
  {Y.}~\bibnamefont {Wu}}, \bibinfo {author} {\bibfnamefont {M.}~\bibnamefont
  {Zaletel}}, \bibinfo {author} {\bibfnamefont {K.}~\bibnamefont {Temme}},\
  and\ \bibinfo {author} {\bibnamefont {et~al.}},\ }\bibfield  {title}
  {\bibinfo {title} {Evidence for the utility of quantum computing before fault
  tolerance},\ }\href {https://doi.org/10.1038/s41586-023-06096-3} {\bibfield
  {journal} {\bibinfo  {journal} {Nature}\ }\textbf {\bibinfo {volume} {618}},\
  \bibinfo {pages} {500–505} (\bibinfo {year} {2023})}\BibitemShut {NoStop}%
\bibitem [{\citenamefont {Farrell}\ \emph
  {et~al.}(2023{\natexlab{a}})\citenamefont {Farrell}, \citenamefont
  {Chernyshev}, \citenamefont {Powell}, \citenamefont {Zemlevskiy},
  \citenamefont {Illa},\ and\ \citenamefont {Savage}}]{PhysRevD.107.054512}%
  \BibitemOpen
  \bibfield  {author} {\bibinfo {author} {\bibfnamefont {R.~C.}\ \bibnamefont
  {Farrell}}, \bibinfo {author} {\bibfnamefont {I.~A.}\ \bibnamefont
  {Chernyshev}}, \bibinfo {author} {\bibfnamefont {S.~J.~M.}\ \bibnamefont
  {Powell}}, \bibinfo {author} {\bibfnamefont {N.~A.}\ \bibnamefont
  {Zemlevskiy}}, \bibinfo {author} {\bibfnamefont {M.}~\bibnamefont {Illa}},\
  and\ \bibinfo {author} {\bibfnamefont {M.~J.}\ \bibnamefont {Savage}},\
  }\bibfield  {title} {\bibinfo {title} {{Preparations for quantum simulations
  of quantum chromodynamics in 1+1 dimensions. I. Axial gauge}},\ }\href
  {https://doi.org/10.1103/PhysRevD.107.054512} {\bibfield  {journal} {\bibinfo
   {journal} {Phys. Rev. D}\ }\textbf {\bibinfo {volume} {107}},\ \bibinfo
  {pages} {054512} (\bibinfo {year} {2023}{\natexlab{a}})},\ \Eprint
  {https://arxiv.org/abs/2207.01731} {arXiv:2207.01731 [quant-ph]} \BibitemShut
  {NoStop}%
\bibitem [{\citenamefont {Farrell}\ \emph
  {et~al.}(2023{\natexlab{b}})\citenamefont {Farrell}, \citenamefont
  {Chernyshev}, \citenamefont {Powell}, \citenamefont {Zemlevskiy},
  \citenamefont {Illa},\ and\ \citenamefont {Savage}}]{PhysRevD.107.054513}%
  \BibitemOpen
  \bibfield  {author} {\bibinfo {author} {\bibfnamefont {R.~C.}\ \bibnamefont
  {Farrell}}, \bibinfo {author} {\bibfnamefont {I.~A.}\ \bibnamefont
  {Chernyshev}}, \bibinfo {author} {\bibfnamefont {S.~J.~M.}\ \bibnamefont
  {Powell}}, \bibinfo {author} {\bibfnamefont {N.~A.}\ \bibnamefont
  {Zemlevskiy}}, \bibinfo {author} {\bibfnamefont {M.}~\bibnamefont {Illa}},\
  and\ \bibinfo {author} {\bibfnamefont {M.~J.}\ \bibnamefont {Savage}},\
  }\bibfield  {title} {\bibinfo {title} {{Preparations for quantum simulations
  of quantum chromodynamics in 1+1 dimensions. II. Single-baryon
  \ensuremath{\beta}-decay in real time}},\ }\href
  {https://doi.org/10.1103/PhysRevD.107.054513} {\bibfield  {journal} {\bibinfo
   {journal} {Phys. Rev. D}\ }\textbf {\bibinfo {volume} {107}},\ \bibinfo
  {pages} {054513} (\bibinfo {year} {2023}{\natexlab{b}})},\ \Eprint
  {https://arxiv.org/abs/2209.10781} {arXiv:2209.10781 [quant-ph]} \BibitemShut
  {NoStop}%
\bibitem [{\citenamefont {Ciavarella}\ \emph {et~al.}(2021)\citenamefont
  {Ciavarella}, \citenamefont {Klco},\ and\ \citenamefont
  {Savage}}]{PhysRevD.103.094501}%
  \BibitemOpen
  \bibfield  {author} {\bibinfo {author} {\bibfnamefont {A.}~\bibnamefont
  {Ciavarella}}, \bibinfo {author} {\bibfnamefont {N.}~\bibnamefont {Klco}},\
  and\ \bibinfo {author} {\bibfnamefont {M.~J.}\ \bibnamefont {Savage}},\
  }\bibfield  {title} {\bibinfo {title} {{Trailhead for quantum simulation of
  SU(3) Yang-Mills lattice gauge theory in the local multiplet basis}},\ }\href
  {https://doi.org/10.1103/PhysRevD.103.094501} {\bibfield  {journal} {\bibinfo
   {journal} {Phys. Rev. D}\ }\textbf {\bibinfo {volume} {103}},\ \bibinfo
  {pages} {094501} (\bibinfo {year} {2021})},\ \Eprint
  {https://arxiv.org/abs/2101.10227} {arXiv:2101.10227 [quant-ph]} \BibitemShut
  {NoStop}%
\bibitem [{\citenamefont {Georgescu}\ \emph {et~al.}(2014)\citenamefont
  {Georgescu}, \citenamefont {Ashhab},\ and\ \citenamefont
  {Nori}}]{RevModPhys.86.153}%
  \BibitemOpen
  \bibfield  {author} {\bibinfo {author} {\bibfnamefont {I.~M.}\ \bibnamefont
  {Georgescu}}, \bibinfo {author} {\bibfnamefont {S.}~\bibnamefont {Ashhab}},\
  and\ \bibinfo {author} {\bibfnamefont {F.}~\bibnamefont {Nori}},\ }\bibfield
  {title} {\bibinfo {title} {{Quantum Simulation}},\ }\href
  {https://doi.org/10.1103/RevModPhys.86.153} {\bibfield  {journal} {\bibinfo
  {journal} {Rev. Mod. Phys.}\ }\textbf {\bibinfo {volume} {86}},\ \bibinfo
  {pages} {153} (\bibinfo {year} {2014})},\ \Eprint
  {https://arxiv.org/abs/1308.6253} {arXiv:1308.6253 [quant-ph]} \BibitemShut
  {NoStop}%
\bibitem [{\citenamefont {Arrazola}\ \emph {et~al.}(2016)\citenamefont
  {Arrazola}, \citenamefont {Pedernales}, \citenamefont {Lamata},\ and\
  \citenamefont {Solano}}]{Arrazola_Pedernales_Lamata_Solano_2016}%
  \BibitemOpen
  \bibfield  {author} {\bibinfo {author} {\bibfnamefont {I.}~\bibnamefont
  {Arrazola}}, \bibinfo {author} {\bibfnamefont {J.~S.}\ \bibnamefont
  {Pedernales}}, \bibinfo {author} {\bibfnamefont {L.}~\bibnamefont {Lamata}},\
  and\ \bibinfo {author} {\bibfnamefont {E.}~\bibnamefont {Solano}},\
  }\bibfield  {title} {\bibinfo {title} {Digital-analog quantum simulation of
  spin models in trapped ions},\ }\bibfield  {journal} {\bibinfo  {journal}
  {Scientific Reports}\ }\textbf {\bibinfo {volume} {6}},\ \href
  {https://doi.org/10.1038/srep30534} {10.1038/srep30534} (\bibinfo {year}
  {2016})\BibitemShut {NoStop}%
\bibitem [{\citenamefont {Lamata}\ \emph {et~al.}(2018)\citenamefont {Lamata},
  \citenamefont {Parra-Rodriguez}, \citenamefont {Sanz},\ and\ \citenamefont
  {Solano}}]{Lamata_2018}%
  \BibitemOpen
  \bibfield  {author} {\bibinfo {author} {\bibfnamefont {L.}~\bibnamefont
  {Lamata}}, \bibinfo {author} {\bibfnamefont {A.}~\bibnamefont
  {Parra-Rodriguez}}, \bibinfo {author} {\bibfnamefont {M.}~\bibnamefont
  {Sanz}},\ and\ \bibinfo {author} {\bibfnamefont {E.}~\bibnamefont {Solano}},\
  }\bibfield  {title} {\bibinfo {title} {Digital-analog quantum simulations
  with superconducting circuits},\ }\href
  {https://doi.org/10.1080/23746149.2018.1457981} {\bibfield  {journal}
  {\bibinfo  {journal} {Advances in Physics: X}\ }\textbf {\bibinfo {volume}
  {3}},\ \bibinfo {pages} {1457981} (\bibinfo {year} {2018})}\BibitemShut
  {NoStop}%
\bibitem [{\citenamefont {Flannigan}\ \emph {et~al.}(2022)\citenamefont
  {Flannigan}, \citenamefont {Pearson}, \citenamefont {Low}, \citenamefont
  {Buyskikh}, \citenamefont {Bloch}, \citenamefont {Zoller}, \citenamefont
  {Troyer},\ and\ \citenamefont
  {Daley}}]{https://doi.org/10.48550/arxiv.2204.13644}%
  \BibitemOpen
  \bibfield  {author} {\bibinfo {author} {\bibfnamefont {S.}~\bibnamefont
  {Flannigan}}, \bibinfo {author} {\bibfnamefont {N.}~\bibnamefont {Pearson}},
  \bibinfo {author} {\bibfnamefont {G.~H.}\ \bibnamefont {Low}}, \bibinfo
  {author} {\bibfnamefont {A.}~\bibnamefont {Buyskikh}}, \bibinfo {author}
  {\bibfnamefont {I.}~\bibnamefont {Bloch}}, \bibinfo {author} {\bibfnamefont
  {P.}~\bibnamefont {Zoller}}, \bibinfo {author} {\bibfnamefont
  {M.}~\bibnamefont {Troyer}},\ and\ \bibinfo {author} {\bibfnamefont {A.~J.}\
  \bibnamefont {Daley}},\ }\bibfield  {title} {\bibinfo {title} {{Propagation
  of errors and quantitative quantum simulation with quantum advantage}},\
  }\href {https://doi.org/10.1088/2058-9565/ac88f5} {\bibfield  {journal}
  {\bibinfo  {journal} {Quantum Sci. Technol.}\ }\textbf {\bibinfo {volume}
  {7}},\ \bibinfo {pages} {045025} (\bibinfo {year} {2022})},\ \Eprint
  {https://arxiv.org/abs/2204.13644} {arXiv:2204.13644 [quant-ph]} \BibitemShut
  {NoStop}%
\bibitem [{\citenamefont {Scholl}\ \emph {et~al.}(2021)\citenamefont {Scholl},
  \citenamefont {Schuler}, \citenamefont {Williams}, \citenamefont
  {Eberharter}, \citenamefont {Barredo}, \citenamefont {Schymik}, \citenamefont
  {Lienhard}, \citenamefont {Henry}, \citenamefont {Lang}, \citenamefont
  {Lahaye},\ and\ \citenamefont
  {et~al.}}]{scholl_schuler_williams_eberharter_barredo_schymik_lienhard_henry_lang_lahaye_et_al._2021}%
  \BibitemOpen
  \bibfield  {author} {\bibinfo {author} {\bibfnamefont {P.}~\bibnamefont
  {Scholl}}, \bibinfo {author} {\bibfnamefont {M.}~\bibnamefont {Schuler}},
  \bibinfo {author} {\bibfnamefont {H.~J.}\ \bibnamefont {Williams}}, \bibinfo
  {author} {\bibfnamefont {A.~A.}\ \bibnamefont {Eberharter}}, \bibinfo
  {author} {\bibfnamefont {D.}~\bibnamefont {Barredo}}, \bibinfo {author}
  {\bibfnamefont {K.-N.}\ \bibnamefont {Schymik}}, \bibinfo {author}
  {\bibfnamefont {V.}~\bibnamefont {Lienhard}}, \bibinfo {author}
  {\bibfnamefont {L.-P.}\ \bibnamefont {Henry}}, \bibinfo {author}
  {\bibfnamefont {T.~C.}\ \bibnamefont {Lang}}, \bibinfo {author}
  {\bibfnamefont {T.}~\bibnamefont {Lahaye}},\ and\ \bibinfo {author}
  {\bibnamefont {et~al.}},\ }\bibfield  {title} {\bibinfo {title} {Quantum
  simulation of 2d antiferromagnets with hundreds of rydberg atoms},\ }\href
  {https://doi.org/10.1038/s41586-021-03585-1} {\bibfield  {journal} {\bibinfo
  {journal} {Nature}\ }\textbf {\bibinfo {volume} {595}},\ \bibinfo {pages}
  {233–238} (\bibinfo {year} {2021})}\BibitemShut {NoStop}%
\bibitem [{\citenamefont {Ebadi}\ \emph {et~al.}(2021)\citenamefont {Ebadi}
  \emph
  {et~al.}}]{ebadi_wang_levine_keesling_semeghini_omran_bluvstein_samajdar_pichler_ho_et_al._2021}%
  \BibitemOpen
  \bibfield  {author} {\bibinfo {author} {\bibfnamefont {S.}~\bibnamefont
  {Ebadi}} \emph {et~al.},\ }\bibfield  {title} {\bibinfo {title} {{Quantum
  phases of matter on a 256-atom programmable quantum simulator}},\ }\href
  {https://doi.org/10.1038/s41586-021-03582-4} {\bibfield  {journal} {\bibinfo
  {journal} {Nature}\ }\textbf {\bibinfo {volume} {595}},\ \bibinfo {pages}
  {227} (\bibinfo {year} {2021})},\ \Eprint {https://arxiv.org/abs/2012.12281}
  {arXiv:2012.12281 [quant-ph]} \BibitemShut {NoStop}%
\bibitem [{\citenamefont {Daley}\ \emph {et~al.}(2022)\citenamefont {Daley},
  \citenamefont {Bloch}, \citenamefont {Kokail}, \citenamefont {Flannigan},
  \citenamefont {Pearson}, \citenamefont {Troyer},\ and\ \citenamefont
  {Zoller}}]{Daley_Bloch_Kokail_Flannigan_Pearson_Troyer_Zoller_2022}%
  \BibitemOpen
  \bibfield  {author} {\bibinfo {author} {\bibfnamefont {A.~J.}\ \bibnamefont
  {Daley}}, \bibinfo {author} {\bibfnamefont {I.}~\bibnamefont {Bloch}},
  \bibinfo {author} {\bibfnamefont {C.}~\bibnamefont {Kokail}}, \bibinfo
  {author} {\bibfnamefont {S.}~\bibnamefont {Flannigan}}, \bibinfo {author}
  {\bibfnamefont {N.}~\bibnamefont {Pearson}}, \bibinfo {author} {\bibfnamefont
  {M.}~\bibnamefont {Troyer}},\ and\ \bibinfo {author} {\bibfnamefont
  {P.}~\bibnamefont {Zoller}},\ }\bibfield  {title} {\bibinfo {title}
  {{Practical quantum advantage in quantum simulation}},\ }\href
  {https://doi.org/10.1038/s41586-022-04940-6} {\bibfield  {journal} {\bibinfo
  {journal} {Nature}\ }\textbf {\bibinfo {volume} {607}},\ \bibinfo {pages}
  {667} (\bibinfo {year} {2022})}\BibitemShut {NoStop}%
\bibitem [{\citenamefont {Monroe}\ \emph {et~al.}(2021)\citenamefont {Monroe}
  \emph {et~al.}}]{RevModPhys.93.025001}%
  \BibitemOpen
  \bibfield  {author} {\bibinfo {author} {\bibfnamefont {C.}~\bibnamefont
  {Monroe}} \emph {et~al.},\ }\bibfield  {title} {\bibinfo {title}
  {{Programmable quantum simulations of spin systems with trapped ions}},\
  }\href {https://doi.org/10.1103/RevModPhys.93.025001} {\bibfield  {journal}
  {\bibinfo  {journal} {Rev. Mod. Phys.}\ }\textbf {\bibinfo {volume} {93}},\
  \bibinfo {pages} {025001} (\bibinfo {year} {2021})},\ \Eprint
  {https://arxiv.org/abs/1912.07845} {arXiv:1912.07845 [quant-ph]} \BibitemShut
  {NoStop}%
\bibitem [{\citenamefont {Heras}\ \emph {et~al.}(2014)\citenamefont {Heras},
  \citenamefont {Mezzacapo}, \citenamefont {Lamata}, \citenamefont {Filipp},
  \citenamefont {Wallraff},\ and\ \citenamefont
  {Solano}}]{PhysRevLett.112.200501}%
  \BibitemOpen
  \bibfield  {author} {\bibinfo {author} {\bibfnamefont {U.~L.}\ \bibnamefont
  {Heras}}, \bibinfo {author} {\bibfnamefont {A.}~\bibnamefont {Mezzacapo}},
  \bibinfo {author} {\bibfnamefont {L.}~\bibnamefont {Lamata}}, \bibinfo
  {author} {\bibfnamefont {S.}~\bibnamefont {Filipp}}, \bibinfo {author}
  {\bibfnamefont {A.}~\bibnamefont {Wallraff}},\ and\ \bibinfo {author}
  {\bibfnamefont {E.}~\bibnamefont {Solano}},\ }\bibfield  {title} {\bibinfo
  {title} {Digital quantum simulation of spin systems in superconducting
  circuits},\ }\href {https://doi.org/10.1103/PhysRevLett.112.200501}
  {\bibfield  {journal} {\bibinfo  {journal} {Phys. Rev. Lett.}\ }\textbf
  {\bibinfo {volume} {112}},\ \bibinfo {pages} {200501} (\bibinfo {year}
  {2014})}\BibitemShut {NoStop}%
\bibitem [{\citenamefont {Andrade}\ \emph {et~al.}(2022)\citenamefont
  {Andrade}, \citenamefont {Davoudi}, \citenamefont {Gra\ss{}}, \citenamefont
  {Hafezi}, \citenamefont {Pagano},\ and\ \citenamefont {Seif}}]{Andrade_2022}%
  \BibitemOpen
  \bibfield  {author} {\bibinfo {author} {\bibfnamefont {B.}~\bibnamefont
  {Andrade}}, \bibinfo {author} {\bibfnamefont {Z.}~\bibnamefont {Davoudi}},
  \bibinfo {author} {\bibfnamefont {T.}~\bibnamefont {Gra\ss{}}}, \bibinfo
  {author} {\bibfnamefont {M.}~\bibnamefont {Hafezi}}, \bibinfo {author}
  {\bibfnamefont {G.}~\bibnamefont {Pagano}},\ and\ \bibinfo {author}
  {\bibfnamefont {A.}~\bibnamefont {Seif}},\ }\bibfield  {title} {\bibinfo
  {title} {{Engineering an effective three-spin Hamiltonian in trapped-ion
  systems for applications in quantum simulation}},\ }\href
  {https://doi.org/10.1088/2058-9565/ac5f5b} {\bibfield  {journal} {\bibinfo
  {journal} {Quantum Sci. Technol.}\ }\textbf {\bibinfo {volume} {7}},\
  \bibinfo {pages} {034001} (\bibinfo {year} {2022})},\ \Eprint
  {https://arxiv.org/abs/2108.01022} {arXiv:2108.01022 [quant-ph]} \BibitemShut
  {NoStop}%
\bibitem [{\citenamefont {Gong}\ \emph {et~al.}(2014)\citenamefont {Gong},
  \citenamefont {Zhu},\ and\ \citenamefont {Sheng}}]{gong_zhu_sheng_2014}%
  \BibitemOpen
  \bibfield  {author} {\bibinfo {author} {\bibfnamefont {S.-S.}\ \bibnamefont
  {Gong}}, \bibinfo {author} {\bibfnamefont {W.}~\bibnamefont {Zhu}},\ and\
  \bibinfo {author} {\bibfnamefont {D.~N.}\ \bibnamefont {Sheng}},\ }\bibfield
  {title} {\bibinfo {title} {Emergent chiral spin liquid: Fractional quantum
  hall effect in a kagome heisenberg model},\ }\bibfield  {journal} {\bibinfo
  {journal} {Scientific Reports}\ }\textbf {\bibinfo {volume} {4}},\ \href
  {https://doi.org/10.1038/srep06317} {10.1038/srep06317} (\bibinfo {year}
  {2014})\BibitemShut {NoStop}%
\bibitem [{\citenamefont {Ma}\ \emph {et~al.}(2011)\citenamefont {Ma},
  \citenamefont {Dakic}, \citenamefont {Naylor}, \citenamefont {Zeilinger},\
  and\ \citenamefont {Walther}}]{ma_dakic_naylor_zeilinger_walther_2011}%
  \BibitemOpen
  \bibfield  {author} {\bibinfo {author} {\bibfnamefont {X.-s.}\ \bibnamefont
  {Ma}}, \bibinfo {author} {\bibfnamefont {B.}~\bibnamefont {Dakic}}, \bibinfo
  {author} {\bibfnamefont {W.}~\bibnamefont {Naylor}}, \bibinfo {author}
  {\bibfnamefont {A.}~\bibnamefont {Zeilinger}},\ and\ \bibinfo {author}
  {\bibfnamefont {P.}~\bibnamefont {Walther}},\ }\bibfield  {title} {\bibinfo
  {title} {Quantum simulation of the wavefunction to probe frustrated
  heisenberg spin systems},\ }\href {https://doi.org/10.1038/nphys1919}
  {\bibfield  {journal} {\bibinfo  {journal} {Nature Physics}\ }\textbf
  {\bibinfo {volume} {7}},\ \bibinfo {pages} {399–405} (\bibinfo {year}
  {2011})}\BibitemShut {NoStop}%
\bibitem [{\citenamefont {Cubitt}\ \emph {et~al.}(2018)\citenamefont {Cubitt},
  \citenamefont {Montanaro},\ and\ \citenamefont
  {Piddock}}]{cubitt_montanaro_piddock_2018}%
  \BibitemOpen
  \bibfield  {author} {\bibinfo {author} {\bibfnamefont {T.~S.}\ \bibnamefont
  {Cubitt}}, \bibinfo {author} {\bibfnamefont {A.}~\bibnamefont {Montanaro}},\
  and\ \bibinfo {author} {\bibfnamefont {S.}~\bibnamefont {Piddock}},\
  }\bibfield  {title} {\bibinfo {title} {Universal quantum hamiltonians},\
  }\href {https://doi.org/10.1073/pnas.1804949115} {\bibfield  {journal}
  {\bibinfo  {journal} {Proceedings of the National Academy of Sciences}\
  }\textbf {\bibinfo {volume} {115}},\ \bibinfo {pages} {9497–9502} (\bibinfo
  {year} {2018})}\BibitemShut {NoStop}%
\bibitem [{\citenamefont {Childs}\ \emph {et~al.}(2018)\citenamefont {Childs},
  \citenamefont {Maslov}, \citenamefont {Nam}, \citenamefont {Ross},\ and\
  \citenamefont {Su}}]{childs_maslov_nam_ross_su_2018}%
  \BibitemOpen
  \bibfield  {author} {\bibinfo {author} {\bibfnamefont {A.~M.}\ \bibnamefont
  {Childs}}, \bibinfo {author} {\bibfnamefont {D.}~\bibnamefont {Maslov}},
  \bibinfo {author} {\bibfnamefont {Y.}~\bibnamefont {Nam}}, \bibinfo {author}
  {\bibfnamefont {N.~J.}\ \bibnamefont {Ross}},\ and\ \bibinfo {author}
  {\bibfnamefont {Y.}~\bibnamefont {Su}},\ }\bibfield  {title} {\bibinfo
  {title} {Toward the first quantum simulation with quantum speedup},\ }\href
  {https://doi.org/10.1073/pnas.1801723115} {\bibfield  {journal} {\bibinfo
  {journal} {Proceedings of the National Academy of Sciences}\ }\textbf
  {\bibinfo {volume} {115}},\ \bibinfo {pages} {9456–9461} (\bibinfo {year}
  {2018})}\BibitemShut {NoStop}%
\bibitem [{\citenamefont {Bauer}\ \emph {et~al.}(2021)\citenamefont {Bauer},
  \citenamefont {de~Jong}, \citenamefont {Nachman},\ and\ \citenamefont
  {Provasoli}}]{Nachman_2021}%
  \BibitemOpen
  \bibfield  {author} {\bibinfo {author} {\bibfnamefont {C.~W.}\ \bibnamefont
  {Bauer}}, \bibinfo {author} {\bibfnamefont {W.~A.}\ \bibnamefont {de~Jong}},
  \bibinfo {author} {\bibfnamefont {B.}~\bibnamefont {Nachman}},\ and\ \bibinfo
  {author} {\bibfnamefont {D.}~\bibnamefont {Provasoli}},\ }\bibfield  {title}
  {\bibinfo {title} {{Quantum Algorithm for High Energy Physics Simulations}},\
  }\href {https://doi.org/10.1103/PhysRevLett.126.062001} {\bibfield  {journal}
  {\bibinfo  {journal} {Phys. Rev. Lett.}\ }\textbf {\bibinfo {volume} {126}},\
  \bibinfo {pages} {062001} (\bibinfo {year} {2021})},\ \Eprint
  {https://arxiv.org/abs/1904.03196} {arXiv:1904.03196 [hep-ph]} \BibitemShut
  {NoStop}%
\bibitem [{\citenamefont {Bauer}\ \emph
  {et~al.}(2023{\natexlab{b}})\citenamefont {Bauer} \emph
  {et~al.}}]{Bauer:2022hpo}%
  \BibitemOpen
  \bibfield  {author} {\bibinfo {author} {\bibfnamefont {C.~W.}\ \bibnamefont
  {Bauer}} \emph {et~al.},\ }\bibfield  {title} {\bibinfo {title} {{Quantum
  Simulation for High-Energy Physics}},\ }\href
  {https://doi.org/10.1103/PRXQuantum.4.027001} {\bibfield  {journal} {\bibinfo
   {journal} {PRX Quantum}\ }\textbf {\bibinfo {volume} {4}},\ \bibinfo {pages}
  {027001} (\bibinfo {year} {2023}{\natexlab{b}})},\ \Eprint
  {https://arxiv.org/abs/2204.03381} {arXiv:2204.03381 [quant-ph]} \BibitemShut
  {NoStop}%
\bibitem [{\citenamefont {Caspar}\ and\ \citenamefont
  {Singh}(2022)}]{Caspar:2022llo}%
  \BibitemOpen
  \bibfield  {author} {\bibinfo {author} {\bibfnamefont {S.}~\bibnamefont
  {Caspar}}\ and\ \bibinfo {author} {\bibfnamefont {H.}~\bibnamefont {Singh}},\
  }\bibfield  {title} {\bibinfo {title} {{From Asymptotic Freedom to
  \ensuremath{\theta} Vacua: Qubit Embeddings of the O(3) Nonlinear
  \ensuremath{\sigma} Model}},\ }\href
  {https://doi.org/10.1103/PhysRevLett.129.022003} {\bibfield  {journal}
  {\bibinfo  {journal} {Phys. Rev. Lett.}\ }\textbf {\bibinfo {volume} {129}},\
  \bibinfo {pages} {022003} (\bibinfo {year} {2022})},\ \Eprint
  {https://arxiv.org/abs/2203.15766} {arXiv:2203.15766 [hep-lat]} \BibitemShut
  {NoStop}%
\bibitem [{\citenamefont {Ciavarella}\ \emph
  {et~al.}(2023{\natexlab{a}})\citenamefont {Ciavarella}, \citenamefont
  {Caspar}, \citenamefont {Singh},\ and\ \citenamefont {Savage}}]{2a}%
  \BibitemOpen
  \bibfield  {author} {\bibinfo {author} {\bibfnamefont {A.~N.}\ \bibnamefont
  {Ciavarella}}, \bibinfo {author} {\bibfnamefont {S.}~\bibnamefont {Caspar}},
  \bibinfo {author} {\bibfnamefont {H.}~\bibnamefont {Singh}},\ and\ \bibinfo
  {author} {\bibfnamefont {M.~J.}\ \bibnamefont {Savage}},\ }\bibfield  {title}
  {\bibinfo {title} {{Preparation for quantum simulation of the
  (1+1)-dimensional O(3) nonlinear \ensuremath{\sigma} model using cold
  atoms}},\ }\href {https://doi.org/10.1103/PhysRevA.107.042404} {\bibfield
  {journal} {\bibinfo  {journal} {Phys. Rev. A}\ }\textbf {\bibinfo {volume}
  {107}},\ \bibinfo {pages} {042404} (\bibinfo {year} {2023}{\natexlab{a}})},\
  \Eprint {https://arxiv.org/abs/2211.07684} {arXiv:2211.07684 [quant-ph]}
  \BibitemShut {NoStop}%
\bibitem [{\citenamefont {Maldacena}(2023)}]{maldacena2023simple}%
  \BibitemOpen
  \bibfield  {author} {\bibinfo {author} {\bibfnamefont {J.}~\bibnamefont
  {Maldacena}},\ }\href@noop {} {\bibinfo {title} {A simple quantum system that
  describes a black hole}} (\bibinfo {year} {2023}),\ \Eprint
  {https://arxiv.org/abs/2303.11534} {arXiv:2303.11534 [hep-th]} \BibitemShut
  {NoStop}%
\bibitem [{\citenamefont {Florio}\ \emph {et~al.}(2023)\citenamefont {Florio},
  \citenamefont {Frenklakh}, \citenamefont {Ikeda}, \citenamefont {Kharzeev},
  \citenamefont {Korepin}, \citenamefont {Shi},\ and\ \citenamefont
  {Yu}}]{florio2023realtime}%
  \BibitemOpen
  \bibfield  {author} {\bibinfo {author} {\bibfnamefont {A.}~\bibnamefont
  {Florio}}, \bibinfo {author} {\bibfnamefont {D.}~\bibnamefont {Frenklakh}},
  \bibinfo {author} {\bibfnamefont {K.}~\bibnamefont {Ikeda}}, \bibinfo
  {author} {\bibfnamefont {D.}~\bibnamefont {Kharzeev}}, \bibinfo {author}
  {\bibfnamefont {V.}~\bibnamefont {Korepin}}, \bibinfo {author} {\bibfnamefont
  {S.}~\bibnamefont {Shi}},\ and\ \bibinfo {author} {\bibfnamefont
  {K.}~\bibnamefont {Yu}},\ }\bibfield  {title} {\bibinfo {title} {{Real-Time
  Nonperturbative Dynamics of Jet Production in Schwinger Model: Quantum
  Entanglement and Vacuum Modification}},\ }\href
  {https://doi.org/10.1103/PhysRevLett.131.021902} {\bibfield  {journal}
  {\bibinfo  {journal} {Phys. Rev. Lett.}\ }\textbf {\bibinfo {volume} {131}},\
  \bibinfo {pages} {021902} (\bibinfo {year} {2023})},\ \Eprint
  {https://arxiv.org/abs/2301.11991} {arXiv:2301.11991 [hep-ph]} \BibitemShut
  {NoStop}%
\bibitem [{\citenamefont {Martin}\ \emph {et~al.}(2022)\citenamefont {Martin},
  \citenamefont {Roggero}, \citenamefont {Duan}, \citenamefont {Carlson},\ and\
  \citenamefont {Cirigliano}}]{PhysRevD.105.083020}%
  \BibitemOpen
  \bibfield  {author} {\bibinfo {author} {\bibfnamefont {J.~D.}\ \bibnamefont
  {Martin}}, \bibinfo {author} {\bibfnamefont {A.}~\bibnamefont {Roggero}},
  \bibinfo {author} {\bibfnamefont {H.}~\bibnamefont {Duan}}, \bibinfo {author}
  {\bibfnamefont {J.}~\bibnamefont {Carlson}},\ and\ \bibinfo {author}
  {\bibfnamefont {V.}~\bibnamefont {Cirigliano}},\ }\bibfield  {title}
  {\bibinfo {title} {{Classical and quantum evolution in a simple coherent
  neutrino problem}},\ }\href {https://doi.org/10.1103/PhysRevD.105.083020}
  {\bibfield  {journal} {\bibinfo  {journal} {Phys. Rev. D}\ }\textbf {\bibinfo
  {volume} {105}},\ \bibinfo {pages} {083020} (\bibinfo {year} {2022})},\
  \Eprint {https://arxiv.org/abs/2112.12686} {arXiv:2112.12686 [hep-ph]}
  \BibitemShut {NoStop}%
\bibitem [{\citenamefont {Verresen}(2023)}]{verresen2023quantum}%
  \BibitemOpen
  \bibfield  {author} {\bibinfo {author} {\bibfnamefont {R.}~\bibnamefont
  {Verresen}},\ }\href@noop {} {\bibinfo {title} {Everything is a quantum ising
  model}} (\bibinfo {year} {2023}),\ \Eprint {https://arxiv.org/abs/2301.11917}
  {arXiv:2301.11917 [quant-ph]} \BibitemShut {NoStop}%
\bibitem [{\citenamefont {Martin}\ \emph {et~al.}(2023)\citenamefont {Martin}
  \emph {et~al.}}]{https://doi.org/10.48550/arxiv.2209.09297}%
  \BibitemOpen
  \bibfield  {author} {\bibinfo {author} {\bibfnamefont {L.~S.}\ \bibnamefont
  {Martin}} \emph {et~al.},\ }\bibfield  {title} {\bibinfo {title}
  {{Controlling Local Thermalization Dynamics in a Floquet-Engineered Dipolar
  Ensemble}},\ }\href {https://doi.org/10.1103/PhysRevLett.130.210403}
  {\bibfield  {journal} {\bibinfo  {journal} {Phys. Rev. Lett.}\ }\textbf
  {\bibinfo {volume} {130}},\ \bibinfo {pages} {210403} (\bibinfo {year}
  {2023})},\ \Eprint {https://arxiv.org/abs/2209.09297} {arXiv:2209.09297
  [quant-ph]} \BibitemShut {NoStop}%
\bibitem [{\citenamefont {Zhou}\ \emph {et~al.}(2020)\citenamefont {Zhou},
  \citenamefont {Choi}, \citenamefont {Choi}, \citenamefont {Landig},
  \citenamefont {Douglas}, \citenamefont {Isoya}, \citenamefont {Jelezko},
  \citenamefont {Onoda}, \citenamefont {Sumiya}, \citenamefont {Cappellaro},\
  and\ \citenamefont {et~al.}}]{lukin_metrology_2020}%
  \BibitemOpen
  \bibfield  {author} {\bibinfo {author} {\bibfnamefont {H.}~\bibnamefont
  {Zhou}}, \bibinfo {author} {\bibfnamefont {J.}~\bibnamefont {Choi}}, \bibinfo
  {author} {\bibfnamefont {S.}~\bibnamefont {Choi}}, \bibinfo {author}
  {\bibfnamefont {R.}~\bibnamefont {Landig}}, \bibinfo {author} {\bibfnamefont
  {A.~M.}\ \bibnamefont {Douglas}}, \bibinfo {author} {\bibfnamefont
  {J.}~\bibnamefont {Isoya}}, \bibinfo {author} {\bibfnamefont
  {F.}~\bibnamefont {Jelezko}}, \bibinfo {author} {\bibfnamefont
  {S.}~\bibnamefont {Onoda}}, \bibinfo {author} {\bibfnamefont
  {H.}~\bibnamefont {Sumiya}}, \bibinfo {author} {\bibfnamefont
  {P.}~\bibnamefont {Cappellaro}},\ and\ \bibinfo {author} {\bibnamefont
  {et~al.}},\ }\bibfield  {title} {\bibinfo {title} {Quantum metrology with
  strongly interacting spin systems},\ }\bibfield  {journal} {\bibinfo
  {journal} {Physical Review X}\ }\textbf {\bibinfo {volume} {10}},\ \href
  {https://doi.org/10.1103/physrevx.10.031003} {10.1103/physrevx.10.031003}
  (\bibinfo {year} {2020})\BibitemShut {NoStop}%
\bibitem [{\citenamefont {Tyler}\ \emph {et~al.}(2023)\citenamefont {Tyler},
  \citenamefont {Zhou}, \citenamefont {Martin}, \citenamefont {Leitao},\ and\
  \citenamefont {Lukin}}]{tyler2023higherorder}%
  \BibitemOpen
  \bibfield  {author} {\bibinfo {author} {\bibfnamefont {M.}~\bibnamefont
  {Tyler}}, \bibinfo {author} {\bibfnamefont {H.}~\bibnamefont {Zhou}},
  \bibinfo {author} {\bibfnamefont {L.~S.}\ \bibnamefont {Martin}}, \bibinfo
  {author} {\bibfnamefont {N.}~\bibnamefont {Leitao}},\ and\ \bibinfo {author}
  {\bibfnamefont {M.~D.}\ \bibnamefont {Lukin}},\ }\bibfield  {title} {\bibinfo
  {title} {{Higher-Order Methods for Hamiltonian Engineering Pulse Sequence
  Design}},\ }\href@noop {} {\  (\bibinfo {year} {2023})},\ \Eprint
  {https://arxiv.org/abs/2303.07374} {arXiv:2303.07374 [quant-ph]} \BibitemShut
  {NoStop}%
\bibitem [{\citenamefont {Zhou}\ \emph
  {et~al.}(2023{\natexlab{a}})\citenamefont {Zhou}, \citenamefont {Martin},
  \citenamefont {Tyler}, \citenamefont {Makarova}, \citenamefont {Leitao},
  \citenamefont {Park},\ and\ \citenamefont {Lukin}}]{zhou2023robust}%
  \BibitemOpen
  \bibfield  {author} {\bibinfo {author} {\bibfnamefont {H.}~\bibnamefont
  {Zhou}}, \bibinfo {author} {\bibfnamefont {L.~S.}\ \bibnamefont {Martin}},
  \bibinfo {author} {\bibfnamefont {M.}~\bibnamefont {Tyler}}, \bibinfo
  {author} {\bibfnamefont {O.}~\bibnamefont {Makarova}}, \bibinfo {author}
  {\bibfnamefont {N.}~\bibnamefont {Leitao}}, \bibinfo {author} {\bibfnamefont
  {H.}~\bibnamefont {Park}},\ and\ \bibinfo {author} {\bibfnamefont {M.~D.}\
  \bibnamefont {Lukin}},\ }\href@noop {} {\bibinfo {title} {Robust higher-order
  hamiltonian engineering for quantum sensing with strongly interacting
  systems}} (\bibinfo {year} {2023}{\natexlab{a}}),\ \Eprint
  {https://arxiv.org/abs/2303.07363} {arXiv:2303.07363 [quant-ph]} \BibitemShut
  {NoStop}%
\bibitem [{\citenamefont {Zhou}\ \emph
  {et~al.}(2023{\natexlab{b}})\citenamefont {Zhou}, \citenamefont {Gao},
  \citenamefont {Leitao}, \citenamefont {Makarova}, \citenamefont {Cong},
  \citenamefont {Douglas}, \citenamefont {Martin},\ and\ \citenamefont
  {Lukin}}]{Zhou:2023xnx}%
  \BibitemOpen
  \bibfield  {author} {\bibinfo {author} {\bibfnamefont {H.}~\bibnamefont
  {Zhou}}, \bibinfo {author} {\bibfnamefont {H.}~\bibnamefont {Gao}}, \bibinfo
  {author} {\bibfnamefont {N.~T.}\ \bibnamefont {Leitao}}, \bibinfo {author}
  {\bibfnamefont {O.}~\bibnamefont {Makarova}}, \bibinfo {author}
  {\bibfnamefont {I.}~\bibnamefont {Cong}}, \bibinfo {author} {\bibfnamefont
  {A.~M.}\ \bibnamefont {Douglas}}, \bibinfo {author} {\bibfnamefont {L.~S.}\
  \bibnamefont {Martin}},\ and\ \bibinfo {author} {\bibfnamefont {M.~D.}\
  \bibnamefont {Lukin}},\ }\bibfield  {title} {\bibinfo {title} {{Robust
  Hamiltonian Engineering for Interacting Qudit Systems}},\ }\href@noop {} {\
  (\bibinfo {year} {2023}{\natexlab{b}})},\ \Eprint
  {https://arxiv.org/abs/2305.09757} {arXiv:2305.09757 [quant-ph]} \BibitemShut
  {NoStop}%
\bibitem [{\citenamefont {Geier}\ \emph {et~al.}(2021)\citenamefont {Geier},
  \citenamefont {Thaicharoen}, \citenamefont {Hainaut}, \citenamefont {Franz},
  \citenamefont {Salzinger}, \citenamefont {Tebben}, \citenamefont
  {Grimshandl}, \citenamefont {Zürn},\ and\ \citenamefont
  {Weidemüller}}]{Geier:2021uxg}%
  \BibitemOpen
  \bibfield  {author} {\bibinfo {author} {\bibfnamefont {S.}~\bibnamefont
  {Geier}}, \bibinfo {author} {\bibfnamefont {N.}~\bibnamefont {Thaicharoen}},
  \bibinfo {author} {\bibfnamefont {C.}~\bibnamefont {Hainaut}}, \bibinfo
  {author} {\bibfnamefont {T.}~\bibnamefont {Franz}}, \bibinfo {author}
  {\bibfnamefont {A.}~\bibnamefont {Salzinger}}, \bibinfo {author}
  {\bibfnamefont {A.}~\bibnamefont {Tebben}}, \bibinfo {author} {\bibfnamefont
  {D.}~\bibnamefont {Grimshandl}}, \bibinfo {author} {\bibfnamefont
  {G.}~\bibnamefont {Zürn}},\ and\ \bibinfo {author} {\bibfnamefont
  {M.}~\bibnamefont {Weidemüller}},\ }\bibfield  {title} {\bibinfo {title}
  {Floquet hamiltonian engineering of an isolated many-body spin system},\
  }\href {https://doi.org/10.1126/science.abd9547} {\bibfield  {journal}
  {\bibinfo  {journal} {Science}\ }\textbf {\bibinfo {volume} {374}},\ \bibinfo
  {pages} {1149} (\bibinfo {year} {2021})},\ \Eprint
  {https://arxiv.org/abs/https://www.science.org/doi/pdf/10.1126/science.abd9547}
  {https://www.science.org/doi/pdf/10.1126/science.abd9547} \BibitemShut
  {NoStop}%
\bibitem [{\citenamefont {Scholl}\ \emph {et~al.}(2022)\citenamefont {Scholl}
  \emph {et~al.}}]{PRXQuantum.3.020303}%
  \BibitemOpen
  \bibfield  {author} {\bibinfo {author} {\bibfnamefont {P.}~\bibnamefont
  {Scholl}} \emph {et~al.},\ }\bibfield  {title} {\bibinfo {title} {{Microwave
  Engineering of Programmable XXZ Hamiltonians in Arrays of Rydberg Atoms}},\
  }\href {https://doi.org/10.1103/PRXQuantum.3.020303} {\bibfield  {journal}
  {\bibinfo  {journal} {PRX Quantum}\ }\textbf {\bibinfo {volume} {3}},\
  \bibinfo {pages} {020303} (\bibinfo {year} {2022})},\ \Eprint
  {https://arxiv.org/abs/2107.14459} {arXiv:2107.14459 [quant-ph]} \BibitemShut
  {NoStop}%
\bibitem [{\citenamefont {Choi}\ \emph {et~al.}(2020)\citenamefont {Choi},
  \citenamefont {Zhou}, \citenamefont {Knowles}, \citenamefont {Landig},
  \citenamefont {Choi},\ and\ \citenamefont
  {Lukin}}]{choi_zhou_knowles_landig_choi_lukin_2020}%
  \BibitemOpen
  \bibfield  {author} {\bibinfo {author} {\bibfnamefont {J.}~\bibnamefont
  {Choi}}, \bibinfo {author} {\bibfnamefont {H.}~\bibnamefont {Zhou}}, \bibinfo
  {author} {\bibfnamefont {H.~S.}\ \bibnamefont {Knowles}}, \bibinfo {author}
  {\bibfnamefont {R.}~\bibnamefont {Landig}}, \bibinfo {author} {\bibfnamefont
  {S.}~\bibnamefont {Choi}},\ and\ \bibinfo {author} {\bibfnamefont {M.~D.}\
  \bibnamefont {Lukin}},\ }\bibfield  {title} {\bibinfo {title} {Robust dynamic
  hamiltonian engineering of many-body spin systems},\ }\bibfield  {journal}
  {\bibinfo  {journal} {Physical Review X}\ }\textbf {\bibinfo {volume} {10}},\
  \href {https://doi.org/10.1103/physrevx.10.031002}
  {10.1103/physrevx.10.031002} (\bibinfo {year} {2020})\BibitemShut {NoStop}%
\bibitem [{\citenamefont {Richerme}\ \emph {et~al.}(2014)\citenamefont
  {Richerme}, \citenamefont {Gong}, \citenamefont {Lee}, \citenamefont {Senko},
  \citenamefont {Smith}, \citenamefont {Foss-Feig}, \citenamefont {Michalakis},
  \citenamefont {Gorshkov},\ and\ \citenamefont
  {Monroe}}]{richerme_gong_lee_senko_smith_foss-feig_michalakis_gorshkov_monroe_2014}%
  \BibitemOpen
  \bibfield  {author} {\bibinfo {author} {\bibfnamefont {P.}~\bibnamefont
  {Richerme}}, \bibinfo {author} {\bibfnamefont {Z.-X.}\ \bibnamefont {Gong}},
  \bibinfo {author} {\bibfnamefont {A.}~\bibnamefont {Lee}}, \bibinfo {author}
  {\bibfnamefont {C.}~\bibnamefont {Senko}}, \bibinfo {author} {\bibfnamefont
  {J.}~\bibnamefont {Smith}}, \bibinfo {author} {\bibfnamefont
  {M.}~\bibnamefont {Foss-Feig}}, \bibinfo {author} {\bibfnamefont
  {S.}~\bibnamefont {Michalakis}}, \bibinfo {author} {\bibfnamefont {A.~V.}\
  \bibnamefont {Gorshkov}},\ and\ \bibinfo {author} {\bibfnamefont
  {C.}~\bibnamefont {Monroe}},\ }\bibfield  {title} {\bibinfo {title}
  {Non-local propagation of correlations in quantum systems with long-range
  interactions},\ }\href {https://doi.org/10.1038/nature13450} {\bibfield
  {journal} {\bibinfo  {journal} {Nature}\ }\textbf {\bibinfo {volume} {511}},\
  \bibinfo {pages} {198–201} (\bibinfo {year} {2014})}\BibitemShut {NoStop}%
\bibitem [{\citenamefont {Jurcevic}\ \emph {et~al.}(2014)\citenamefont
  {Jurcevic}, \citenamefont {Lanyon}, \citenamefont {Hauke}, \citenamefont
  {Hempel}, \citenamefont {Zoller}, \citenamefont {Blatt},\ and\ \citenamefont
  {Roos}}]{jurcevic_lanyon_hauke_hempel_zoller_blatt_roos_2014}%
  \BibitemOpen
  \bibfield  {author} {\bibinfo {author} {\bibfnamefont {P.}~\bibnamefont
  {Jurcevic}}, \bibinfo {author} {\bibfnamefont {B.~P.}\ \bibnamefont
  {Lanyon}}, \bibinfo {author} {\bibfnamefont {P.}~\bibnamefont {Hauke}},
  \bibinfo {author} {\bibfnamefont {C.}~\bibnamefont {Hempel}}, \bibinfo
  {author} {\bibfnamefont {P.}~\bibnamefont {Zoller}}, \bibinfo {author}
  {\bibfnamefont {R.}~\bibnamefont {Blatt}},\ and\ \bibinfo {author}
  {\bibfnamefont {C.~F.}\ \bibnamefont {Roos}},\ }\bibfield  {title} {\bibinfo
  {title} {Quasiparticle engineering and entanglement propagation in a quantum
  many-body system},\ }\href {https://doi.org/10.1038/nature13461} {\bibfield
  {journal} {\bibinfo  {journal} {Nature}\ }\textbf {\bibinfo {volume} {511}},\
  \bibinfo {pages} {202–205} (\bibinfo {year} {2014})}\BibitemShut {NoStop}%
\bibitem [{\citenamefont {Wall}\ \emph {et~al.}(2017)\citenamefont {Wall},
  \citenamefont {Safavi-Naini},\ and\ \citenamefont
  {Rey}}]{PhysRevA.95.013602}%
  \BibitemOpen
  \bibfield  {author} {\bibinfo {author} {\bibfnamefont {M.~L.}\ \bibnamefont
  {Wall}}, \bibinfo {author} {\bibfnamefont {A.}~\bibnamefont {Safavi-Naini}},\
  and\ \bibinfo {author} {\bibfnamefont {A.~M.}\ \bibnamefont {Rey}},\
  }\bibfield  {title} {\bibinfo {title} {Boson-mediated quantum spin simulators
  in transverse fields: $xy$ model and spin-boson entanglement},\ }\href
  {https://doi.org/10.1103/PhysRevA.95.013602} {\bibfield  {journal} {\bibinfo
  {journal} {Phys. Rev. A}\ }\textbf {\bibinfo {volume} {95}},\ \bibinfo
  {pages} {013602} (\bibinfo {year} {2017})}\BibitemShut {NoStop}%
\bibitem [{\citenamefont {Kiely}\ and\ \citenamefont
  {Freericks}(2018)}]{PhysRevA.97.023611}%
  \BibitemOpen
  \bibfield  {author} {\bibinfo {author} {\bibfnamefont {T.~G.}\ \bibnamefont
  {Kiely}}\ and\ \bibinfo {author} {\bibfnamefont {J.~K.}\ \bibnamefont
  {Freericks}},\ }\bibfield  {title} {\bibinfo {title} {Relationship between
  the transverse-field ising model and the $xy$ model via the rotating-wave
  approximation},\ }\href {https://doi.org/10.1103/PhysRevA.97.023611}
  {\bibfield  {journal} {\bibinfo  {journal} {Phys. Rev. A}\ }\textbf {\bibinfo
  {volume} {97}},\ \bibinfo {pages} {023611} (\bibinfo {year}
  {2018})}\BibitemShut {NoStop}%
\bibitem [{\citenamefont {Ciavarella}\ \emph
  {et~al.}(2023{\natexlab{b}})\citenamefont {Ciavarella}, \citenamefont
  {Caspar}, \citenamefont {Singh}, \citenamefont {Savage},\ and\ \citenamefont
  {Lougovski}}]{1a}%
  \BibitemOpen
  \bibfield  {author} {\bibinfo {author} {\bibfnamefont {A.~N.}\ \bibnamefont
  {Ciavarella}}, \bibinfo {author} {\bibfnamefont {S.}~\bibnamefont {Caspar}},
  \bibinfo {author} {\bibfnamefont {H.}~\bibnamefont {Singh}}, \bibinfo
  {author} {\bibfnamefont {M.~J.}\ \bibnamefont {Savage}},\ and\ \bibinfo
  {author} {\bibfnamefont {P.}~\bibnamefont {Lougovski}},\ }\bibfield  {title}
  {\bibinfo {title} {{Simulating Heisenberg interactions in the Ising model
  with strong drive fields}},\ }\href
  {https://doi.org/10.1103/PhysRevA.108.042216} {\bibfield  {journal} {\bibinfo
   {journal} {Phys. Rev. A}\ }\textbf {\bibinfo {volume} {108}},\ \bibinfo
  {pages} {042216} (\bibinfo {year} {2023}{\natexlab{b}})},\ \Eprint
  {https://arxiv.org/abs/2207.09438} {arXiv:2207.09438 [quant-ph]} \BibitemShut
  {NoStop}%
\bibitem [{\citenamefont {Bermudez}\ \emph {et~al.}(2017)\citenamefont
  {Bermudez}, \citenamefont {Tagliacozzo}, \citenamefont {Sierra},\ and\
  \citenamefont {Richerme}}]{PhysRevB.95.024431}%
  \BibitemOpen
  \bibfield  {author} {\bibinfo {author} {\bibfnamefont {A.}~\bibnamefont
  {Bermudez}}, \bibinfo {author} {\bibfnamefont {L.}~\bibnamefont
  {Tagliacozzo}}, \bibinfo {author} {\bibfnamefont {G.}~\bibnamefont
  {Sierra}},\ and\ \bibinfo {author} {\bibfnamefont {P.}~\bibnamefont
  {Richerme}},\ }\bibfield  {title} {\bibinfo {title} {{Long-range Heisenberg
  models in quasiperiodically driven crystals of trapped ions}},\ }\href
  {https://doi.org/10.1103/PhysRevB.95.024431} {\bibfield  {journal} {\bibinfo
  {journal} {Phys. Rev. B}\ }\textbf {\bibinfo {volume} {95}},\ \bibinfo
  {pages} {024431} (\bibinfo {year} {2017})},\ \Eprint
  {https://arxiv.org/abs/1607.03337} {arXiv:1607.03337 [quant-ph]} \BibitemShut
  {NoStop}%
\bibitem [{\citenamefont {Gonzalez-Raya}\ \emph {et~al.}(2021)\citenamefont
  {Gonzalez-Raya}, \citenamefont {Asensio-Perea}, \citenamefont {Martin},
  \citenamefont {C\'eleri}, \citenamefont {Sanz}, \citenamefont {Lougovski},\
  and\ \citenamefont {Dumitrescu}}]{PRXQuantum.2.020328}%
  \BibitemOpen
  \bibfield  {author} {\bibinfo {author} {\bibfnamefont {T.}~\bibnamefont
  {Gonzalez-Raya}}, \bibinfo {author} {\bibfnamefont {R.}~\bibnamefont
  {Asensio-Perea}}, \bibinfo {author} {\bibfnamefont {A.}~\bibnamefont
  {Martin}}, \bibinfo {author} {\bibfnamefont {L.~C.}\ \bibnamefont
  {C\'eleri}}, \bibinfo {author} {\bibfnamefont {M.}~\bibnamefont {Sanz}},
  \bibinfo {author} {\bibfnamefont {P.}~\bibnamefont {Lougovski}},\ and\
  \bibinfo {author} {\bibfnamefont {E.~F.}\ \bibnamefont {Dumitrescu}},\
  }\bibfield  {title} {\bibinfo {title} {Digital-analog quantum simulations
  using the cross-resonance effect},\ }\href
  {https://doi.org/10.1103/PRXQuantum.2.020328} {\bibfield  {journal} {\bibinfo
   {journal} {PRX Quantum}\ }\textbf {\bibinfo {volume} {2}},\ \bibinfo {pages}
  {020328} (\bibinfo {year} {2021})}\BibitemShut {NoStop}%
\bibitem [{\citenamefont {Henriet}\ \emph {et~al.}(2020)\citenamefont
  {Henriet}, \citenamefont {Beguin}, \citenamefont {Signoles}, \citenamefont
  {Lahaye}, \citenamefont {Browaeys}, \citenamefont {Reymond},\ and\
  \citenamefont
  {Jurczak}}]{Henriet_Beguin_Signoles_Lahaye_Browaeys_Reymond_Jurczak_2020}%
  \BibitemOpen
  \bibfield  {author} {\bibinfo {author} {\bibfnamefont {L.}~\bibnamefont
  {Henriet}}, \bibinfo {author} {\bibfnamefont {L.}~\bibnamefont {Beguin}},
  \bibinfo {author} {\bibfnamefont {A.}~\bibnamefont {Signoles}}, \bibinfo
  {author} {\bibfnamefont {T.}~\bibnamefont {Lahaye}}, \bibinfo {author}
  {\bibfnamefont {A.}~\bibnamefont {Browaeys}}, \bibinfo {author}
  {\bibfnamefont {G.-O.}\ \bibnamefont {Reymond}},\ and\ \bibinfo {author}
  {\bibfnamefont {C.}~\bibnamefont {Jurczak}},\ }\bibfield  {title} {\bibinfo
  {title} {Quantum computing with neutral atoms},\ }\href
  {https://doi.org/10.22331/q-2020-09-21-327} {\bibfield  {journal} {\bibinfo
  {journal} {Quantum}\ }\textbf {\bibinfo {volume} {4}},\ \bibinfo {pages}
  {327} (\bibinfo {year} {2020})}\BibitemShut {NoStop}%
\bibitem [{\citenamefont {Browaeys}\ and\ \citenamefont
  {Lahaye}(2020)}]{Browaeys_Lahaye_2020}%
  \BibitemOpen
  \bibfield  {author} {\bibinfo {author} {\bibfnamefont {A.}~\bibnamefont
  {Browaeys}}\ and\ \bibinfo {author} {\bibfnamefont {T.}~\bibnamefont
  {Lahaye}},\ }\bibfield  {title} {\bibinfo {title} {Many-body physics with
  individually controlled rydberg atoms},\ }\href
  {https://doi.org/10.1038/s41567-019-0733-z} {\bibfield  {journal} {\bibinfo
  {journal} {Nature Physics}\ }\textbf {\bibinfo {volume} {16}},\ \bibinfo
  {pages} {132–142} (\bibinfo {year} {2020})}\BibitemShut {NoStop}%
\bibitem [{\citenamefont {Mádi}\ \emph {et~al.}(1997)\citenamefont {Mádi},
  \citenamefont {Brutscher}, \citenamefont {Schulte-Herbrüggen}, \citenamefont
  {Brüschweiler},\ and\ \citenamefont {Ernst}}]{MADI1997300}%
  \BibitemOpen
  \bibfield  {author} {\bibinfo {author} {\bibfnamefont {Z.}~\bibnamefont
  {Mádi}}, \bibinfo {author} {\bibfnamefont {B.}~\bibnamefont {Brutscher}},
  \bibinfo {author} {\bibfnamefont {T.}~\bibnamefont {Schulte-Herbrüggen}},
  \bibinfo {author} {\bibfnamefont {R.}~\bibnamefont {Brüschweiler}},\ and\
  \bibinfo {author} {\bibfnamefont {R.}~\bibnamefont {Ernst}},\ }\bibfield
  {title} {\bibinfo {title} {Time-resolved observation of spin waves in a
  linear chain of nuclear spins},\ }\href
  {https://doi.org/https://doi.org/10.1016/S0009-2614(97)00194-2} {\bibfield
  {journal} {\bibinfo  {journal} {Chemical Physics Letters}\ }\textbf {\bibinfo
  {volume} {268}},\ \bibinfo {pages} {300} (\bibinfo {year}
  {1997})}\BibitemShut {NoStop}%
\bibitem [{\citenamefont {Roumpos}\ \emph {et~al.}(2007)\citenamefont
  {Roumpos}, \citenamefont {Master},\ and\ \citenamefont
  {Yamamoto}}]{PhysRevB.75.094415}%
  \BibitemOpen
  \bibfield  {author} {\bibinfo {author} {\bibfnamefont {G.}~\bibnamefont
  {Roumpos}}, \bibinfo {author} {\bibfnamefont {C.~P.}\ \bibnamefont
  {Master}},\ and\ \bibinfo {author} {\bibfnamefont {Y.}~\bibnamefont
  {Yamamoto}},\ }\bibfield  {title} {\bibinfo {title} {Quantum simulation of
  spin ordering with nuclear spins in a solid-state lattice},\ }\href
  {https://doi.org/10.1103/PhysRevB.75.094415} {\bibfield  {journal} {\bibinfo
  {journal} {Phys. Rev. B}\ }\textbf {\bibinfo {volume} {75}},\ \bibinfo
  {pages} {094415} (\bibinfo {year} {2007})}\BibitemShut {NoStop}%
\bibitem [{\citenamefont {Salath\'e}\ \emph {et~al.}(2015)\citenamefont
  {Salath\'e}, \citenamefont {Mondal}, \citenamefont {Oppliger}, \citenamefont
  {Heinsoo}, \citenamefont {Kurpiers}, \citenamefont
  {Poto\ifmmode~\check{c}\else \v{c}\fi{}nik}, \citenamefont {Mezzacapo},
  \citenamefont {Las~Heras}, \citenamefont {Lamata}, \citenamefont {Solano},
  \citenamefont {Filipp},\ and\ \citenamefont {Wallraff}}]{PhysRevX.5.021027}%
  \BibitemOpen
  \bibfield  {author} {\bibinfo {author} {\bibfnamefont {Y.}~\bibnamefont
  {Salath\'e}}, \bibinfo {author} {\bibfnamefont {M.}~\bibnamefont {Mondal}},
  \bibinfo {author} {\bibfnamefont {M.}~\bibnamefont {Oppliger}}, \bibinfo
  {author} {\bibfnamefont {J.}~\bibnamefont {Heinsoo}}, \bibinfo {author}
  {\bibfnamefont {P.}~\bibnamefont {Kurpiers}}, \bibinfo {author}
  {\bibfnamefont {A.}~\bibnamefont {Poto\ifmmode~\check{c}\else
  \v{c}\fi{}nik}}, \bibinfo {author} {\bibfnamefont {A.}~\bibnamefont
  {Mezzacapo}}, \bibinfo {author} {\bibfnamefont {U.}~\bibnamefont
  {Las~Heras}}, \bibinfo {author} {\bibfnamefont {L.}~\bibnamefont {Lamata}},
  \bibinfo {author} {\bibfnamefont {E.}~\bibnamefont {Solano}}, \bibinfo
  {author} {\bibfnamefont {S.}~\bibnamefont {Filipp}},\ and\ \bibinfo {author}
  {\bibfnamefont {A.}~\bibnamefont {Wallraff}},\ }\bibfield  {title} {\bibinfo
  {title} {Digital quantum simulation of spin models with circuit quantum
  electrodynamics},\ }\href {https://doi.org/10.1103/PhysRevX.5.021027}
  {\bibfield  {journal} {\bibinfo  {journal} {Phys. Rev. X}\ }\textbf {\bibinfo
  {volume} {5}},\ \bibinfo {pages} {021027} (\bibinfo {year}
  {2015})}\BibitemShut {NoStop}%
\bibitem [{\citenamefont {Johnson}\ \emph {et~al.}(2011)\citenamefont
  {Johnson}, \citenamefont {Amin}, \citenamefont {Gildert}, \citenamefont
  {Lanting}, \citenamefont {Hamze}, \citenamefont {Dickson}, \citenamefont
  {Harris}, \citenamefont {Berkley}, \citenamefont {Johansson}, \citenamefont
  {Bunyk},\ and\ \citenamefont
  {et~al.}}]{johnson_amin_gildert_lanting_hamze_dickson_harris_berkley_johansson_bunyk_et_al._2011}%
  \BibitemOpen
  \bibfield  {author} {\bibinfo {author} {\bibfnamefont {M.~W.}\ \bibnamefont
  {Johnson}}, \bibinfo {author} {\bibfnamefont {M.~H.}\ \bibnamefont {Amin}},
  \bibinfo {author} {\bibfnamefont {S.}~\bibnamefont {Gildert}}, \bibinfo
  {author} {\bibfnamefont {T.}~\bibnamefont {Lanting}}, \bibinfo {author}
  {\bibfnamefont {F.}~\bibnamefont {Hamze}}, \bibinfo {author} {\bibfnamefont
  {N.}~\bibnamefont {Dickson}}, \bibinfo {author} {\bibfnamefont
  {R.}~\bibnamefont {Harris}}, \bibinfo {author} {\bibfnamefont {A.~J.}\
  \bibnamefont {Berkley}}, \bibinfo {author} {\bibfnamefont {J.}~\bibnamefont
  {Johansson}}, \bibinfo {author} {\bibfnamefont {P.}~\bibnamefont {Bunyk}},\
  and\ \bibinfo {author} {\bibnamefont {et~al.}},\ }\bibfield  {title}
  {\bibinfo {title} {Quantum annealing with manufactured spins},\ }\href
  {https://doi.org/10.1038/nature10012} {\bibfield  {journal} {\bibinfo
  {journal} {Nature}\ }\textbf {\bibinfo {volume} {473}},\ \bibinfo {pages}
  {194–198} (\bibinfo {year} {2011})}\BibitemShut {NoStop}%
\bibitem [{\citenamefont {Lloyd}\ and\ \citenamefont
  {Slotine}(1998)}]{PhysRevLett.80.4088}%
  \BibitemOpen
  \bibfield  {author} {\bibinfo {author} {\bibfnamefont {S.}~\bibnamefont
  {Lloyd}}\ and\ \bibinfo {author} {\bibfnamefont {J.-J.~E.}\ \bibnamefont
  {Slotine}},\ }\bibfield  {title} {\bibinfo {title} {Analog quantum error
  correction},\ }\href {https://doi.org/10.1103/PhysRevLett.80.4088} {\bibfield
   {journal} {\bibinfo  {journal} {Phys. Rev. Lett.}\ }\textbf {\bibinfo
  {volume} {80}},\ \bibinfo {pages} {4088} (\bibinfo {year}
  {1998})}\BibitemShut {NoStop}%
\bibitem [{\citenamefont {Fukui}\ \emph {et~al.}(2017)\citenamefont {Fukui},
  \citenamefont {Tomita},\ and\ \citenamefont
  {Okamoto}}]{PhysRevLett.119.180507}%
  \BibitemOpen
  \bibfield  {author} {\bibinfo {author} {\bibfnamefont {K.}~\bibnamefont
  {Fukui}}, \bibinfo {author} {\bibfnamefont {A.}~\bibnamefont {Tomita}},\ and\
  \bibinfo {author} {\bibfnamefont {A.}~\bibnamefont {Okamoto}},\ }\bibfield
  {title} {\bibinfo {title} {Analog quantum error correction with encoding a
  qubit into an oscillator},\ }\href
  {https://doi.org/10.1103/PhysRevLett.119.180507} {\bibfield  {journal}
  {\bibinfo  {journal} {Phys. Rev. Lett.}\ }\textbf {\bibinfo {volume} {119}},\
  \bibinfo {pages} {180507} (\bibinfo {year} {2017})}\BibitemShut {NoStop}%
\bibitem [{\citenamefont {Klco}\ and\ \citenamefont
  {Savage}(2019)}]{PhysRevA.99.052335}%
  \BibitemOpen
  \bibfield  {author} {\bibinfo {author} {\bibfnamefont {N.}~\bibnamefont
  {Klco}}\ and\ \bibinfo {author} {\bibfnamefont {M.~J.}\ \bibnamefont
  {Savage}},\ }\bibfield  {title} {\bibinfo {title} {{Digitization of scalar
  fields for quantum computing}},\ }\href
  {https://doi.org/10.1103/PhysRevA.99.052335} {\bibfield  {journal} {\bibinfo
  {journal} {Phys. Rev. A}\ }\textbf {\bibinfo {volume} {99}},\ \bibinfo
  {pages} {052335} (\bibinfo {year} {2019})},\ \Eprint
  {https://arxiv.org/abs/1808.10378} {arXiv:1808.10378 [quant-ph]} \BibitemShut
  {NoStop}%
\bibitem [{\citenamefont {Wiebe}\ \emph {et~al.}(2010)\citenamefont {Wiebe},
  \citenamefont {Berry}, \citenamefont {H{\o}yer},\ and\ \citenamefont
  {Sanders}}]{Wiebe_2010}%
  \BibitemOpen
  \bibfield  {author} {\bibinfo {author} {\bibfnamefont {N.}~\bibnamefont
  {Wiebe}}, \bibinfo {author} {\bibfnamefont {D.}~\bibnamefont {Berry}},
  \bibinfo {author} {\bibfnamefont {P.}~\bibnamefont {H{\o}yer}},\ and\
  \bibinfo {author} {\bibfnamefont {B.~C.}\ \bibnamefont {Sanders}},\
  }\bibfield  {title} {\bibinfo {title} {Higher order decompositions of ordered
  operator exponentials},\ }\href
  {https://doi.org/10.1088/1751-8113/43/6/065203} {\bibfield  {journal}
  {\bibinfo  {journal} {Journal of Physics A: Mathematical and Theoretical}\
  }\textbf {\bibinfo {volume} {43}},\ \bibinfo {pages} {065203} (\bibinfo
  {year} {2010})}\BibitemShut {NoStop}%
\bibitem [{\citenamefont {Childs}\ and\ \citenamefont
  {Su}(2019)}]{Childs_2019}%
  \BibitemOpen
  \bibfield  {author} {\bibinfo {author} {\bibfnamefont {A.~M.}\ \bibnamefont
  {Childs}}\ and\ \bibinfo {author} {\bibfnamefont {Y.}~\bibnamefont {Su}},\
  }\bibfield  {title} {\bibinfo {title} {Nearly optimal lattice simulation by
  product formulas},\ }\bibfield  {journal} {\bibinfo  {journal} {Physical
  Review Letters}\ }\textbf {\bibinfo {volume} {123}},\ \href
  {https://doi.org/10.1103/physrevlett.123.050503}
  {10.1103/physrevlett.123.050503} (\bibinfo {year} {2019})\BibitemShut
  {NoStop}%
\bibitem [{\citenamefont {Childs}\ \emph {et~al.}(2021)\citenamefont {Childs},
  \citenamefont {Su}, \citenamefont {Tran}, \citenamefont {Wiebe},\ and\
  \citenamefont {Zhu}}]{childs_wiebe_2021}%
  \BibitemOpen
  \bibfield  {author} {\bibinfo {author} {\bibfnamefont {A.~M.}\ \bibnamefont
  {Childs}}, \bibinfo {author} {\bibfnamefont {Y.}~\bibnamefont {Su}}, \bibinfo
  {author} {\bibfnamefont {M.~C.}\ \bibnamefont {Tran}}, \bibinfo {author}
  {\bibfnamefont {N.}~\bibnamefont {Wiebe}},\ and\ \bibinfo {author}
  {\bibfnamefont {S.}~\bibnamefont {Zhu}},\ }\bibfield  {title} {\bibinfo
  {title} {Theory of trotter error with commutator scaling},\ }\href
  {https://doi.org/10.1103/PhysRevX.11.011020} {\bibfield  {journal} {\bibinfo
  {journal} {Phys. Rev. X}\ }\textbf {\bibinfo {volume} {11}},\ \bibinfo
  {pages} {011020} (\bibinfo {year} {2021})}\BibitemShut {NoStop}%
\bibitem [{\citenamefont {Cai}\ \emph {et~al.}(2022)\citenamefont {Cai},
  \citenamefont {Babbush}, \citenamefont {Benjamin}, \citenamefont {Endo},
  \citenamefont {Huggins}, \citenamefont {Li}, \citenamefont {McClean},\ and\
  \citenamefont {O'Brien}}]{cai2022quantum}%
  \BibitemOpen
  \bibfield  {author} {\bibinfo {author} {\bibfnamefont {Z.}~\bibnamefont
  {Cai}}, \bibinfo {author} {\bibfnamefont {R.}~\bibnamefont {Babbush}},
  \bibinfo {author} {\bibfnamefont {S.~C.}\ \bibnamefont {Benjamin}}, \bibinfo
  {author} {\bibfnamefont {S.}~\bibnamefont {Endo}}, \bibinfo {author}
  {\bibfnamefont {W.~J.}\ \bibnamefont {Huggins}}, \bibinfo {author}
  {\bibfnamefont {Y.}~\bibnamefont {Li}}, \bibinfo {author} {\bibfnamefont
  {J.~R.}\ \bibnamefont {McClean}},\ and\ \bibinfo {author} {\bibfnamefont
  {T.~E.}\ \bibnamefont {O'Brien}},\ }\bibfield  {title} {\bibinfo {title}
  {{Quantum Error Mitigation}},\ }\href@noop {} {\  (\bibinfo {year} {2022})},\
  \Eprint {https://arxiv.org/abs/2210.00921} {arXiv:2210.00921 [quant-ph]}
  \BibitemShut {NoStop}%
\bibitem [{\citenamefont {Endo}\ \emph {et~al.}(2018)\citenamefont {Endo},
  \citenamefont {Benjamin},\ and\ \citenamefont {Li}}]{Endo_2018}%
  \BibitemOpen
  \bibfield  {author} {\bibinfo {author} {\bibfnamefont {S.}~\bibnamefont
  {Endo}}, \bibinfo {author} {\bibfnamefont {S.~C.}\ \bibnamefont {Benjamin}},\
  and\ \bibinfo {author} {\bibfnamefont {Y.}~\bibnamefont {Li}},\ }\bibfield
  {title} {\bibinfo {title} {Practical quantum error mitigation for near-future
  applications},\ }\bibfield  {journal} {\bibinfo  {journal} {Physical Review
  X}\ }\textbf {\bibinfo {volume} {8}},\ \href
  {https://doi.org/10.1103/physrevx.8.031027} {10.1103/physrevx.8.031027}
  (\bibinfo {year} {2018})\BibitemShut {NoStop}%
\bibitem [{\citenamefont {Shaffer}\ \emph {et~al.}(2021)\citenamefont
  {Shaffer}, \citenamefont {Megidish}, \citenamefont {Broz}, \citenamefont
  {Chen},\ and\ \citenamefont
  {Häffner}}]{Shaffer_Megidish_Broz_Chen_Häffner_2021}%
  \BibitemOpen
  \bibfield  {author} {\bibinfo {author} {\bibfnamefont {R.}~\bibnamefont
  {Shaffer}}, \bibinfo {author} {\bibfnamefont {E.}~\bibnamefont {Megidish}},
  \bibinfo {author} {\bibfnamefont {J.}~\bibnamefont {Broz}}, \bibinfo {author}
  {\bibfnamefont {W.-T.}\ \bibnamefont {Chen}},\ and\ \bibinfo {author}
  {\bibfnamefont {H.}~\bibnamefont {Häffner}},\ }\bibfield  {title} {\bibinfo
  {title} {Practical verification protocols for analog quantum simulators},\
  }\bibfield  {journal} {\bibinfo  {journal} {npj Quantum Information}\
  }\textbf {\bibinfo {volume} {7}},\ \href
  {https://doi.org/10.1038/s41534-021-00380-8} {10.1038/s41534-021-00380-8}
  (\bibinfo {year} {2021})\BibitemShut {NoStop}%
\bibitem [{\citenamefont {Trivedi}\ \emph {et~al.}(2022)\citenamefont
  {Trivedi}, \citenamefont {Rubio},\ and\ \citenamefont
  {Cirac}}]{trivedi2022quantum}%
  \BibitemOpen
  \bibfield  {author} {\bibinfo {author} {\bibfnamefont {R.}~\bibnamefont
  {Trivedi}}, \bibinfo {author} {\bibfnamefont {A.~F.}\ \bibnamefont {Rubio}},\
  and\ \bibinfo {author} {\bibfnamefont {J.~I.}\ \bibnamefont {Cirac}},\
  }\href@noop {} {\bibinfo {title} {Quantum advantage and stability to errors
  in analogue quantum simulators}} (\bibinfo {year} {2022}),\ \Eprint
  {https://arxiv.org/abs/2212.04924} {arXiv:2212.04924 [quant-ph]} \BibitemShut
  {NoStop}%
\bibitem [{\citenamefont {Poggi}\ \emph {et~al.}(2020)\citenamefont {Poggi},
  \citenamefont {Lysne}, \citenamefont {Kuper}, \citenamefont {Deutsch},\ and\
  \citenamefont {Jessen}}]{PRXQuantum.1.020308}%
  \BibitemOpen
  \bibfield  {author} {\bibinfo {author} {\bibfnamefont {P.~M.}\ \bibnamefont
  {Poggi}}, \bibinfo {author} {\bibfnamefont {N.~K.}\ \bibnamefont {Lysne}},
  \bibinfo {author} {\bibfnamefont {K.~W.}\ \bibnamefont {Kuper}}, \bibinfo
  {author} {\bibfnamefont {I.~H.}\ \bibnamefont {Deutsch}},\ and\ \bibinfo
  {author} {\bibfnamefont {P.~S.}\ \bibnamefont {Jessen}},\ }\bibfield  {title}
  {\bibinfo {title} {Quantifying the sensitivity to errors in analog quantum
  simulation},\ }\href {https://doi.org/10.1103/PRXQuantum.1.020308} {\bibfield
   {journal} {\bibinfo  {journal} {PRX Quantum}\ }\textbf {\bibinfo {volume}
  {1}},\ \bibinfo {pages} {020308} (\bibinfo {year} {2020})}\BibitemShut
  {NoStop}%
\bibitem [{\citenamefont {Cong}\ \emph {et~al.}(2022)\citenamefont {Cong},
  \citenamefont {Levine}, \citenamefont {Keesling}, \citenamefont {Bluvstein},
  \citenamefont {Wang},\ and\ \citenamefont {Lukin}}]{PhysRevX.12.021049}%
  \BibitemOpen
  \bibfield  {author} {\bibinfo {author} {\bibfnamefont {I.}~\bibnamefont
  {Cong}}, \bibinfo {author} {\bibfnamefont {H.}~\bibnamefont {Levine}},
  \bibinfo {author} {\bibfnamefont {A.}~\bibnamefont {Keesling}}, \bibinfo
  {author} {\bibfnamefont {D.}~\bibnamefont {Bluvstein}}, \bibinfo {author}
  {\bibfnamefont {S.-T.}\ \bibnamefont {Wang}},\ and\ \bibinfo {author}
  {\bibfnamefont {M.~D.}\ \bibnamefont {Lukin}},\ }\bibfield  {title} {\bibinfo
  {title} {{Hardware-Efficient, Fault-Tolerant Quantum Computation with Rydberg
  Atoms}},\ }\href {https://doi.org/10.1103/PhysRevX.12.021049} {\bibfield
  {journal} {\bibinfo  {journal} {Phys. Rev. X}\ }\textbf {\bibinfo {volume}
  {12}},\ \bibinfo {pages} {021049} (\bibinfo {year} {2022})},\ \Eprint
  {https://arxiv.org/abs/2105.13501} {arXiv:2105.13501 [quant-ph]} \BibitemShut
  {NoStop}%
\bibitem [{\citenamefont {Masanes}\ \emph {et~al.}(2002)\citenamefont
  {Masanes}, \citenamefont {Vidal},\ and\ \citenamefont
  {Latorre}}]{masanes2002timeoptimal}%
  \BibitemOpen
  \bibfield  {author} {\bibinfo {author} {\bibfnamefont {L.}~\bibnamefont
  {Masanes}}, \bibinfo {author} {\bibfnamefont {G.}~\bibnamefont {Vidal}},\
  and\ \bibinfo {author} {\bibfnamefont {J.~I.}\ \bibnamefont {Latorre}},\
  }\href@noop {} {\bibinfo {title} {Time-optimal hamiltonian simulation and
  gate synthesis using homogeneous local unitaries}} (\bibinfo {year} {2002}),\
  \Eprint {https://arxiv.org/abs/quant-ph/0202042} {arXiv:quant-ph/0202042
  [quant-ph]} \BibitemShut {NoStop}%
\bibitem [{\citenamefont {Suzuki}(1976)}]{suzuki_1976}%
  \BibitemOpen
  \bibfield  {author} {\bibinfo {author} {\bibfnamefont {M.}~\bibnamefont
  {Suzuki}},\ }\bibfield  {title} {\bibinfo {title} {Generalized trotter's
  formula and systematic approximants of exponential operators and inner
  derivations with applications to many-body problems},\ }\href
  {https://doi.org/10.1007/bf01609348} {\bibfield  {journal} {\bibinfo
  {journal} {Communications in Mathematical Physics}\ }\textbf {\bibinfo
  {volume} {51}},\ \bibinfo {pages} {183–190} (\bibinfo {year}
  {1976})}\BibitemShut {NoStop}%
\bibitem [{\citenamefont {Ciavarella}\ \emph
  {et~al.}(2023{\natexlab{c}})\citenamefont {Ciavarella}, \citenamefont
  {Caspar}, \citenamefont {Illa},\ and\ \citenamefont {Savage}}]{1c}%
  \BibitemOpen
  \bibfield  {author} {\bibinfo {author} {\bibfnamefont {A.~N.}\ \bibnamefont
  {Ciavarella}}, \bibinfo {author} {\bibfnamefont {S.}~\bibnamefont {Caspar}},
  \bibinfo {author} {\bibfnamefont {M.}~\bibnamefont {Illa}},\ and\ \bibinfo
  {author} {\bibfnamefont {M.~J.}\ \bibnamefont {Savage}},\ }\bibfield  {title}
  {\bibinfo {title} {{State Preparation in the Heisenberg Model through
  Adiabatic Spiraling}},\ }\href {https://doi.org/10.22331/q-2023-04-06-970}
  {\bibfield  {journal} {\bibinfo  {journal} {Quantum}\ }\textbf {\bibinfo
  {volume} {7}},\ \bibinfo {pages} {970} (\bibinfo {year}
  {2023}{\natexlab{c}})},\ \Eprint {https://arxiv.org/abs/2210.04965}
  {arXiv:2210.04965 [quant-ph]} \BibitemShut {NoStop}%
\bibitem [{\citenamefont {Eckstein}\ \emph {et~al.}(2023)\citenamefont
  {Eckstein}, \citenamefont {Mansuroglu}, \citenamefont {Czarnik},
  \citenamefont {Zhu}, \citenamefont {Hartmann}, \citenamefont {Cincio},
  \citenamefont {Sornborger},\ and\ \citenamefont
  {Holmes}}]{eckstein2023largescale}%
  \BibitemOpen
  \bibfield  {author} {\bibinfo {author} {\bibfnamefont {T.}~\bibnamefont
  {Eckstein}}, \bibinfo {author} {\bibfnamefont {R.}~\bibnamefont
  {Mansuroglu}}, \bibinfo {author} {\bibfnamefont {P.}~\bibnamefont {Czarnik}},
  \bibinfo {author} {\bibfnamefont {J.-X.}\ \bibnamefont {Zhu}}, \bibinfo
  {author} {\bibfnamefont {M.~J.}\ \bibnamefont {Hartmann}}, \bibinfo {author}
  {\bibfnamefont {L.}~\bibnamefont {Cincio}}, \bibinfo {author} {\bibfnamefont
  {A.~T.}\ \bibnamefont {Sornborger}},\ and\ \bibinfo {author} {\bibfnamefont
  {Z.}~\bibnamefont {Holmes}},\ }\href@noop {} {\bibinfo {title} {Large-scale
  simulations of floquet physics on near-term quantum computers}} (\bibinfo
  {year} {2023}),\ \Eprint {https://arxiv.org/abs/2303.02209} {arXiv:2303.02209
  [quant-ph]} \BibitemShut {NoStop}%
\bibitem [{\citenamefont {Saffman}\ \emph {et~al.}(2010)\citenamefont
  {Saffman}, \citenamefont {Walker},\ and\ \citenamefont
  {M\o{}lmer}}]{RevModPhys.82.2313}%
  \BibitemOpen
  \bibfield  {author} {\bibinfo {author} {\bibfnamefont {M.}~\bibnamefont
  {Saffman}}, \bibinfo {author} {\bibfnamefont {T.~G.}\ \bibnamefont
  {Walker}},\ and\ \bibinfo {author} {\bibfnamefont {K.}~\bibnamefont
  {M\o{}lmer}},\ }\bibfield  {title} {\bibinfo {title} {Quantum information
  with rydberg atoms},\ }\href {https://doi.org/10.1103/RevModPhys.82.2313}
  {\bibfield  {journal} {\bibinfo  {journal} {Rev. Mod. Phys.}\ }\textbf
  {\bibinfo {volume} {82}},\ \bibinfo {pages} {2313} (\bibinfo {year}
  {2010})}\BibitemShut {NoStop}%
\bibitem [{\citenamefont {Wu}\ \emph {et~al.}(2021)\citenamefont {Wu},
  \citenamefont {Liang}, \citenamefont {Tian}, \citenamefont {Yang},
  \citenamefont {Chen}, \citenamefont {Liu}, \citenamefont {Tey},\ and\
  \citenamefont {You}}]{Wu_2021}%
  \BibitemOpen
  \bibfield  {author} {\bibinfo {author} {\bibfnamefont {X.}~\bibnamefont
  {Wu}}, \bibinfo {author} {\bibfnamefont {X.}~\bibnamefont {Liang}}, \bibinfo
  {author} {\bibfnamefont {Y.}~\bibnamefont {Tian}}, \bibinfo {author}
  {\bibfnamefont {F.}~\bibnamefont {Yang}}, \bibinfo {author} {\bibfnamefont
  {C.}~\bibnamefont {Chen}}, \bibinfo {author} {\bibfnamefont {Y.-C.}\
  \bibnamefont {Liu}}, \bibinfo {author} {\bibfnamefont {M.~K.}\ \bibnamefont
  {Tey}},\ and\ \bibinfo {author} {\bibfnamefont {L.}~\bibnamefont {You}},\
  }\bibfield  {title} {\bibinfo {title} {{A concise review of Rydberg atom
  based quantum computation and quantum simulation}},\ }\href
  {https://doi.org/10.1088/1674-1056/abd76f} {\bibfield  {journal} {\bibinfo
  {journal} {Chin. Phys. B}\ }\textbf {\bibinfo {volume} {30}},\ \bibinfo
  {pages} {020305} (\bibinfo {year} {2021})},\ \Eprint
  {https://arxiv.org/abs/2012.10614} {arXiv:2012.10614 [quant-ph]} \BibitemShut
  {NoStop}%
\bibitem [{\citenamefont {Wurtz}\ \emph {et~al.}(2023)\citenamefont {Wurtz},
  \citenamefont {Bylinskii}, \citenamefont {Braverman}, \citenamefont
  {Amato-Grill}, \citenamefont {Cantu}, \citenamefont {Huber}, \citenamefont
  {Lukin}, \citenamefont {Liu}, \citenamefont {Weinberg}, \citenamefont {Long},
  \citenamefont {Wang}, \citenamefont {Gemelke},\ and\ \citenamefont
  {Keesling}}]{wurtz2023aquila}%
  \BibitemOpen
  \bibfield  {author} {\bibinfo {author} {\bibfnamefont {J.}~\bibnamefont
  {Wurtz}}, \bibinfo {author} {\bibfnamefont {A.}~\bibnamefont {Bylinskii}},
  \bibinfo {author} {\bibfnamefont {B.}~\bibnamefont {Braverman}}, \bibinfo
  {author} {\bibfnamefont {J.}~\bibnamefont {Amato-Grill}}, \bibinfo {author}
  {\bibfnamefont {S.~H.}\ \bibnamefont {Cantu}}, \bibinfo {author}
  {\bibfnamefont {F.}~\bibnamefont {Huber}}, \bibinfo {author} {\bibfnamefont
  {A.}~\bibnamefont {Lukin}}, \bibinfo {author} {\bibfnamefont
  {F.}~\bibnamefont {Liu}}, \bibinfo {author} {\bibfnamefont {P.}~\bibnamefont
  {Weinberg}}, \bibinfo {author} {\bibfnamefont {J.}~\bibnamefont {Long}},
  \bibinfo {author} {\bibfnamefont {S.-T.}\ \bibnamefont {Wang}}, \bibinfo
  {author} {\bibfnamefont {N.}~\bibnamefont {Gemelke}},\ and\ \bibinfo {author}
  {\bibfnamefont {A.}~\bibnamefont {Keesling}},\ }\href@noop {} {\bibinfo
  {title} {Aquila: Quera's 256-qubit neutral-atom quantum computer}} (\bibinfo
  {year} {2023}),\ \Eprint {https://arxiv.org/abs/2306.11727} {arXiv:2306.11727
  [quant-ph]} \BibitemShut {NoStop}%
\bibitem [{\citenamefont {{\c{S}}ahino{\u{g}}lu}\ and\ \citenamefont
  {Somma}(2021)}]{_ahino_lu_2021}%
  \BibitemOpen
  \bibfield  {author} {\bibinfo {author} {\bibfnamefont {B.}~\bibnamefont
  {{\c{S}}ahino{\u{g}}lu}}\ and\ \bibinfo {author} {\bibfnamefont {R.~D.}\
  \bibnamefont {Somma}},\ }\bibfield  {title} {\bibinfo {title} {Hamiltonian
  simulation in the low-energy subspace},\ }\bibfield  {journal} {\bibinfo
  {journal} {npj Quantum Information}\ }\textbf {\bibinfo {volume} {7}},\ \href
  {https://doi.org/10.1038/s41534-021-00451-w} {10.1038/s41534-021-00451-w}
  (\bibinfo {year} {2021})\BibitemShut {NoStop}%
\bibitem [{\citenamefont {Heyl}\ \emph {et~al.}(2019)\citenamefont {Heyl},
  \citenamefont {Hauke},\ and\ \citenamefont
  {Zoller}}]{doi:10.1126/sciadv.aau8342}%
  \BibitemOpen
  \bibfield  {author} {\bibinfo {author} {\bibfnamefont {M.}~\bibnamefont
  {Heyl}}, \bibinfo {author} {\bibfnamefont {P.}~\bibnamefont {Hauke}},\ and\
  \bibinfo {author} {\bibfnamefont {P.}~\bibnamefont {Zoller}},\ }\bibfield
  {title} {\bibinfo {title} {Quantum localization bounds trotter errors in
  digital quantum simulation},\ }\href {https://doi.org/10.1126/sciadv.aau8342}
  {\bibfield  {journal} {\bibinfo  {journal} {Science Advances}\ }\textbf
  {\bibinfo {volume} {5}},\ \bibinfo {pages} {eaau8342} (\bibinfo {year}
  {2019})},\ \Eprint
  {https://arxiv.org/abs/https://www.science.org/doi/pdf/10.1126/sciadv.aau8342}
  {https://www.science.org/doi/pdf/10.1126/sciadv.aau8342} \BibitemShut
  {NoStop}%
\bibitem [{\citenamefont {Chinni}\ \emph {et~al.}(2022)\citenamefont {Chinni},
  \citenamefont {Mu\~noz Arias}, \citenamefont {Deutsch},\ and\ \citenamefont
  {Poggi}}]{Chinni_2022}%
  \BibitemOpen
  \bibfield  {author} {\bibinfo {author} {\bibfnamefont {K.}~\bibnamefont
  {Chinni}}, \bibinfo {author} {\bibfnamefont {M.~H.}\ \bibnamefont {Mu\~noz
  Arias}}, \bibinfo {author} {\bibfnamefont {I.~H.}\ \bibnamefont {Deutsch}},\
  and\ \bibinfo {author} {\bibfnamefont {P.~M.}\ \bibnamefont {Poggi}},\
  }\bibfield  {title} {\bibinfo {title} {{Trotter Errors from Dynamical
  Structural Instabilities of Floquet Maps in Quantum Simulation}},\ }\href
  {https://doi.org/10.1103/PRXQuantum.3.010351} {\bibfield  {journal} {\bibinfo
   {journal} {PRX Quantum}\ }\textbf {\bibinfo {volume} {3}},\ \bibinfo {pages}
  {010351} (\bibinfo {year} {2022})}\BibitemShut {NoStop}%
\bibitem [{\citenamefont {Schlosser}\ \emph {et~al.}(2020)\citenamefont
  {Schlosser}, \citenamefont {Ohl~de Mello}, \citenamefont {Schäffner},
  \citenamefont {Preuschoff}, \citenamefont {Kohfahl},\ and\ \citenamefont
  {Birkl}}]{schlosser_ohl_demello_schaffner_preuschoff_kohfahl_birkl_2020}%
  \BibitemOpen
  \bibfield  {author} {\bibinfo {author} {\bibfnamefont {M.}~\bibnamefont
  {Schlosser}}, \bibinfo {author} {\bibfnamefont {D.}~\bibnamefont {Ohl~de
  Mello}}, \bibinfo {author} {\bibfnamefont {D.}~\bibnamefont {Schäffner}},
  \bibinfo {author} {\bibfnamefont {T.}~\bibnamefont {Preuschoff}}, \bibinfo
  {author} {\bibfnamefont {L.}~\bibnamefont {Kohfahl}},\ and\ \bibinfo {author}
  {\bibfnamefont {G.}~\bibnamefont {Birkl}},\ }\bibfield  {title} {\bibinfo
  {title} {Assembled arrays of rydberg-interacting atoms},\ }\href
  {https://doi.org/10.1088/1361-6455/ab8b46} {\bibfield  {journal} {\bibinfo
  {journal} {Journal of Physics B: Atomic, Molecular and Optical Physics}\
  }\textbf {\bibinfo {volume} {53}},\ \bibinfo {pages} {144001} (\bibinfo
  {year} {2020})}\BibitemShut {NoStop}%
\bibitem [{\citenamefont {Zache}\ \emph {et~al.}(2019)\citenamefont {Zache},
  \citenamefont {Mueller}, \citenamefont {Schneider}, \citenamefont
  {Jendrzejewski}, \citenamefont {Berges},\ and\ \citenamefont
  {Hauke}}]{PhysRevLett.122.050403}%
  \BibitemOpen
  \bibfield  {author} {\bibinfo {author} {\bibfnamefont {T.~V.}\ \bibnamefont
  {Zache}}, \bibinfo {author} {\bibfnamefont {N.}~\bibnamefont {Mueller}},
  \bibinfo {author} {\bibfnamefont {J.~T.}\ \bibnamefont {Schneider}}, \bibinfo
  {author} {\bibfnamefont {F.}~\bibnamefont {Jendrzejewski}}, \bibinfo {author}
  {\bibfnamefont {J.}~\bibnamefont {Berges}},\ and\ \bibinfo {author}
  {\bibfnamefont {P.}~\bibnamefont {Hauke}},\ }\bibfield  {title} {\bibinfo
  {title} {{Dynamical Topological Transitions in the Massive Schwinger Model
  with a $\theta$ Term}},\ }\href
  {https://doi.org/10.1103/PhysRevLett.122.050403} {\bibfield  {journal}
  {\bibinfo  {journal} {Phys. Rev. Lett.}\ }\textbf {\bibinfo {volume} {122}},\
  \bibinfo {pages} {050403} (\bibinfo {year} {2019})},\ \Eprint
  {https://arxiv.org/abs/1808.07885} {arXiv:1808.07885 [quant-ph]} \BibitemShut
  {NoStop}%
\bibitem [{\citenamefont {Samajdar}\ \emph {et~al.}(2021)\citenamefont
  {Samajdar}, \citenamefont {Ho}, \citenamefont {Pichler}, \citenamefont
  {Lukin},\ and\ \citenamefont {Sachdev}}]{Samajdar_2021}%
  \BibitemOpen
  \bibfield  {author} {\bibinfo {author} {\bibfnamefont {R.}~\bibnamefont
  {Samajdar}}, \bibinfo {author} {\bibfnamefont {W.~W.}\ \bibnamefont {Ho}},
  \bibinfo {author} {\bibfnamefont {H.}~\bibnamefont {Pichler}}, \bibinfo
  {author} {\bibfnamefont {M.~D.}\ \bibnamefont {Lukin}},\ and\ \bibinfo
  {author} {\bibfnamefont {S.}~\bibnamefont {Sachdev}},\ }\bibfield  {title}
  {\bibinfo {title} {{Quantum phases of Rydberg atoms on a kagome lattice}},\
  }\href {https://doi.org/10.1073/pnas.2015785118} {\bibfield  {journal}
  {\bibinfo  {journal} {Proc. Nat. Acad. Sci.}\ }\textbf {\bibinfo {volume}
  {118}},\ \bibinfo {pages} {e2015785118} (\bibinfo {year} {2021})},\ \Eprint
  {https://arxiv.org/abs/2011.12295} {arXiv:2011.12295 [cond-mat.quant-gas]}
  \BibitemShut {NoStop}%
\bibitem [{\citenamefont {Müller}\ and\ \citenamefont
  {Yao}(2023)}]{müller2023simple}%
  \BibitemOpen
  \bibfield  {author} {\bibinfo {author} {\bibfnamefont {B.}~\bibnamefont
  {Müller}}\ and\ \bibinfo {author} {\bibfnamefont {X.}~\bibnamefont {Yao}},\
  }\href@noop {} {\bibinfo {title} {Simple hamiltonian for quantum simulation
  of strongly coupled 2+1d su(2) lattice gauge theory on a honeycomb lattice}}
  (\bibinfo {year} {2023}),\ \Eprint {https://arxiv.org/abs/2307.00045}
  {arXiv:2307.00045 [quant-ph]} \BibitemShut {NoStop}%
\bibitem [{iqu(2023)}]{iqus}%
  \BibitemOpen
  \href {https://iqus.uw.edu} {\bibinfo {title} {https://iqus.uw.edu}}
  (\bibinfo {year} {2023})\BibitemShut {NoStop}%
\bibitem [{uw_(2023{\natexlab{a}})}]{uw_phys}%
  \BibitemOpen
  \href {https://phys.washington.edu} {\bibinfo {title}
  {https://phys.washington.edu}} (\bibinfo {year}
  {2023}{\natexlab{a}})\BibitemShut {NoStop}%
\bibitem [{uw_(2023{\natexlab{b}})}]{uw_artsci}%
  \BibitemOpen
  \href {https://artsci.washington.edu} {\bibinfo {title}
  {https://artsci.washington.edu}} (\bibinfo {year}
  {2023}{\natexlab{b}})\BibitemShut {NoStop}%
\bibitem [{\citenamefont {Inc.}()}]{Mathematica}%
  \BibitemOpen
  \bibfield  {author} {\bibinfo {author} {\bibfnamefont {W.~R.}\ \bibnamefont
  {Inc.}},\ }\href {https://www.wolfram.com/mathematica} {\bibinfo {title}
  {Mathematica, {V}ersion 13.2}},\ \bibinfo {note} {champaign, IL,
  2022}\BibitemShut {NoStop}%
\bibitem [{\citenamefont {Van~Rossum}\ and\ \citenamefont
  {Drake}(2009)}]{python}%
  \BibitemOpen
  \bibfield  {author} {\bibinfo {author} {\bibfnamefont {G.}~\bibnamefont
  {Van~Rossum}}\ and\ \bibinfo {author} {\bibfnamefont {F.~L.}\ \bibnamefont
  {Drake}},\ }\href@noop {} {\emph {\bibinfo {title} {Python 3 Reference
  Manual}}}\ (\bibinfo  {publisher} {CreateSpace},\ \bibinfo {address} {Scotts
  Valley, CA},\ \bibinfo {year} {2009})\BibitemShut {NoStop}%
\end{thebibliography}%

\clearpage
\appendix

\section{Constant-Field Method Error Rate Derivation}
\label{app:constant_field_drive_appendix}

For the choice of $\vec{B}(t)$ given by \eqref{eq:bspiral}, time evolution in the interaction picture is given by 
\begin{equation*}
    U_\text{Ising}(t,0)=\text{exp}_{\mathcal{T}}\left\{-i\int_0^tdt' H_\text{int}(t')\right\},
\end{equation*}
where $H_\text{int}(t)=\frac{1}{2}\sum_{i\neq j}J_{ij}Z_i(t)Z_{j}(t)$, $Z_i(t)=2\vec{e}(t)\cdot\vec{S}_i$, and
\begin{equation}
    \vec{e}(t) = \frac{1}{3}
    \begin{pmatrix}
        2\sqrt{2}\sin^2\left(\frac{\Omega t}{2}\right)\\
        -\sqrt{6}\sin{\Omega t}\\
        1+2\cos{\Omega t}\\
    \end{pmatrix}.
\end{equation}
The error rate at Floquet periods $\tau_{\mathcal{C}_1}=\frac{2\pi}{\Omega}$ (i.e., the period of $\vec{e}(t)$) is given by
\begin{equation}
    ER_{\mathcal{C}_1}(\tau_{\mathcal{C}_1}) = \frac{1}{\tau_{\mathcal{C}_1}}\specnorm{\text{exp}_{\mathcal{T}}\left\{-i\int_0^{\tau_{\mathcal{C}_1}}dt H_\text{int}(t)\right\} - \text{exp}\left\{-i\frac{\tau_{\mathcal{C}_1}}{3}H_{XXX}\right\}}.
\end{equation}
To get the tightest possible bounds on $ER_{\mathcal{C}_1}$ with minimal application of triangle inequalities, the two propagators are organized in powers of $\Omega$ (equivalently powers of $\tau_{\mathcal{C}_1}$). For the target evolution,
\begin{equation*}
    \text{exp}\left\{-i\frac{\tau_{\mathcal{C}_1}}{3}H_{XXX}\right\} = 1-i\frac{\tau_{\mathcal{C}_1}}{3}H_{XXX}-\frac{\tau_{\mathcal{C}_1}^2}{18}H_{XXX}^2+\mathit{O}(\tau_{\mathcal{C}_1}^3).
\end{equation*}
Working up to second-order in a Magnus expansion, the Floquet evolution gives:
\begin{align}
    \text{exp}_{\mathcal{T}}\left\{-i\int_0^{\tau_{\mathcal{C}_1}} dt H_\text{int}(t)\right\} &= e^{-i\sum_{k=1}^{\infty}\Omega_k(\tau_{\mathcal{C}_1})}\\
    &= 1-\left(i\Omega_1(\tau_{\mathcal{C}_1})-i\Omega_2(\tau_{\mathcal{C}_1}) + \frac{1}{2}\Omega^2_1(\tau_{\mathcal{C}_1})\right)+\mathit{O}(\tau_{\mathcal{C}_1}^3),\label{eq:magConst}
\end{align}
where
\begin{align}
    \Omega_1(\tau_{\mathcal{C}_1}) &= \int_0^{\tau_{\mathcal{C}_1}}dt H_\text{int}(t)=\frac{\tau_{\mathcal{C}_1}}{3}H_{XXX},\\
    \Omega_2(\tau_{\mathcal{C}_1}) &= \frac{1}{2}\int_0^{\tau_{\mathcal{C}_1}}dt_1\int_0^{t_1}dt_2 [H_\text{int}(t_1),H_\text{int}(t_2)].\label{eq:magConstSecondOrder}
\end{align}

Evaluating Eq. \eqref{eq:magConstSecondOrder} and defining $\vec{e}_i=\vec{e}(t_i)$
\begin{multline}
    \Omega_2(\tau_{\mathcal{C}_1}) = \\
   -i\int_0^{\tau_{\mathcal{C}_1}}dt_1\int_0^{t_1}dt_2 \left(8\sum_{i\neq j\neq k}J_{ij}J_{jk}\left(\vec{e}_1\cdot\vec{S}_i\right)\left((\vec{e}_1\times\vec{e}_2)\cdot\vec{S}_j\right)\left(\vec{e}_2\cdot\vec{S}_m\right)+2\sum_{i\neq j}J_{ij}^2\left(\vec{e}_1\cdot\vec{e}_2\right)\left(\vec{e}_1\times\vec{e}_2\right)\cdot\vec{S}_j\right),
\end{multline}
and the following definitions to simplify expressions,
\begin{align}
    \Lambda_{\alpha\beta\gamma}(t_1,t_2) &= \left(\vec{e}_1\right)_{\alpha}\left(\vec{e}_1\times\vec{e}_2\right)_{\beta}\left(\vec{e}_2\right)_{\gamma},\\
    \tilde{\Lambda}_{\alpha}(t_1,t_2) &= \left(\vec{e}_1\cdot\vec{e}_2\right)\left(\vec{e}_1\times\vec{e}_2\right)_{\alpha},
\end{align}
the $\Omega_1$ terms exactly cancel the target Heisenberg evolution, giving the error rate at leading-order in $\tau_{\mathcal{C}_1}$ as:
\begin{align}
    ER_{\mathcal{C}_1}(\tau_{\mathcal{C}_1}) &= \frac{1}{\tau_{\mathcal{C}_1}}\specnorm{\Omega_2(\tau_{\mathcal{C}_1})} + \mathit{O}(\tau_{\mathcal{C}_1}^3)\\
    &\leq \frac{1}{2 \pi \Omega} \left(\specnorm{\int_0^1\int_0^{t_1}dt_1dt_2\Lambda(t_1,t_2)}\left(\sum_{i\neq j\neq k}J_{ij}J_{jk}\right)+\specnorm{\int_0^1\int_0^{t_1}dt_1dt_2\tilde{\Lambda}(t_1,t_2)}\left(\sum_{i\neq j}J_{ij}^2\right)\right)\\
    &= \frac{2\epsilon}{\pi}\left(\frac{\sqrt{142+24\pi^2}}{18}\left(\sum_{i\neq j\neq k}J_{ij}J_{jk}\right)+\frac{1}{3}\left(\sum_{i\neq j}J_{ij}^2\right)\right)\label{eq:e_mag}\\
    &= \mathit{O}(\epsilon N).
\end{align}

\section{Derivation of the $\frac{\pi}{2}$-Pulse Error}\label{app:pi_2_pulse_error}
As noted in the main body of the paper, the global rotation gates cannot be perfectly implemented due to the persistent Ising interaction. The deviation from the ideal behavior is found using a Magnus expansion, following Appendix A of Ref. \cite{1c}:
\begin{align}
    \Upsilon^{\pm}_X(\epsilon) &= R_X\left(\mp\frac{\pi}{2}\right)R_X^{\pm}(\epsilon) = \text{exp}\left(-i\sum_{k=1}{\chi^{\pm}_k\epsilon^k}\right)\label{secondordergates1},\\
    \Upsilon^{\pm}_Y(\epsilon) &= R_Y\left(\mp\frac{\pi}{2}\right)R_Y^{\pm}(\epsilon) = \text{exp}\left(-i\sum_{k=1}{\nu^{\pm}_k\epsilon^k}\right)\label{secondordergates2}.
\end{align}
To derive the error rate bounds, it is only necessary to work to $\mathit{O}(\epsilon^2)$ in $ \Upsilon^{\pm}_{X,Y}$. The expressions for $\chi_{1,2}^{\pm}$ and $\nu_{1,2}^{\pm}$ are:
\begin{align}
    \chi^{\pm}_1 &= \int_0^1dt e^{\pm\frac{i\pi t}{4}\sum_iX_i}\left(\sum_{i<j}J_{ij}Z_iZ_j\right)e^{\mp\frac{i\pi t}{4}\sum_iX_i}\nonumber\\
    &= \sum_{i<j}J_{ij}\left(\frac{Y_iY_j+Z_iZ_j}{2}\pm\frac{Y_iZ_j+Z_iY_j}{\pi}\right),\\
    \nu^{\pm}_1 &= \sum_{i<j}J_{ij}\left(\frac{X_iX_j+Z_iZ_j}{2}\mp\frac{X_iZ_j+Z_iX_j}{\pi}\right),
\end{align}
and
\begin{align}
    \chi^{\pm}_2 &= \frac{-i}{2}\int_0^1\int_0^{t_1}dt_1dt_2 \left[e^{\pm\frac{i\pi t_1}{4}\sum_iX_i}\left(\sum_{i<j}J_{ij}Z_iZ_j\right)e^{\mp\frac{i\pi t_1}{4}\sum_iX_i},e^{\pm\frac{i\pi t_2}{4}\sum_iX_i}\left(\sum_{i<j}J_{ij}Z_iZ_j\right)e^{\mp\frac{i\pi t_2}{4}\sum_iX_i}\right] \nonumber\\
    &= -i\sum_{i<j,m<n}J_{ij}J_{mn}\left(\frac{1}{\pi^2}\left[Y_iY_j,Z_mZ_n\right]\pm\frac{1}{8\pi}\left[Y_iY_j-Z_iZ_j,Z_mY_n+Y_mZ_n\right]\right)\nonumber\\
    &= \frac{1}{\pi}\sum_{i,m\neq j}J_{ij}J_{jm}\frac{Y_iX_jZ_m+X_jZ_mY_i}{\pi}\pm\frac{Y_iY_mX_j+Z_iX_jZ_m+Y_mX_jY_i+X_jZ_mZ_i}{8},\\
    \nu_2^{\pm} &= -\frac{1}{\pi}\sum_{i,m\neq j}J_{ij}J_{jm}\frac{X_iY_jZ_m+Y_jZ_mX_i}{\pi}\mp\frac{X_iX_mY_j+Z_iY_jZ_m+X_mY_jX_i+Y_jZ_mZ_i}{8}.
\end{align}

\section{First-Order Product Formula Error}
\label{app:first_order_error_analytics}
As an example, the analysis of Section \ref{sec:error_analysis} is applied to the product formulas $\mathcal{S}_{1/2}$ and $\mathcal{S}_1$, with the specific choice $H_\text{tar}=H_{XXX}$. The result of this analysis will be Eqs. \eqref{eq:approximate_optimal_step_sizes} and \eqref{eq:approximate_Trotter_error_rates}. The error rate formulas are
\begin{align}
    ER_{\mathcal{S}_{1/2}} &\leq \frac{1}{\tau_{\mathcal{S}_{1/2}}(t,\epsilon)}\left(t^2\specnorm{\mathcal{E}_{\mathcal{S}_{1/2};t^2}}+\epsilon\specnorm{\mathcal{E}_{\mathcal{S}_{1/2};\epsilon}}\right)\label{eq:ER_S1t},\\
    ER_{\mathcal{S}_1} &\leq \frac{1}{\tau_{\mathcal{S}_1}(t,\epsilon)}\left(t^2\specnorm{\mathcal{E}_{\mathcal{S}_1;t^2}}+\epsilon^2\specnorm{\mathcal{E}_{\mathcal{S}_1;\epsilon^2}}+t\epsilon\specnorm{\mathcal{E}_{\mathcal{S}_1;t\epsilon}}\right).\label{eq:ER_S1}
\end{align}
The choices of $t$ that optimizes the error rates are
\begin{align}
    t_{\mathcal{S}_{1/2}*} &= \epsilon^{\frac{1}{2}}\sqrt{\frac{\specnorm{\mathcal{E}_{\mathcal{S}_{1/2};\epsilon}}}{\specnorm{\mathcal{E}_{\mathcal{S}_{1/2};t^2}}}}+\mathit{O}(\epsilon),\label{eq:topt_S1t}\\
    t_{\mathcal{S}_1*} &= \epsilon\left(\sqrt{4-\frac{2\specnorm{\mathcal{E}_{\mathcal{S}_1;t\epsilon}}-\specnorm{\mathcal{E}_{\mathcal{S}_1;\epsilon^2}}}{\specnorm{\mathcal{E}_{\mathcal{S}_1;t^2}}}}-2\right).\label{eq:topt_S1}
\end{align}
The corresponding error rate bounds are (valid for $t_{\mathcal{S}_{1/2}*},t_{\mathcal{S}_1*}>0$) are given by
\begin{align}
    ER_{\mathcal{S}_{1/2}} &\leq 2\sqrt{\specnorm{\mathcal{E}_{\mathcal{S}_{1/2};t^2}}\specnorm{\mathcal{E}_{\mathcal{S}_{1/2};\epsilon}}}\epsilon^{\frac{1}{2}}+\mathit{O}(\epsilon),\\
    ER_{\mathcal{S}_1} &\leq \left(2\sqrt{\specnorm{\mathcal{E}_{\mathcal{S}_1;t^2}}\left(4\specnorm{\mathcal{E}_{\mathcal{S}_1;t^2}}+\specnorm{\mathcal{E}_{\mathcal{S}_1;\epsilon^2}}-2\specnorm{\mathcal{E}_{\mathcal{S}_1;t\epsilon}}\right)}+\specnorm{\mathcal{E}_{\mathcal{S}_1;t\epsilon}}-4\specnorm{\mathcal{E}_{\mathcal{S}_1;t^2}}\right)\epsilon.
\end{align}

An important point to consider is that Eq. \eqref{eq:topt_S1} suggests that $t_{\mathcal{S}_1*}$ may be negative, while in both Eq. \eqref{eq:ER_S1t} and \eqref{eq:ER_S1}, $t_{\mathcal{S}_{1/2}*}$ and $t_{\mathcal{S}_1*}$ were assumed to be positive numbers. This apparent contradiction simply reflects the fact that this is an optimization problem whose solution is on the boundary $t_{\mathcal{S}_1*}=0$. It is now clear which error terms must be calculated to recover the optimal step sizes and the error rates, which is done below.

\subsection{First-Order Error in $\epsilon$}
At leading-order in $\epsilon$ and $t$, Eqs. \eqref{eq:firstpseudo} and \eqref{eq:firsttrue} take the form
\begin{align}
    \mathcal{S}_{1/2} &= \mathbbm{1} - i\sum_{i<j}J_{ij}\left(t_xX_iX_j+t_yY_iY_j+t_zZ_iZ_j\right) - i\epsilon\sum_{i<j}J_{ij}\left(X_iX_j+Y_iY_j+2Z_iZ_j+\frac{2}{\pi}\left(Y_iZ_j+Z_iY_j-X_iZ_j-Z_iX_j\right)\right),\\
    \mathcal{S}_1 &=R_{Z}(\pi)\left(\mathbbm{1} - i\sum_{i<j}J_{ij}\left(t_xX_iX_j+t_yY_iY_j+t_zZ_iZ_j\right)- i\epsilon\sum_{i<j}J_{ij}\left(X_iX_j+Y_iY_j+2Z_iZ_j\right)\right).
\end{align}
The leading-order target evolution is $\mathbbm{1}-i\tau_U(t,\epsilon)H_{XXX}$, and matching these evolutions requires $t_x,t_y=\epsilon+t$ and $t_z=t$. These choices imply that $\tau_U(t,\epsilon)=2\epsilon+t$. With these choices, the leading-order Heisenberg evolution is the same for $\mathcal{S}_{1/2}$ and $\mathcal{S}_1$, implying that $\tau_{\mathcal{S}_{1/2}}=\tau_{\mathcal{S}_1}$. The individual terms in the expansion \eqref{eq:pftaylor} are
\begin{align}  
    \mathcal{S}_{1/2;t} &= -i\sum_{i<j}J_{ij}\left(X_iX_j+Y_iY_j+Z_iZ_j\right),\\
    \mathcal{S}_{1;t} &= R_{Z}(\pi)\left(-i\sum_{i<j}J_{ij}\left(X_iX_j+Y_iY_j+Z_iZ_j\right)\right),\\
    \mathcal{S}_{1/2;\epsilon} &=-2i\sum_{i<j}J_{ij}\left(X_iX_j+Y_iY_j+Z_iZ_j+\frac{Y_iZ_j+Z_iY_j-X_iZ_j-Z_iX_j}{\pi}\right),\\
    \mathcal{S}_{1;\epsilon} &= R_{Z}(\pi)\left(-2i\sum_{i<j}J_{ij}\left(X_iX_j+Y_iY_j+Z_iZ_j\right)\right).
\end{align}
The error terms that contribute to the error rates at this order are
\begin{align}
    \mathcal{E}_{\mathcal{S}_1;t},\mathcal{E}_{\mathcal{S}_1;\epsilon} &= 0,\\
    \mathcal{E}_{\mathcal{S}_{1/2};t} &= 0,\\
    \mathcal{E}_{\mathcal{S}_{1/2};\epsilon} &= -\frac{2i}{\pi}\sum_{i<j}J_{ij}\left((Y_i-X_i)Z_j+Z_i(Y_j-X_j)\right).
\end{align}
There is $\mathit{O}(\epsilon)$ error appearing for $\mathcal{S}_{1/2}$ which causes the $t\propto\sqrt{\epsilon}$ scaling. To find the optimal $t_*$ for $\mathcal{S}_1$, higher-order terms must be calculated.

\subsection{Higher-Order Error in $\epsilon, t$}
The error terms that contribute to the error rates are (up to a factor of $R_{Z}(\pi)$)
\begin{align}
    \mathcal{E}_{\mathcal{S}_1;t^2} &= 8i\sum_{i\neq j\neq k}J_{ij}J_{jk}\left(X_iY_jZ_k-Y_iX_jZ_k-Y_iZ_jX_k\right),\\
    \mathcal{E}_{\mathcal{S}_1;\epsilon^2} &= 2i\sum_{i\neq j\neq k}J_{ij}J_{jk}\left(8Y_iZ_jX_k-4(Y_iX_j+X_iY_j)Z_k+\frac{7(Y_iX_jY_k-X_iY_jX_k)+Z_i(Y_j-X_j)Z_k}{\pi}\right)\nonumber\\
    &\quad+\frac{12 i}{\pi}\sum_{i\neq j}J_{ij}^2(X_i-Y_i),\\
    \mathcal{E}_{\mathcal{S}_1;t\epsilon} &= 8i\sum_{i\neq j\neq k}J_{ij}J_{jk}\left(2Y_iZ_jX_k-\frac{3}{2}X_iY_jZ_k+\frac{1}{2}Y_iX_jZ_k+\frac{Y_iX_jY_k-X_iY_jX_k}{\pi}\right)+\frac{8i}{\pi}\sum_{i\neq j}J_{ij}^2(X_i-Y_i).
\end{align}

\subsection{Evaluation of Prefactors}
\label{app:prefactor_eval}
\begin{figure*}
\centering
\begin{minipage}{0.84\linewidth}
\subfloat[]{
\includegraphics[width=0.49\linewidth]{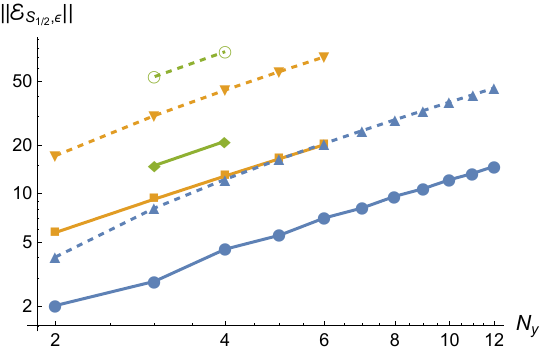}
}
\subfloat[]{
\includegraphics[width=0.49\linewidth]{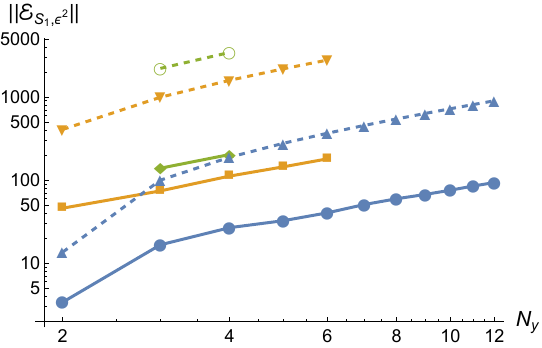}
}
\hspace{0mm}
\subfloat[]{
\includegraphics[width=0.49\linewidth]{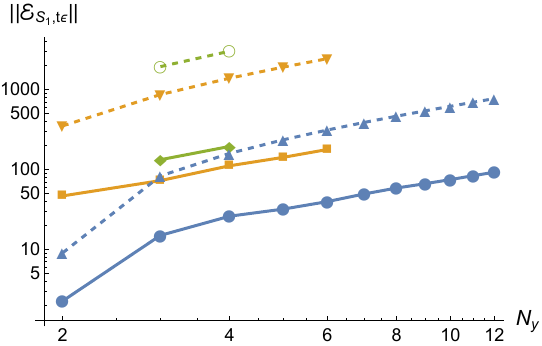}
}
\subfloat[]{
\includegraphics[width=0.49\linewidth]{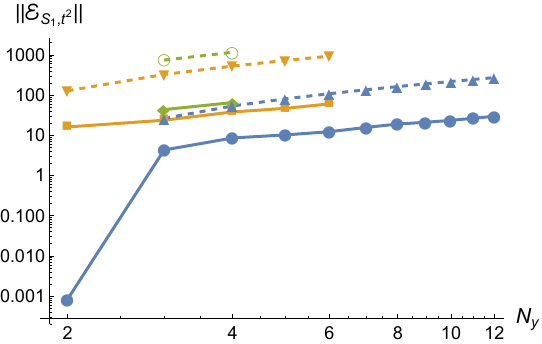}
}
\end{minipage}
\begin{minipage}{0.14\linewidth}

\subfloat{
\includegraphics[width=\linewidth]{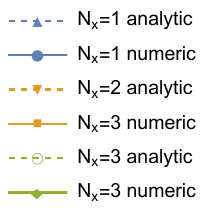}
}
\end{minipage}
\caption{Scaling of the contributions $\mathcal{E}$ to the error from different sources in a single Trotter step of the pulse sequences $\mathcal{S}_{1/2}$ and $\mathcal{S}_1$ as a function of system geometry. $N_y$ is varied on the $x$ axis and each line represents the contribution to the error with a different value of $N_x$. Here the analytic bounds of Eqs. \eqref{eq:E1_tilde_eps_norm} - \eqref{eq:E1_t2_norm} are compared to values obtained by numerically evaluating the norm of Eq. \eqref{eq:deriv_terms} for each type of contribution. Note that in the latter three panels, the $N_x=1, N_y=1$ points are much lower than the rest of the $N_x=1$ lines because of the fact that there are three-body contributions to the error for these terms; when the system has less than three sites, these contributions are absent, leading to a much smaller error.}
\label{fig:individual_term_scalings}
\end{figure*}

The scaling prefactors in Eqs. \eqref{eq:topt_S1t} and \eqref{eq:topt_S1} require evaluation of the following quantities:
\begin{align}
    \specnorm{\mathcal{E}_{\mathcal{S}_{1/2};\epsilon}} &\leq \frac{4\sqrt{2}}{\pi}J_1,\label{eq:E1_tilde_eps_norm}\\
    \specnorm{\mathcal{E}_{\mathcal{S}_1;\epsilon^2}} &\leq 28J_3+\frac{12\sqrt{2}}{\pi}J_2,\label{eq:E1_eps2_norm}\\
    \specnorm{\mathcal{E}_{\mathcal{S}_1;t\epsilon}} &\leq 25J_3+\frac{8\sqrt{2}}{\pi}J_2,\label{eq:E1_teps_norm}\\
    \specnorm{\mathcal{E}_{\mathcal{S}_{1/2};t^2}} = \specnorm{\mathcal{E}_{\mathcal{S}_1;t^2}} &\leq 8\sqrt{3+2\sqrt{3}}J_3,\label{eq:E1_t2_norm}
\end{align}
where the factors $J_1,J_2,J_3$ are defined in Eqs. \eqref{eq:jsums1} - \eqref{eq:jsums3}. As explained in Section \ref{sec:results}, numerics are used to see the values of the different $\mathcal{E}$ terms in practice. Eqs. \eqref{eq:E1_tilde_eps_norm} - \eqref{eq:E1_t2_norm} are compared to values obtained numerically as a function of system size in Fig. \ref{fig:individual_term_scalings}. 

With these expressions, calculating Eqs. \eqref{eq:approximate_optimal_step_sizes} and \eqref{eq:approximate_Trotter_error_rates} is now straightforward.
\begin{align}
    t_{\mathcal{S}_{1/2}*} &\approx .3\sqrt{\frac{J_1}{J_3}}\epsilon^{\frac{1}{2}}+\mathit{O}(\epsilon)\label{eq:topt_S1t_with_constraint},\\
    t_{\mathcal{S}_1*} &= 0\label{eq:topt_S1_with_constraint}.
\end{align}
Note that calculating $t_{\mathcal{S}_1*}$ using the bounds derived gives a negative number, which is why Eq. \eqref{eq:topt_S1_with_constraint} is zero; this conclusion is supported by numerics.
The corresponding optimal error rate bounds are
\begin{align}
    ER_{\mathcal{S}_{1/2}} &\leq 12.2\sqrt{J_1J_3}\epsilon^{\frac{1}{2}}+\mathit{O}(\epsilon),\label{eq:ER_S1t_with_constraint}\\
    ER_{\mathcal{S}_1} &\leq (14J_3+2.7J_2)\epsilon.\label{eq:ER_S1_with_constraint}
\end{align}

\section{Analysis of the Interaction Sums}
Throughout this paper, various sums over the Ising interaction terms appear. These sums encapsulate the geometry of the device used. $N_x$ is the horizontal extent of the atoms and $N_y$ is the vertical extent. These sums depend very cleanly on total number of atoms $N=N_xN_y$. In the following analysis it is assumed that the configuration of atoms in the experimental apparatus is two-dimensional and rectangular (i.e. $N_x,N_y\geq 2$). The sums analyzed are:
\begin{align}
    J_1(N) &= \sum_{i\neq j}J_{ij} = \tilde{J}_{11}N-\tilde{J}_{12},\label{eq:jsums1}\\
    J_2(N) &= \sum_{i\neq j}J_{ij}^2 = \tilde{J}_{21}N-\tilde{J}_{22},\label{eq:jsums2}\\
    J_3(N) &= \sum_{i\neq j\neq k}J_{ij}J_{jk} = \tilde{J}_{31}N-\tilde{J}_{32}.\label{eq:jsums3}
\end{align}
By explicit numerical evaluation of the sums for the considered system sizes, the following fits are found, which are plotted in Fig. \ref{fig:jplots}:

\begin{figure*}
\centering
\subfloat{
\includegraphics[width=0.49\linewidth]{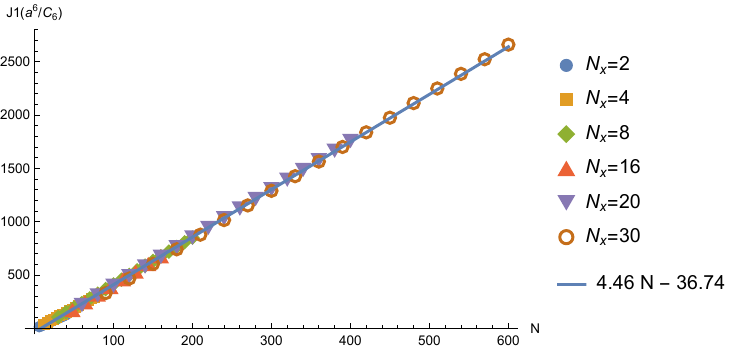}
\label{fig:appE1}
}
\subfloat{
\includegraphics[width=0.49\linewidth]{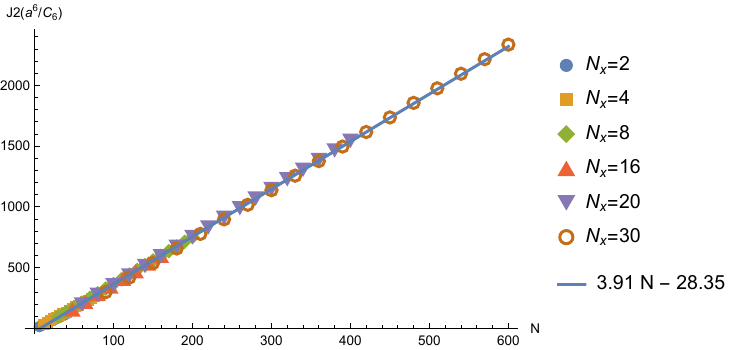}
\label{fig:appE2}
}
\hspace{0mm}
\subfloat{
\includegraphics[width=0.49\linewidth]{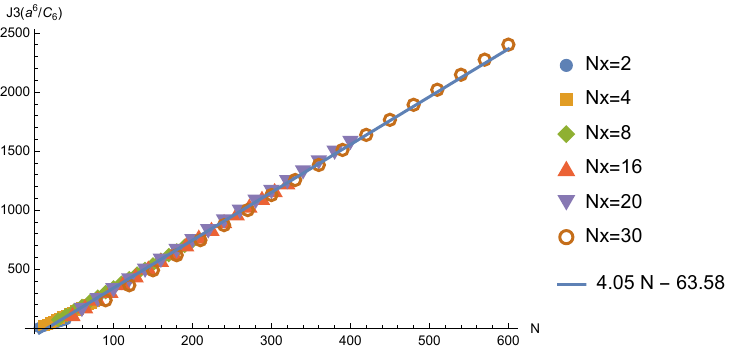}
\label{fig:appE3}
}
\caption{$J_1,J_2,J_3$ sums evaluated for explicit array configurations with fits. The extent of the array in the $x$ direction is $N_x=20$ atoms. While there is a deviation for smaller array sizes, a linear scaling is still seen.}
\label{fig:jplots}
\end{figure*}

\begin{align*}
    \tilde{J}_{11} &= 4.46,\qquad\tilde{J}_{12}=-36.74,\\
    \tilde{J}_{21} &= 3.91,\qquad\tilde{J}_{22}=28.35,\\
    \tilde{J}_{31} &= 4.05,\qquad\tilde{J}_{32}=-63.58.\\
\end{align*}

To see how this scaling arises, consider the $J_1$ sum, which may be written as:
\begin{align*}
    J_1 &= \frac{C_6}{a^6}\sum^{N_x,N_y}_{\substack{x_1=1 \\ y_1=1}}\sum^{N_x,N_y}_{\substack{x_2=x_1+1\\y_2=y_1+1}}\frac{1}{\left((x_2-x_1)^2+(y_2-y_1)^2\right)^3}\\
    &=\frac{C_6}{a^6}\sum^{N_x,N_y}_{\substack{x_1=1 \\ y_1=1}}\sum^{\substack{N_x-x_1 \\ N_y-y_1}}_{\substack{x'_2=1\\y'_2=1}}\frac{1}{\left((x'_2)^2+(y'_2)^2\right)^3}\\
    &\leq \frac{C_6}{a^6}\sum^{N_x,N_y}_{\substack{x_1=1 \\ y_1=1}}\int_1^{\infty}\frac{dx'_2dy'_2}{\left((x'_2)^2+(y'_2)^2\right)^3}.
\end{align*}
The integral converges to some constant, so we see (for $N=N_xN_y$) that
\begin{equation*}
    J_1 = \mathit{O}(Na^{-6}).
\end{equation*}
Scalings for $J_2$ and $J_3$ may be shown in a similar fashion. This scaling may be understood intuitively in the following manner. The power law interactions are rapidly decaying and so the dominant contribution to these sums will be given by a single sum over an effective nearest-neighbor interaction. For a square array, this implies a $\sim 4N$ scaling, which is seen in Fig. \ref{fig:jplots}. The deviation in the fit values from the ideal $4N$ scaling stem from the fact that the fits include values from geometrically thin arrays (e.g., $2\times 20$), where boundary interactions play a larger role.
\end{document}